\documentclass{aa}  % [useAMS]

\makeatletter
\renewcommand*\aa@pageof{, page \thepage{} of \pageref*{LastPage}}
\makeatother

\usepackage{times}
\usepackage{footmisc}
\usepackage{graphicx,multicol}
\usepackage{natbib}

\usepackage{amsmath,amsfonts,bm}
\usepackage{amsthm}
\bibpunct{(}{)}{;}{a}{}{,} % to follow the A&A style
\usepackage{txfonts}
\usepackage{relsize}
\usepackage[draft=False, colorlinks=true, citecolor=MidnightBlue, urlcolor=MidnightBlue, linkcolor=MidnightBlue, linktoc=all]{hyperref}
\usepackage{cleveref}
\usepackage[dvipsnames]{xcolor}

\newcommand\Tstrut{\rule{0pt}{2.9ex}}         % "top" strut
\newcommand\Bstrut{\rule[-1.2ex]{0pt}{0pt}}   % "bottom" strut
\newcommand\TBstrut{\Tstrut\Bstrut}           % "top and bottom" strut

\begin{document}

\title{Mixing is easy: New insights for cosmochemical evolution from pre-stellar core collapse}

\author{Asmita~Bhandare\inst{\ref{lyon},\ref{lmu}}
\fnmsep \thanks{\href{asmita.bhandare@lmu.de}{asmita.bhandare@lmu.de}}\orcid{0000-0002-1197-3946}
\and Beno\^{i}t~Commer\c{c}on\inst{\ref{lyon}} \orcid{0000-0003-2407-1025}
\and Guillaume~Laibe\inst{\ref{lyon}}\orcid{0000-0002-3116-3166}
\and Mario~Flock\inst{\ref{mpia}} \orcid{0000-0002-9298-3029}
\and Rolf~Kuiper\inst{\ref{duisburg}} \orcid{0000-0003-2309-8963}
\and Thomas~Henning\inst{\ref{mpia}} \orcid{0000-0002-1493-300X}
\and Andrea~Mignone\inst{\ref{torino}}\orcid{0000-0002-8352-6635}
\and Gabriel-Dominique~Marleau\inst{\ref{duisburg},\ref{tuebingen},\ref{mpia},\ref{bern}} \orcid{0000-0002-2919-7500}
}

\institute{
ENS de Lyon, CRAL UMR5574, Universit\'e Claude Bernard Lyon 1, CNRS, Lyon, 69007, France
\label{lyon}
  \and
Universit\"ats-Sternwarte, Fakult\"at f\"ur Physik, Ludwig-Maximilians-Universit\"at M\"unchen, Scheinerstr.~1, 81679 M\"unchen, Germany \hspace{-1em} %%to fix the empty line following otherwise
\label{lmu}
  \and
Max-Planck-Institut f\"ur Astronomie, K\"onigstuhl 17, 69117 Heidelberg, Germany
\label{mpia}
  \and
Fakult\"at f\"ur Physik, Universit\"at Duisburg--Essen, Lotharstra{\ss}e 1, 47057 Duisburg, Germany
\label{duisburg}
  \and
Dipartimento di Fisica, Universit\`a di Torino via Pietro Giuria 1, 10125 Torino, Italy
\label{torino}
  \and
Instit\"ut f\"ur Astronomie und Astrophysik, Universit\"at T\"ubingen, Auf der Morgenstelle 10, 72076 T\"ubingen, Germany
\label{tuebingen}
  \and
Physikalisches Institut, Universit\"at Bern, Gesellschaftsstr.~6, 3012 Bern, Switzerland
\label{bern}
}

%\date{\today}
\date{Submitted: February 13, 2024; Accepted: April 13, 2024}

\abstract{Signposts of early planet formation are ubiquitous in substructured young discs. Dense, hot, and high-pressure regions that formed during the gravitational collapse process, integral to star formation,  facilitate dynamical mixing of dust within the protostellar disc. This provides an incentive to constrain the role of gas and dust interaction and resolve potential zones of dust concentration during star and disc formation stages.}
{We explore whether the thermal and dynamical conditions that developed during protostellar disc formation can generate gas flows that efficiently mix and transport the well-coupled gas and dust components.}
{We simulated the collapse of dusty molecular cloud cores with the hydrodynamics code \texttt{PLUTO} augmented with radiation transport and self-gravity. We used a two-dimensional axisymmetric geometry and followed the azimuthal component of the velocity. The dust was treated as Lagrangian particles that are subject to drag from the gas, whose motion is computed on a Eulerian grid. We considered 1, 10, and 100~micron-sized neutral, spherical dust grains. Importantly, the equation of state accurately includes molecular hydrogen dissociation. We focus on molecular cloud core masses of 1 and 3~$M_{\odot}$ and explore the effects of different initial rotation rates and cloud core sizes.} 
{Our study underlines mechanisms for the early transport of dust from the inner hot disc regions via the occurrence of two transient gas motions, namely meridional flow and outflow. The vortical flow fosters dynamical mixing and retention of dust, while the thermal pressure driven outflow replenishes dust in the outer disc. Notably, these phenomena occur regardless of the initial cloud core mass, size, and rotation rate.}
{Young dynamical precursors to planet-forming discs exhibit regions with complex hydrodynamical gas features and high-temperature structures. These can play a crucial role in concentrating dust for subsequent growth into protoplanets. Dust transport, especially, from sub-au scales surrounding the protostar to the outer relatively cooler parts, offers an efficient pathway for thermal reprocessing during pre-stellar core collapse.}

\keywords{Stars: formation -- Methods: numerical -- Hydrodynamics -- Radiative transfer -- Gravitation -- Equation of state}

\titlerunning{Gas \& dust dynamics during pre-stellar core collapse}

\authorrunning{A. Bhandare et al.}

\maketitle

\graphicspath{{./Plots/}}

%%%%%%%%%%%%%%%%%%%%%%%%%%%%%%%%%%%%%%%%%%%%%%%%%%%%%%%%%%%%%%%%%%
\section{Introduction}
\label{sec:intro}

The advent of unprecedented circumstellar disc surveys with the Atacama Large Millimetre/submillimetre Array (\texttt{ALMA}) and with facilities like the Spectro-Polarimetric High-contrast Exoplanet REsearch on the Very Large Telescope \citep[\texttt{SPHERE/VLT}, e.g.][]{Garufi2022} as well as observations with the NOrthern Extended Millimetre Array (\texttt{NOEMA}) highlighting environmental effects \citep[e.g.][]{Pineda2020} are starting to shift the focus from `how' planets form to `when' they form. Over the last decade, with dedicated disc surveys such as Disc Substructures at High Angular Resolution Project \citep[\texttt{DSHARP};][]{Andrews2018} and Molecules with \texttt{ALMA} at Planet-forming Scales \citep[\texttt{MAPS};][]{Oberg2021}, there has been growing observational evidence for substructures such as rings, gaps, and spiral arms via dust continuum emission from Class II discs \citep[see also][]{Perez2016, Paneque-Carreno2021}. More recently, the Early Planet Formation in Embedded Discs survey \citep[\texttt{eDisc};][]{Ohashi2023} has revealed substructure even in young Class 0 and~I discs \citep[see also][]{HLTau-ALMA2015, Sheehan2018, Tobin2020, Podio2020, Segura-Cox2020, Sheehan2020}. These data suggest, among other things, ongoing planet formation or the presence of already formed young protoplanets carving gaps in these discs \citep{Keppler2018, Mueller2018, Gratton2019, Haffert2019, Fedele2023, Hammond2023, Pinte2023}. The early formation of Jupiter in our Solar System is also widely proposed as an explanation for the isotopic dichotomy and radial separation between inner non-carbonaceous chondrites (NC) and outer carbonaceous chondrites (CC) \citep[e.g.][]{Kruijer2020}.

Magnetised, cold, dense, gaseous, and dusty molecular cloud cores provide the birth environment for stars, discs, and planets. A protostar and its surrounding protostellar disc are formed concurrently as an outcome of the process of gravitational collapse within these molecular cloud cores with either a non-zero net angular momentum \citep[e.g.][]{Chen2023} or with turbulent perturbations cascading from larger scales \citep{Pineda2023}. Early collapse phases lead to high density, pressure, and temperature regions within young protostellar discs \citep{Zhao2020, Tsukamoto2023PPVII}. These localised zones provide a unique setting for spatial mixing and thermal reprocessing of dust early on. Protostellar and protoplanetary discs are extremely dynamical throughout their evolution due to a multitude of physical and chemical processes \citep{Aikawa2022, Lesur2023, Bae2023}. The inclusion of dust and gas interactions, while treating the microphysics at `all' scales, will play a major role in understanding the earliest stages of star and disc formation and their subsequent evolution. This will also allow the exploration of the high-temperature sub-au disc region, which is known to be an ideal location to form refractory minerals like calcium-aluminium-rich inclusions (CAIs) and amoeboid olivine aggregates (AOAs). Surprisingly, CAIs and AOAs are the oldest ingredients mostly found in carbonaceous chondrites that were accreted in the outer Solar System, implying large-scale transport after their formation close to the proto-Sun \citep{Grossman1988, Krot2004, MacPherson2012, Ruzicka2012, Desch2018, Pignatale2019, Marrocchi2019, Hellmann2023, Piralla2023}.

There remain several open questions for a comprehensive characterisation of the multi-scale gravitational collapse within molecular cloud cores: for example, values of the initial magnetic field strength and orientation, angular momentum, turbulence, and dust size distribution in the cloud cores (i.e.~pre-stellar cores). These unknown initial conditions introduce various caveats in our current understanding of the observed star--disc--planet systems. Filling in the gaps in our current knowledge of formation and evolution of planet-forming discs will also help decipher cosmochemical signatures found in different parts of our Solar System. Zooming in on the smallest sub-au scales to unravel the effects of various complex physical processes, such as hydrodynamics; radiative transfer; phase transition, in particular molecular hydrogen ($\mathrm{H_2}$) dissociation and ionisation; turbulence; chemistry; magnetic fields; and gas--dust (de)coupling, remains a major challenge for both observations and theory \citep{Benisty2023, Tsukamoto2023PPVII}. Despite the recent wealth of high-resolution observations of young discs, high optical depths in the surrounding envelope pose several difficulties to resolve these embedded systems. On the other hand, newly developed theoretical and numerical models help elucidate and quantify these nascent phases of star and disc formation. 

Since the pioneering work by \citet{Larson1969}, there has been significant progress with numerical studies to follow the collapse using both grid-based simulations \citep[e.g.][]{Masunaga2000, Tomida2013, Vaytet2017, Vaytet2018, Bhandare2018, Bhandare2020, Ahmad2023} and smoothed particle hydrodynamics (SPH) methods \citep[e.g.][]{Whitehouse2006, Stamatellos2007, Bate2014, Wurster2018, Tsukamoto2020} as also described in the review by \citet{Teyssier2019}. However, despite the vast amount of theoretical work, several aspects of disc formation and evolution are far from being fully understood, mostly due to numerical and computational constraints in resolving the disc structure in the protostellar neighbourhood. These depend strongly on the initial physical conditions and numerical set-up of simulations \citep[e.g.][]{Machida2014}. There is no consensus on the formation time of the disc with respect to the first and second core formation. The relative importance of the presence of two discs around these two cores, which could eventually merge, is yet to be quantitatively determined. Furthermore, dust grains constitute the main ingredients to form planetary building blocks. Dust properties can also have significant cumulative dynamical effects on the thermal budget and kinematic structure of the disc. Even so, present-day numerical codes using sophisticated techniques rarely account for the role of drag forces and dust decoupling during protostar (second core) and protostellar disc formation stages. The dust size distribution and concentration in very young protostellar discs and its influence on the radial drift remain poorly constrained. These missing pieces are roadblocks on our way to solving the puzzle of early planet formation, which is strongly affected by the properties of the disc, the formation of substructures, and the transport of material within the disc.

Historically, the effects of dust dynamics have been neglected in star and disc formation studies based on the assumption of the diffuse interstellar medium (ISM) dust-to-gas mass ratio being one per cent. Most methods consider the dust and gas to be strongly coupled, and use a maximum dust size of up to 0.25~micron. The dust size distribution (number of dust grains per grain radius $a$) is assumed to be a power law, proportional to $a^{\mathrm{-3.5}}$ from Mathis, Rumpl \& Nordsieck \citep[hereafter MRN,][]{Mathis1977, Weingartner2001}. This has been recently observed in the Orion~A molecular cloud \citep{Uehara2021}. However, discoveries of mid-infrared light scattered from dense core regions \citep{Steinacker2010, Pagani2010, Steinacker2015, McClure2023} and evidence of lower dust opacity spectral index in dense cores compared to the ISM \citep{Henning1995, Miettinen2012, Zhang2021} and in the envelopes of young stellar objects \citep{Kwon2009, Miotello2014, Galametz2019} have shed light on the presence of larger dust grains of up to 1~micron in pre-stellar cores. Dust polarisation detection due to self-scattering processes at (sub)millimetre wavelengths have also provided indirect evidence for up to 100~micron grains in young discs \citep{Kataoka2015, Valdivia2019}. Additionally, a lower spectral slope in the spectral energy distribution derived from longer wavelength observations of spatially resolved young discs has been linked to the presence of millimetre-sized dust in the disc midplane \citep[e.g.][]{Carrasco-Gonzalez2016, Liu2017}. Details on different observational techniques for estimating dust sizes in discs can be found in \citet{Natta2007}, \citet{Testi2014}, and \citet{Birnstiel2023}. Modelling transitions of CO emission lines within the CO snowline suggest local enhancements of dust-to-gas mass ratios, especially in the disc midplane \citep[e.g.][]{Boneberg2016}. Additionally, there has been a growing indirect evidence for up to millimetre-sized dust lifted by outflows, winds, and jets \citep[e.g.][]{Bans2012, Ellerbroek2014, Duchene2024, Cacciapuoti2024}. 

The assumption of an MRN size distribution has also been recently challenged by numerical studies that account for interactions between gas and dust components through the drag force in the Epstein regime where the particle size is much smaller than the mean free path of the gas \citep{Epstein1924}. To date, this has been achieved either by treating the neutral dust as a pressureless fluid \citep[e.g.][]{Lebreuilly2020} or as Lagrangian particles \citep[e.g.][]{Bate2017, Cridland2022, Koga2022, Koga2023}. The current framework used in most simulations excludes an accurate treatment of the thermodynamics, which should be modelled by incorporating radiation transport and hydrogen dissociation and ionisation effects, especially important during the second collapse stage \citep{Teyssier2019}. Most methods evolve the gas pressure using a barotropic equation of state that does not account for shock heating and optical depth effects such as disc cooling in the vertical direction. Comparisons between the influence of a radiative transfer treatment versus a barotropic approximation on the temporal evolution can be found in \citet{Commercon2010}, \citet{Tomida2010}, \citet{Bate2011}, and \citet{Tomida2013}.

Within these numerical restrictions, to date only a handful of research groups have tackled this challenging problem down to protostellar disc scales no smaller than 1~au \citep{Bate2017, Vorobyov2019, Lebreuilly2019, Lebreuilly2020, Tsukamoto2021a, Tsukamoto2021b, Koga2022, Koga2023}. These studies have used a mono- or multi-fluid approach or a Lagrangian particle treatment to capture the momentum exchange between gas and dust. \citet{Hopkins2016}, \citet{Tricco2017}, and \citet{Commercon2023} have pointed out the possible decoupling of micrometre dust grains in molecular clouds, whereas \citet{Bate2017}, \citet{Vorobyov2019}, \citet{Lebreuilly2019, Lebreuilly2020}, and \citet{Tsukamoto2021a, Tsukamoto2021b} have shown the decoupling of $\geq$ 100~$\muup$m grains in collapsing pre-stellar cores resulting in enhancements of the dust-to-gas mass ratio within the young protostellar disc. Numerical studies of disc evolution using a dust-to-gas mass ratio higher than one per cent have indicated a pile-up of dust, which is favourable to enhancing dust growth \citep{Laibe2014, Gonzalez2017, Vericel2020}. A recent three-dimensional magneto-hydrodynamics study accounting for dust growth by \citet{Tsukamoto2021b} indicated that outflows on $>$~1~au scales provide a passage to redistribute and recycle large dust grains ($\geq$~10~$\muup$m) within the disc. The distribution of dust sizes can influence the transfer of angular momentum by altering the coupling between gas and magnetic fields. This influence is attributed to the impact of dust size on resistivity \citep{Kawasaki2022} and opacity \citep{Ormel2009, Ormel2011}. Various studies that include detailed time-dependent chemical networks also underline the importance of dust grains as efficient charge carriers \citep{Marchand2016, Zhao2016, Dzyurkevich2017, Zhao2018b, Tsukamoto2020}. Effects of radiation transport and an accurate gas equation of state to treat the influence of $\mathrm{H_2}$ dissociation and ionisation need to be investigated in combination with resistivity and angular momentum transport induced by turbulence and shear forces, while tracing the evolution of gas and dust separately. 

As a first step, we account for the combined effects of hydrodynamics, self-gravity, and radiation on the gas and the influence of gas drag force on the dust. We monitor the behaviour of 1, 10, and 100 micron-sized neutral, spherical dust particles throughout the collapse. The numerical framework and initial set-up is detailed in Sect.~\ref{sec:method}. In Sect.~\ref{sec:features} and Sect.~\ref{sec:thermodynamics} we present the results from our study aimed at determining the interplay and impact of gas and dust dynamics at pre-stellar core collapse scales, especially within the inner sub-au region. Future improvements for collapse models are mentioned in Sect.~\ref{sec:limitations}. Lastly, we summarise the objectives and outcomes of this project in Sect.~\ref{sec:summary}. 

%%%%%%%%%%%%%%%%%%%%%%%%%%%%%%%%%%%%%%%%%%%%%%%%%%%%%%%%%%%%%%%%%%
\section{Numerics and initial set-up}
\label{sec:method}

In this section we present details of the hybrid scheme where the gas was treated as a Eulerian fluid and the dust was considered as an ensemble of Lagrangian particles. The fluid-dust method accounts for aerodynamic drag forces and ensures momentum conservation. 

%%%%%%%%%%%%%%%%%%%%%%%%%%%%%%%%%%%%%%%%%%%%%%%%%%%%%%%%%%%%%%%%%%
\subsection{Gas treatment}

Two-dimensional radiation hydrodynamic (RHD) simulations were performed using a modified version of the \texttt{PLUTO} code \citep[version 4.4:][]{Mignone2007, Mignone2012}. The equations of hydrodynamics are detailed in \citet{Bhandare2018}. The \texttt{HAUMEA} module was used for the self-gravity Poisson solver \citep{Kuiper2010b, Kuiper2011}. A variable gas equation of state from \citet{Dangelo2013} accounts for effects of $\mathrm{H_2}$ dissociation, ionisation of atomic hydrogen and helium, and molecular vibrations and rotation \citep{Vaidya2015}. 

Tabulated dust opacities from \citet{Ossenkopf1994} and gas opacities from \citet{Malygin2014} are included. We apply a grey, frequency-averaged flux-limited diffusion (FLD) approximation for the radiation transport using the module \texttt{MAKEMAKE} \citep{Kuiper2010a}. This includes updates to consider a time-dependent evolution of the dust dominating at low temperatures as previously described in \citet{KuiperRT2020}. Gas thermodynamics is treated under the approximation of local thermodynamic equilibrium and a two-temperature approach for the gas and radiation. The two-temperature model uses a linearisation method in which the radiation and medium temperatures evolve as two different quantities \citep{Commercon2011}. Thermal coupling between gas and dust is assumed to be perfect so that they have the same temperature \citep{Galli2002}. Note that in the \texttt{MAKEMAKE} module, the dust-to-gas mass ratio is fixed to one per cent meaning that the gas and dust are also dynamically well-coupled. This is seen to be a reasonable assumption since we do not find a significant variation in the dust-to-gas mass ratio (see Sect.~\ref{sec:dust2gas}).

The mass, momentum, and energy transport equations are solved using second-order Godunov-type schemes for the fluid wherein a shock-capturing Riemann solver is used in a conservative finite volume approach and integrated with a second-order Runge--Kutta method. We used the Harten--Lax--van Leer approximate Riemann solver that restores the middle contact discontinuity (HLLC) and a monotonised central difference flux limiter using piecewise linear interpolation for the flux computation. The FLD and Poisson equations are treated implicitly with a standard generalised minimal residual solver with approximations to the error from previous restart cycles. This is used from the Portable, Extensible Toolkit for Scientific Computation open-source solver library \citep[PETSc;][]{Balay1997, Balay2019a, Balay2019}. 

%%%%%%%%%%%%%%%%%%%%%%%%%%%%%%%%%%%%%%%%%%%%%%%%%%%%%%%%%%%%%%%%%%
\subsection{Dust particle treatment}

A particle treatment of dust grains proves to be beneficial to trace their dynamical and thermal history in comparison with the Eulerian method. We treated the dust dynamics only in the Epstein regime and did not account for the back-reaction of the dust on the gas nor the interaction between dust grains. Although the back-reaction can modify the gas evolution by decreasing the inward gas velocity compared to pure viscous effects \citep{Dipierro2018}, these feedback effects are found to be negligible in a low-Stokes-number regime covered during the collapse \citep[][Appendix B]{Lebreuilly2020}.

For the collapse set-up herein, we have modified the dust module to account for the gas self-gravity on the dust. The dust grains are additionally affected by the gas via the drag force defined as 
%\begin{linenomath}
\begin{equation}
\bm F_\mathrm{d} = - \dfrac{m_{\mathrm{dg}}}{t_\mathrm{s}} \bm \Delta \mathrm{\textbf v}, 
\end{equation}
where $m_\mathrm{dg}$ is the mass of the dust grain and $\bm \Delta \mathrm{\textbf v = \textbf v_\mathrm{dg} - \textbf v_\mathrm{g}}$ is the differential velocity between the dust grain and gas. 
%\end{linenomath}

The dust stopping time $t_\mathrm{s}$ is defined as
%\begin{linenomath}
\begin{equation}
t_\mathrm{s} = \sqrt{\dfrac{\pi \Gamma_1}{8}} \dfrac{\rho_\mathrm{dg}}{\rho_\mathrm{g}} \dfrac{s_\mathrm{dg}}{c_\mathrm{s}},
\label{eq:Stokes}
\end{equation}
%\end{linenomath}
where $\Gamma_1$ is the first adiabatic index,  $c_\mathrm{s}$ is the sound speed, $\rho_\mathrm{dg}$ is the intrinsic dust grain density, $s_\mathrm{dg}$ is the dust grain size, and $\rho_\mathrm{g}$ is the gas density. 

The time integration of the dust particles equations of motion is performed using an exponential mid-point method. A logarithmic interpolation in the radial direction and a cloud-in-cell weighting scheme in the $\theta$ (polar) direction is applied for depositing particle quantities such as velocities in the grid cells and interpolating fluid properties at the particle positions as detailed in \citet{Mignone2018, Mignone2019}.

%%%%%%%%%%%%%%%%%%%%%%%%%%%%%%%%%%%%%%%%%%%%%%%%%%%%%%%%%%%%%%%%%%
\subsection{Initial fluid set-up}

The physical and numerical set-up of the collapse is similar to the one previously described in \citet{Bhandare2018, Bhandare2020}. The initial density distribution has a stable Bonnor--Ebert sphere like profile with the outer $\rho_\mathrm{o}$ and central $\rho_\mathrm{c}$ density determined as
%\begin{linenomath}
\begin{equation}
\begin{split}
&\rho_\mathrm{o} = \Bigg (\dfrac{1.18 ~c_\mathrm{s0}^3}{M_0 ~G^{3/2}} \Bigg)^2, \\
&\rho_\mathrm{c} = 14.1 ~\rho_\mathrm{o}, 
\end{split}
\end{equation}
%\end{linenomath}
where $G$ is the gravitational constant and $M_0$ is the initial pre-stellar core mass \citep{Ebert1955, Bonnor1956}. The initial uniform sound speed $c_\mathrm{s0}$ is estimated using
%\begin{linenomath}
\begin{align}
\mbox{$c_\mathrm{s0}^2 = \dfrac{G M_0}{\mathrm{ln}(14.1) ~R_\mathrm{out}} $},
\end{align}
%\end{linenomath}
where $R_\mathrm{out}$ is the pre-stellar core radius. The density contrast between the inner and outer edge corresponds to a dimensionless radius of $\xi$ = 6.45 where $\xi$ is defined as 
%\begin{linenomath}
\begin{align}
\mbox{$ \xi = \sqrt{\dfrac{4 \pi G \rho_\mathrm{o}}{c_\mathrm{s0}^2}} R_\mathrm{out} $}.
\end{align}
%\end{linenomath}
The initial thermal pressure was computed for a fixed temperature $T_\mathrm{0}$ of 10~K. However, the temperature $T_\mathrm{BE}$ for the stable Bonnor--Ebert sphere is slightly higher than 10~K and is computed as
%\begin{linenomath}
\begin{align}
\mbox{$T_\mathrm{BE} = \dfrac{\mu ~c_\mathrm{s0}^2}{\gamma ~\Re}$},
\end{align}
%\end{linenomath}
where the mean molecular weight $\mu$ = 2.353 g/mol, the adiabatic index $\gamma$ = 5/3, and $\Re$ is the universal gas constant. Thus the reduced pressure support of the stable Bonnor--Ebert sphere initiates the collapse. The radiation temperature was initially in equilibrium with the gas temperature.

The outer radius was fixed to 3000~au for the fiducial case of an initial 1~$M_{\odot}$ and also for an initial 3~$M_{\odot}$ pre-stellar core. The initial ratio of thermal-to-gravitational energy was fixed to 0.29 and 0.098 for the 1 and 3~$M_{\odot}$ cases, respectively and computed as
%\begin{linenomath}
\begin{align}
\mbox{$ \dfrac{E_\mathrm{th}}{E_\mathrm{grav}} = \dfrac{5 R_\mathrm{out} ~k_\mathrm{B} T_\mathrm{0}}{2 G M_\mathrm{0} \mu m_\mathrm{H}}$,} 
\end{align}
%\end{linenomath}
where $k_\mathrm{B}$ is Boltzmann constant and $m_\mathrm{H}$ is the hydrogen mass. 
We compare simulations with two different values for the initial solid-body rotation rate of \mbox{$\Omega_\mathrm{0} = 1.77 \times 10^{-13}$ rad~$\mathrm{s}^{-1}$} and \mbox{$\Omega_\mathrm{0} = 2.099 \times 10^{-13}$ rad~$\mathrm{s}^{-1}$}. The ratio of rotational-to-gravitational energy is approximated as 
%\begin{linenomath}
\begin{align}
\mbox{$ \dfrac{E_\mathrm{rot}}{E_\mathrm{grav}} = \dfrac{R_\mathrm{out}^3 \Omega_\mathrm{0}^2}{3 G M_\mathrm{0}}$} 
\end{align}
%\end{linenomath}
and for the two rotation rates is 0.007 and 0.01, respectively. These are below the disc fragmentation limit of \mbox{$E_\mathrm{rot} / E_\mathrm{grav} >$ 0.01} \citep{Matsumoto2003,Bate2011}. The fiducial run with the slower rotation rate is compared to a case with a larger pre-stellar core with an outer radius of 5000~au detailed in Appendix~\ref{sec:discproperties}. The varied initial pre-stellar core properties discussed here are listed in Table~\ref{tab:initialparams}. 

Mechanisms such as gravitational torques exerted from the spiral arms or bar-like structures, magneto-rotational instability, hydrodynamically driven turbulence, outflows, disc winds, and jets are responsible for the angular momentum transport during the collapse and protostellar disc formation. A physical shear viscosity was implemented to mimic some of these effects of angular momentum transport \citep{Kuiper2010b, Kuiper2011}. The shear viscosity is described using the so-called $\alpha$-parametrisation from \citet{Shakura1973} and is given by
%\begin{linenomath}
\begin{align}
\mbox{$ \nu = \alpha ~c_\mathrm{s} ~H ~\rho \approx \alpha ~\Omega_\mathrm{K}(r) ~R^2 \left(\dfrac{H}{R}\right)^2 \rho$}. 
\end{align}
%\end{linenomath}
The local sound speed $c_\mathrm{s} \approx H~\Omega_\mathrm{K}(r)$ with the Keplerian angular velocity $\Omega_\mathrm{K}(r) \approx \sqrt{G M(r) / r^3}$. The local pressure scale height \mbox{$H$ = $(H/R) R$} where the cylindrical radius $R$ = $r\sin(\theta)$. The dimensionless parameters $(H/R)=0.05$ and $\alpha=1.0$ were fixed in space and time for calculating the viscosity and to mimic the net outward angular momentum transport within self-gravitating discs. This shear viscosity is similar to the $\beta$-viscosity prescription for self-gravitating discs by \citet{Duschl2000} with a $\beta$-parameter of $\beta = 2.5 \times 10^{-3}$ for the fiducial case at $\mathrm{time}=0$. This temperature-independent $\beta$-viscosity prescription was also previously tested in the low-mass core collapse studies by \citet{Schoenke2011} and in the high-mass core collapse studies by \citet{Kuiper2013, Kuiper2015, Kuiper2016}, and \citet{Kuiper2018}. 

\begin{table}[!tp]
	\centering
	\caption{Initial pre-stellar core properties.}
	\resizebox{0.5\textwidth}{!}{
	\begin{tabular}[t]{cccccc}
		\hline
        $M_{0} ~(M_{\odot})$ & $R_\mathrm{out}$ (au) & $E_\mathrm{rot} / E_\mathrm{grav}$ & Dust size ($\muup$m) & Discussion  
        \TBstrut\\ \hline \hline
        1  & 3000  & 0.007  & 1           &  ~Sect.~\ref{sec:hydrodisc}, \ref{sec:discproperties}
     \Tstrut \\
        1  & 3000  & 0.01   & 1           &  
        Sect.~\ref{sec:rotation}, \ref{sec:discproperties}     \\
        1  & 5000  & 0.007  & 1           & ~Sect.~\ref{sec:Outerradius}, \ref{sec:discproperties}  \Bstrut \\ \hline 
        3  & 3000  & 0.007  & 1           &  
        ~Sect.~\ref{sec:mass}          \TBstrut\\ \hline        
        1  & 3000  & 0.007  & 1, 10, 100  &  
        ~Sect.~\ref{sec:thermodynamics}      \TBstrut\\ \hline
	\end{tabular}
	}
 	\tablefoot{Listed above are the pre-stellar core properties for runs with different initial pre-stellar core mass $M_{0} ~(M_{\odot})$, outer radius $R_\mathrm{out}$~(au), rotational-to-gravitational energy ratio $E_\mathrm{rot} / E_\mathrm{grav}$, fixed dust size~($\muup$m), and the corresponding discussion section.}
	\label{tab:initialparams}
\end{table}

%%%%%%%%%%%%%%%%%%%%%%%%%%%%%%%%%%%%%%%%%%%%%%%%%%%%%%%%%%%%%%%%%%
\subsection{Initial particle set-up}

We performed simulations using a single fixed size of 1 $\muup$m dust and also a run with three fixed sizes of 1, 10, and 100 $\muup$m initialised at the same location on the grid. We considered spherical, neutral dust with an intrinsic density $\rho_\mathrm{dg}$ of 3~$\mathrm{g~cm^{-3}}$ to mimic a combination of carbonaceous and silicate grains as previously used in \citet{Tricco2017} and \citet{Lebreuilly2019}. The initial dust density was fixed to one per cent of the gas density using the conservative estimate of the averaged dust-to-gas mass ratio of the ISM. This results in a particle mass distribution that depends on the grid cell density and volume, which is reasonable when neglecting the dynamical back-reaction of the dust onto the gas. In this case, the mass of the particle does not play a role for the dynamics. In order to ensure a uniform particle sampling throughout the pre-stellar core, especially in the innermost regions, we used 200 particles per cell for all the simulations with the fixed 1 $\muup$m dust and outer radius of 3000~au. For the multi-size simulation, we assigned 50 particles per cell per size for the three different sizes, that is, a total of 150 particles per cell. The particles are randomly distributed between the cell interfaces and assigned a fraction of the grid cell mass. This leads to a mass variation between the particles also within a grid cell. The local and global mass conservation is ensured. Observational evidence suggests an MRN-like size distribution with a larger population of smaller sized dust within a pre-stellar core. In the simulations presented here, the choice of the size distribution does not affect the results since back-reaction is neglected. As such, we choose an equal number of particles for each size in order to compare their relative behaviour during the collapse.

%%%%%%%%%%%%%%%%%%%%%%%%%%%%%%%%%%%%%%%%%%%%%%%%%%%%%%%%%%%%%%%%%%
%% Figure placement

\begin{figure*}[ht]
    \centering
    \includegraphics[width=0.76\linewidth]{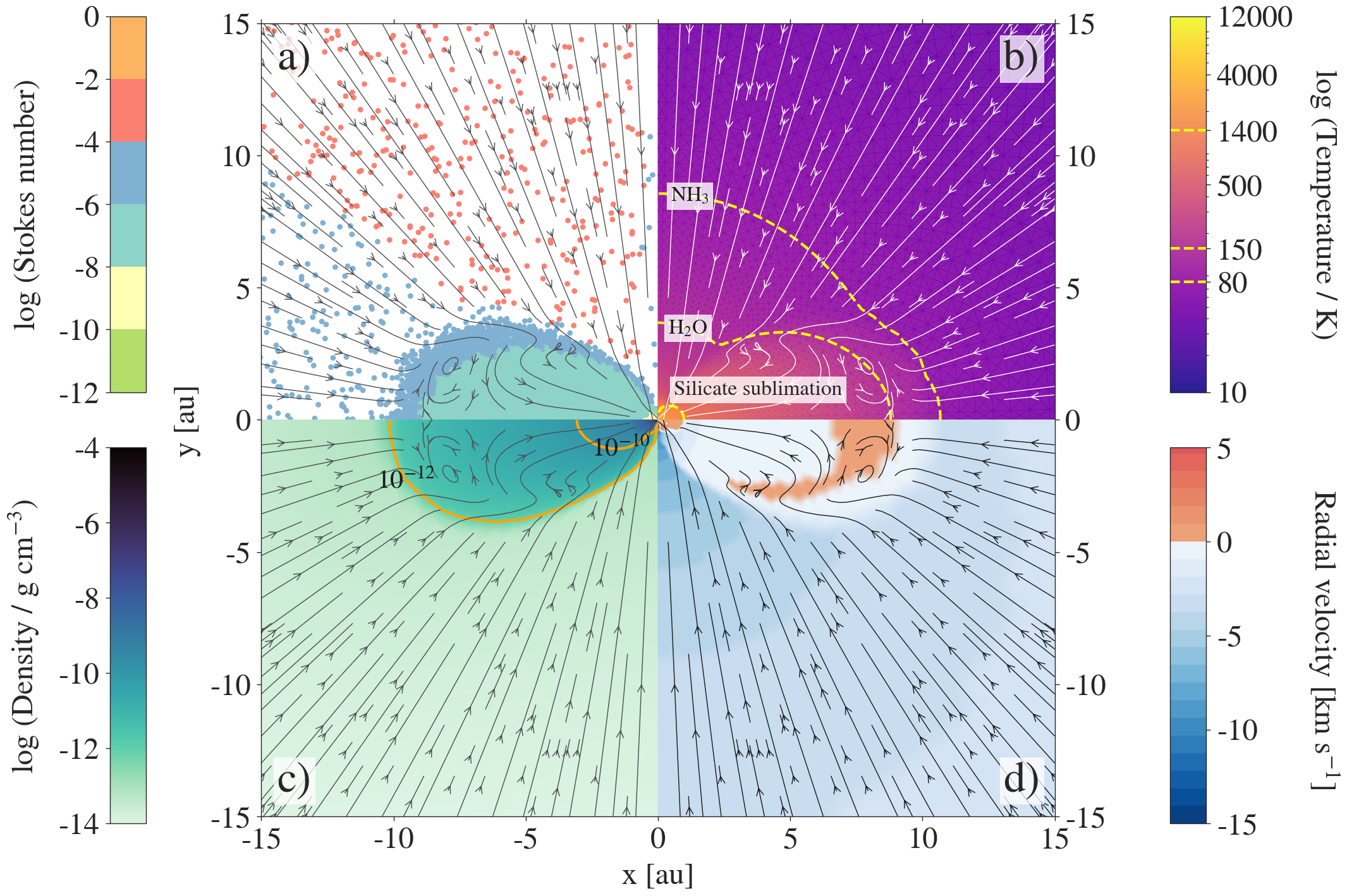}
    \caption{2D view of the first hydrostatic core evolved into a rotationally supported disc at 1221~years after its formation as a result of the fiducial collapse of a 1~$M_{\odot}$ pre-stellar core with an initial rotation rate of \mbox{$\Omega_\mathrm{0} = 1.77 \times 10^{-13}$~rad~$\mathrm{s}^{-1}$} (same time snapshot as Fig.~\ref{fig:HD-seconddisc007}). The dust size is fixed to a constant value of 1~$\muup$m. The four panels show the \mbox{\bf a)}~Stokes number, \mbox{\bf b)}~gas temperature, \mbox{\bf c)}~gas density, and \mbox{\bf d)}~radial gas velocity within the inner 15~au of the 3000~au collapsing pre-stellar core. The gas velocity streamlines indicate the material falling onto the disc and the mixing within. The meridional flow exhibited in the plot between 5--10~au forms in the inner regions and travels outwards. This motion results in an outward transport and retains dust in the outer disc.}
    % \vspace*{0.5cm}
    \label{fig:HDdisc007}
\end{figure*}
\begin{figure*}[!ht]
    \centering
    \includegraphics[width=0.76\linewidth]{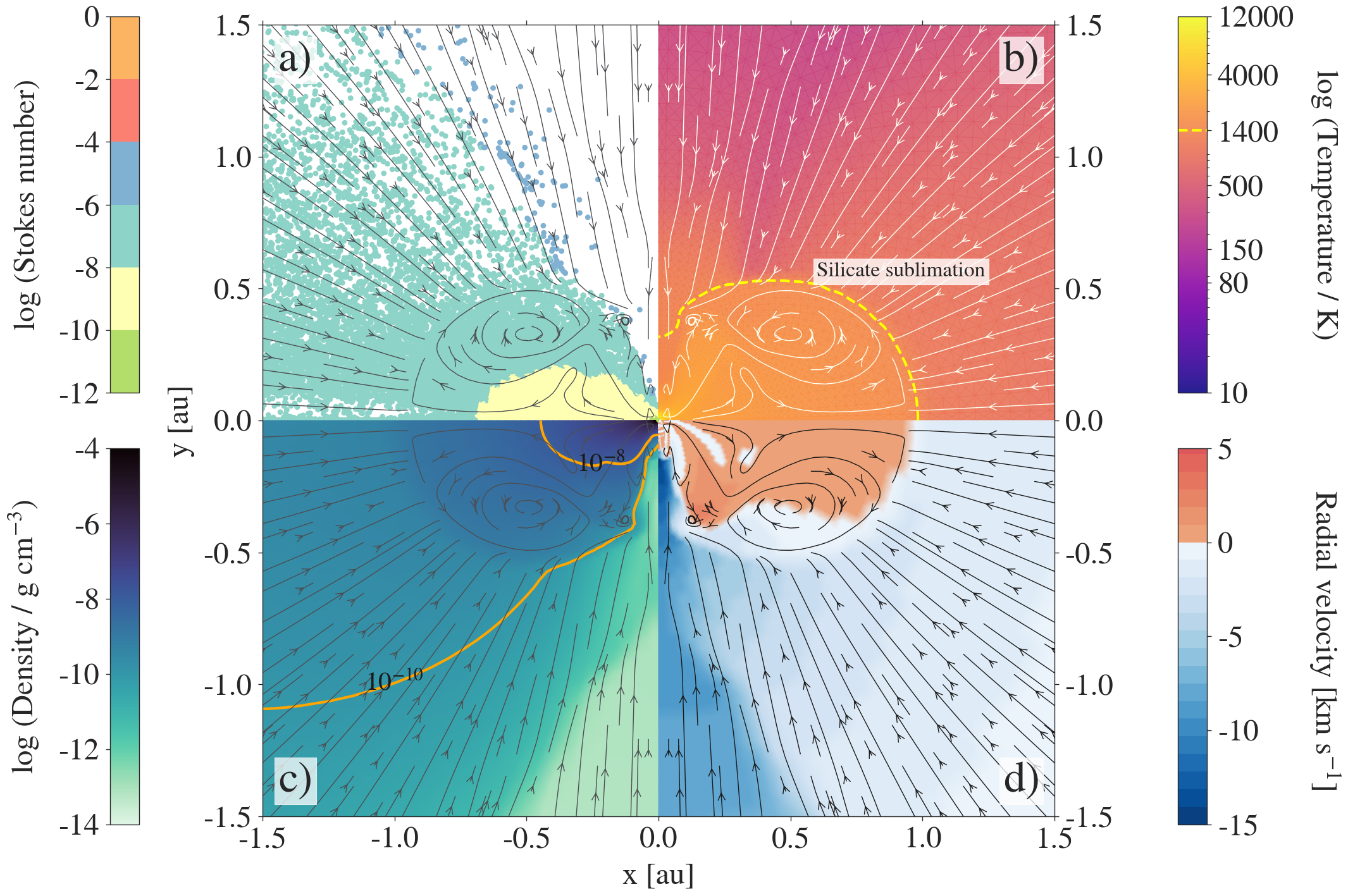}
    \caption{2D view of a pseudo-disc around the second core at three years after its formation resulting from the same initial conditions and at the same time snapshot as Fig.~\ref{fig:HDdisc007}. The vortical feature forming at sub-au scales in the vicinity of the unresolved protostar circulates dust in the high-temperature inner regions.}
    \label{fig:HD-seconddisc007}
\end{figure*}

%%%%%%%%%%%%%%%%%%%%%%%%%%%%%%%%%%%%%%%%%%%%%%%%%%%%%%%%%%%%%%%%%%
\subsection{Grid}

We used a two-dimensional (2D) axially and midplane symmetric Eulerian grid. The radial grid extends from $10^{-2}$~au to 3000~au and consists of 241 logarithmically spaced cells. This resolves better the central parts of the computational domain. We used 30 uniformly spaced cells in the polar direction stretching from the pole ($\theta = 0^\circ$) to the midplane ($\theta = 90^\circ$). The number of grid cells is selected to have the same spatial extent in the radial and polar direction. The smallest cell size is $\Delta x_\mathrm{min} = \Delta r = r \Delta\theta = 5.37\times10^{-4}$~au. This resulted in 84 cells per Jeans length at the highest central density achieved in the course of the fiducial simulation. Otherwise, we used 21 -- 573 cells per Jeans length over the computational domain estimated using the midplane density and sound speed at the final simulation time snapshot.

%%%%%%%%%%%%%%%%%%%%%%%%%%%%%%%%%%%%%%%%%%%%%%%%%%%%%%%%%%%%%%%%%%
\subsection{Boundary conditions}

At the inner radial edge, we adopted reflective boundary conditions for the hydrodynamics, which sets the density, thermal pressure, and radial velocity. A zero gradient condition was adopted for the polar and azimuthal velocity components as well as for the radiation energy, which implies that no radiative flux can cross the inner boundary interface. At the outer radial edge, a Dirichlet boundary condition was applied for the radiation temperature with a constant boundary value of 10~K. Additionally, we imposed an outflow--no-inflow condition for the hydrodynamics, which includes a zero-gradient boundary condition for the thermal pressure, the polar, and the azimuthal velocity components. Axisymmetric boundaries were used at the pole and mirror-symmetric boundaries at the equator.

%%%%%%%%%%%%%%%%%%%%%%%%%%%%%%%%%%%%%%%%%%%%%%%%%%%%%%%%%%%%%%%%%%
\section{Results I: Features formed during the collapse}
\label{sec:features}

Studies over the last $\sim$55 years have established that low-mass stars form via a two-step process involving the formation of first and second hydrostatic cores \citep[see summary in][]{Bhandare2018, Bhandare2020}. Angular momentum conservation in these collapsing pre-stellar cores initiates the formation of protostellar discs as early as the first core stage. This section describes transient gas features, namely meridional flows and outflow, formed during the collapse process for cases with different initial conditions. 

%%%%%%%%%%%%%%%%%%%%%%%%%%%%%%%%%%%%%%%%%%%%%%%%%%%%%%%%%%%%%%%%%%
\subsection[Fiducial case: 1-Msol pre-stellar core collapse]{Fiducial case: 1~$M_{\odot}$ pre-stellar core collapse}
\label{sec:hydrodisc}

The 2D RHD simulation for a 1~$M_{\odot}$ pre-stellar core collapse shows an oblate first hydrostatic core transitioning into a rotationally supported disc over a period of 1218~years even `before' the second collapse, which is the protostellar formation stage. The presence of a disc before the second collapse phase is also noted by previous hydrodynamic studies \citep[e.g.][]{Bate1998, Saigo2008, Tscharnuter2009, Machida2010, Bate2010, Machida2011c, Bate2011}.

We extract the disc based on a combination of definitions from \citet{Joos2012} and \citet{Koga2022}. The disc is defined as the region which satisfies all the following conditions:
\begin{itemize}
\item Azimuthal (rotational) velocity must be larger than twice the radial velocity \mbox{$v_{\phi} > 2~|v_{r}|$}. 
\item Rotational support must be higher than the thermal support \mbox{$(\rho v_{\phi}^2 / 2) > 2~P_\mathrm{th}$} and \mbox{$v_{\phi} > 0.6~v_\mathrm{K}$}, where $v_\mathrm{K}$ is the Keplerian velocity.
\item Gas density must be above the minimum limit set by the midplane density immediately outside of the first (second) shock location for the outer (inner) disc. 
\end{itemize}

Figure~\ref{fig:HDdisc007} is a zoom-in within 15~au of a 3000~au pre-stellar core. The initial temperature was fixed to 10~K and the initial rotation rate to \mbox{$\Omega_\mathrm{0} = 1.77 \times 10^{-13}$~rad~$\mathrm{s}^{-1}$} (or $E_\mathrm{rot} / E_\mathrm{grav} =$~0.007). Comparisons to a faster rotation case ($E_\mathrm{rot} / E_\mathrm{grav} =$~0.01) are detailed in the Appendix~\ref{sec:rotation} and to a larger initial cloud core (5000~au) are discussed in Appendix~\ref{sec:Outerradius}. The dust size was fixed to a constant value of 1~$\muup$m throughout the evolution. The four panels show various dust and gas properties at 1221~years from the start of the first core formation as it transitions into a rotationally supported disc, which is still embedded in its surrounding infalling envelope. Panel~(a) displays the Stokes number at the dust particle location defined using the stopping time and orbital frequency as \mbox{$\mathrm{St} = \mathrm{t_s}~\times~\mathrm{\Omega_{orbital}}$}, where $\mathrm{\Omega_{orbital}} = v_{\phi}/r\sin(\theta)$ and $t_\mathrm{s}$ is defined above in Eq.~\ref{eq:Stokes}. A Stokes number much below unity indicates a strong dust--gas coupling. Gas temperature is shown in panel~(b) with contours at a few specific temperatures for silicate sublimation \citep[1400~K;][]{Isella2005, Izidoro2022}, ice line (150~K), and ammonia snowline \citep[80~K;][]{Pinilla2017}. Panel~(c) shows the non-homologous nature of the collapse with gas density increasing towards the centre. Gas radial velocity highlighting both inward (blue) and outward (red) motion is shown in panel~(d). Gas falls in faster with a higher radial velocity onto the central part via the polar regions compared to the midplane. Streamlines in all four panels help visualise material falling onto the disc and mixing within, especially the meridional flow close to the first accretion shock at around 10~au. The turbulent feature moves radially outwards during the 1221~years of evolution, while trapping some dust and enabling an outward transport. Mixing and outward gas transport via circulation currents within early disc formation stages was also previously reported in \citet{Tscharnuter2009}.

Figure~\ref{fig:HD-seconddisc007} is a further zoom-in showing different properties within 1.5~au of the 3000~au pre-stellar core at the same time snapshot as Fig.~\ref{fig:HDdisc007}. The four panels show the same dust and gas properties at three years after the pseudo-disc\footnote{A pseudo-disc is a dense, non-rotationally supported region.} begins to form around the second core and is embedded within the first disc. Similar examples of nested discs can be found in \citet{Saigo2008} and \citet{Machida2011c}. The second shock location coincides with the silicate sublimation temperature of 1400~K. A second vortical feature seen in the figure, although lasting only for a few years, helps circulate and retain dust in the high-temperature sub-au region. Further discussion on the cause and effects of both the turbulent flows can be found next in Sect.~\ref{sec:baro1Msun} and Sect.~\ref{sec:dustinflow}.

\begin{figure}[t]
\centering
\includegraphics[width=0.9\linewidth]{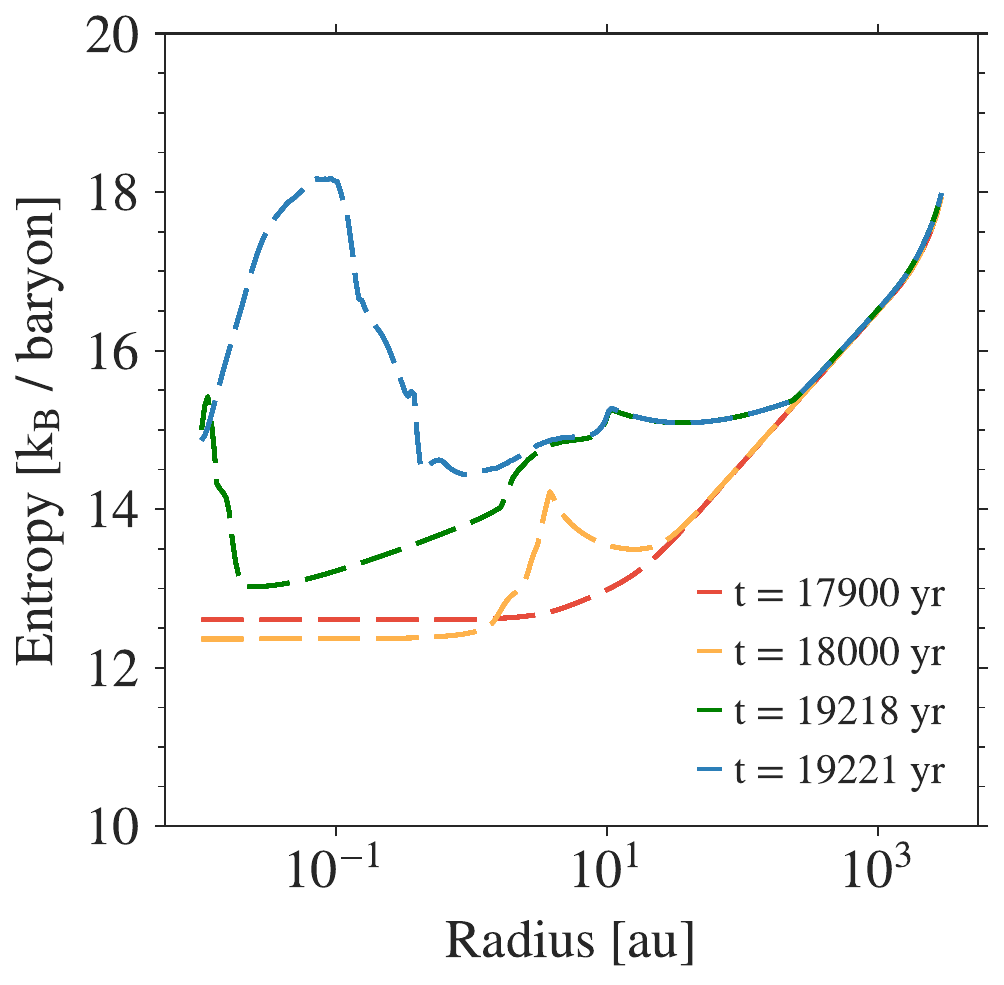}
\caption{Polar-angle averaged radial profile of the entropy at different time snapshots during the 1~$M_{\odot}$ pre-stellar core collapse. A negative radial gradient generates turbulent flows within the two discs.}
\label{fig:entropyvsr}
\end{figure}
\begin{figure}[t]
    \centering
    \includegraphics[width=0.95\linewidth]{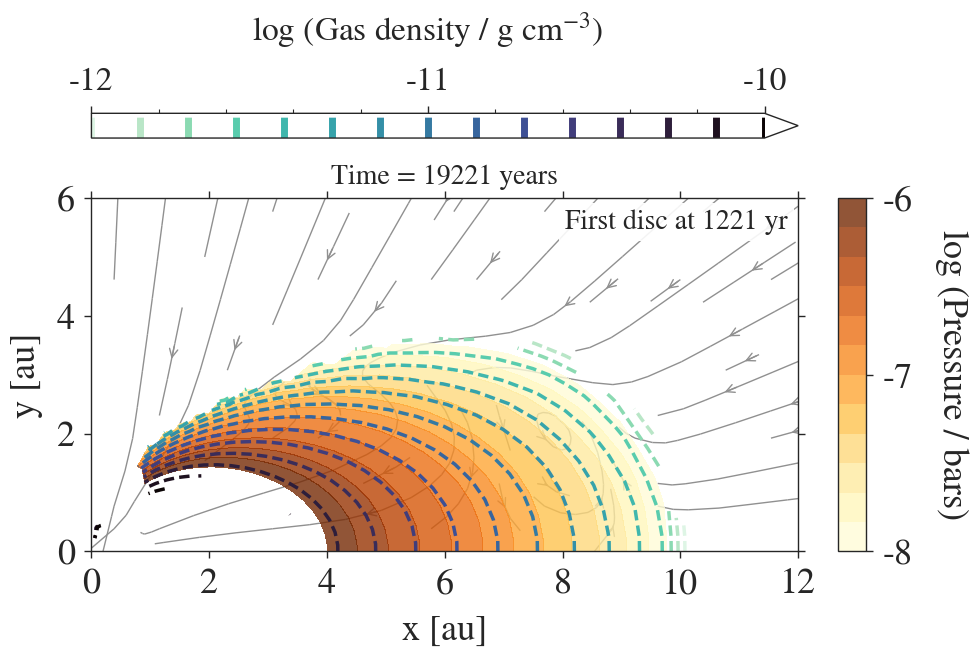}
    \caption{Thermal pressure $P_\mathrm{th}$ in colour and isodensity contours within the first core disc at 1221~years. The infalling envelope has been masked to highlight the non-zero baroclinicity within the disc. The gas velocity streamlines in the background indicate the material falling onto the disc and the circulation within.} 
    \label{fig:pressure-rho}
\end{figure}
\begin{figure}[t]
    \centering
    \includegraphics[width=0.95\linewidth]{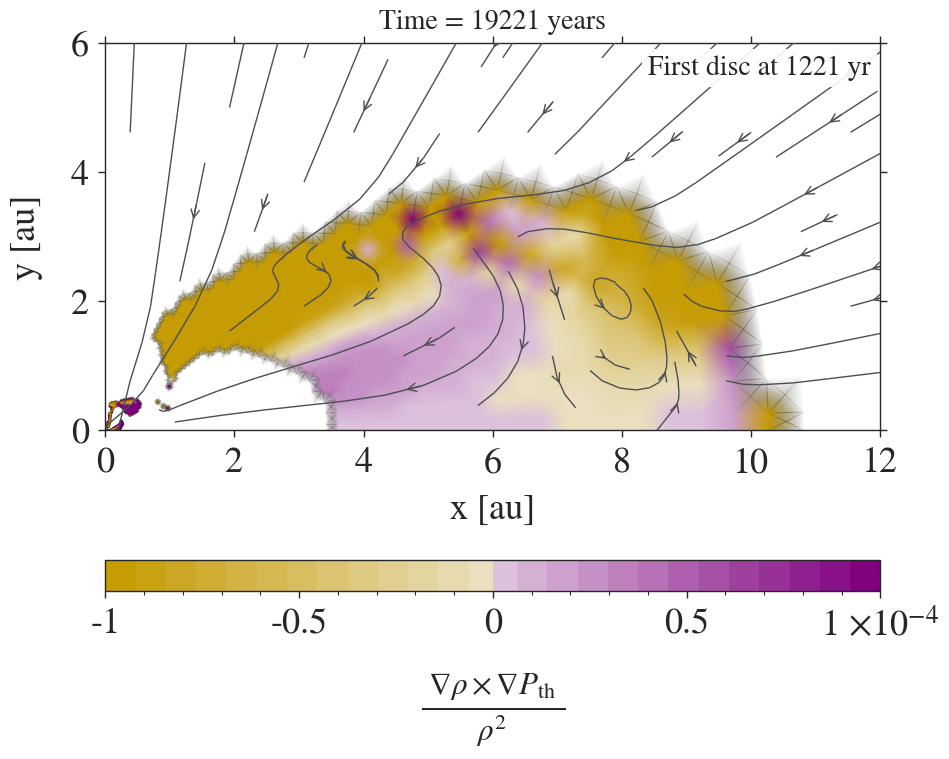}
    \caption{Baroclinicity $\mathrm{(s^{-2})}$ at 1221~years after the transition of the first core into a rotationally supported disc resulting from a 1~$M_{\odot}$ pre-stellar core collapse. The infalling envelope has been masked to highlight the baroclinic contribution in generating vortical flows within the disc. The gas velocity streamlines indicate the material falling onto the disc and the circulation within.} 
    \label{fig:baroclinityentropymap}
\end{figure}
\begin{figure}[ht!]
    \centering
    \includegraphics[width=0.95\linewidth]{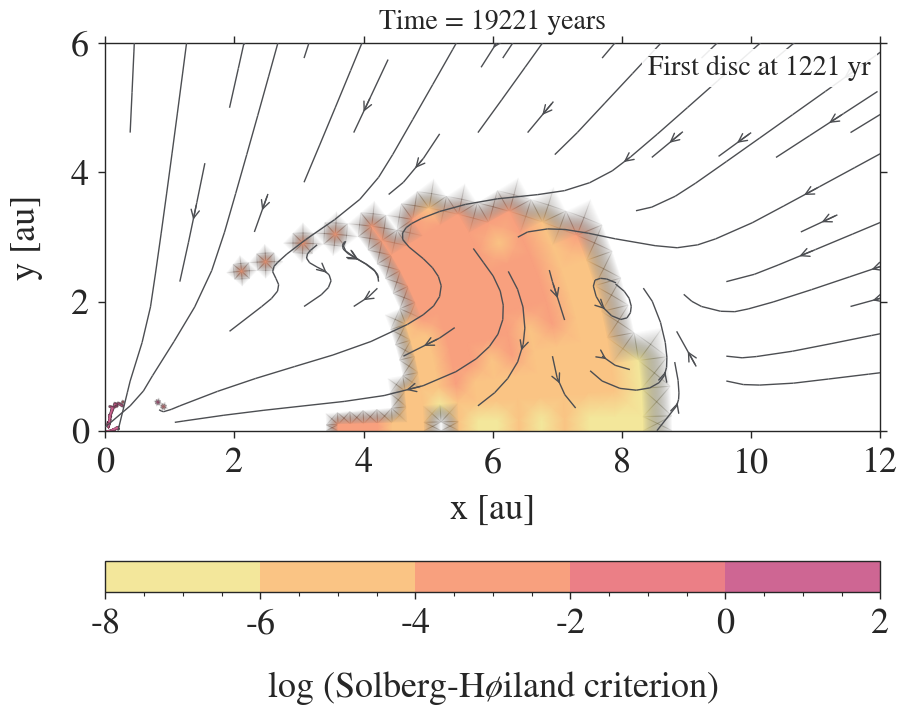}
    \caption{Solberg--H{\o}iland unstable region within the first core disc at 1221~years found mainly around the meridional feature.}
    \label{fig:SHcondition}
\end{figure}

%%%%%%%%%%%%%%%%%%%%%%%%%%%%%%%%%%%%%%%%%%%%%%%%%%%%%%%%%%%%%%%%%%
\subsubsection{\textit{Formation of dust traps in turbulent gas flows}}
\label{sec:baro1Msun}

Our results show the formation of two turbulent gas flows during the 1~$M_{\odot}$ collapse. A meridional flow starts building up within the first core and moves radially outwards as this first core transitions into a rotationally supported disc. Another vortical feature forms within the pseudo-disc around the protostar. Figure~\ref{fig:entropyvsr} shows the radial profile of the polar-angle averaged entropy over different time snapshots. The first core begins to form at 18000~years and the pseudo-disc at 19218~years after the onset of the collapse. At both these times, a negative radial entropy gradient triggers a turbulent flow. In case of the first core disc, this turbulent gas flow eventually generates a baroclinically amplified vortical feature. Formation of such meridional gas flows due to a mismatch in the planes of constant density and pressure (i.e.~non-zero baroclinicity) were previously reported in \citet{Klahr1997, Klahr2003, Klahr2004, Petersen2007}, and \citet{Lesur2010}. This is depicted in Fig.~\ref{fig:pressure-rho} as a slight misalignment between isopressure surfaces and isodensity contours. Figure~\ref{fig:baroclinityentropymap} shows the baroclinic term from the vorticity equation within the first core disc at 1221~years. The infalling envelope surrounding the rotationally supported disc is masked to highlight the disc vortical feature. A flip in the baroclinic term occurs gradually during the radial outward motion of this meridional flow. This change of sign is visible at the locations where there is a deviation from the inward gas flow. Additionally, in Fig.~\ref{fig:SHcondition} we explore the origin of this baroclinicity. For that purpose, we compute the Solberg--H{\o}iland criterion for dynamic stability of an incompressible rotating flow with respect to small isentropic disturbances  \citep{Tassoul2000, Rudiger2002}. In cylindrical coordinates, this condition requires
%\begin{linenomath}
\begin{align}
\mbox{ $ \dfrac{\partial P_\mathrm{th}}{\partial z} \left( \dfrac{\partial j^2}{\partial R} \dfrac{\partial S}{\partial z} - \dfrac{\partial j^2}{\partial z} \dfrac{\partial S}{\partial R} \right) < 0 $},
\end{align}
%\end{linenomath}
where $P_\mathrm{th}$ is the thermal gas pressure and $j = R^2 \Omega$ is the angular momentum per unit mass. The entropy $S$ is computed\footnote{\url{https://github.com/gabrielastro/idealgasentropy}} using the Sackur--Tetrode equation for a perfect gas consistent with the \citet{Dangelo2013} equation of state used in the simulation (see \citealt{Vaidya2015} for the details of the implementation in \texttt{PLUTO}). It accounts for molecular, atomic, and ionised hydrogen as well as the contribution from the electrons and thus represents a straightforward extension of the expressions in \citet[][Appendix~A]{Berardo2017}. We only highlight the unstable regions found within the rotationally supported first core disc and mask out the stable envelope.

On the other hand, the turbulent feature within the pseudo-disc spreads horizontally across the pseudo-disc within its short three-year lifetime, only to be destroyed a year later due to the launch of an outflow discussed next in Sect.~\ref{sec:outflow1Msun}. The short-lived nature of the pseudo-disc makes it difficult to follow the evolution and development of a baroclinic and Solberg--H{\o}iland unstable region.

Deviation from a radial inward motion creates transient dust pockets that trap particles and circulate or sweep them outwards in both the discs before they can either settle at the midplane or drift towards the protostar. Furthermore, the longer-lived dust pocket close to the first core shock can act as a viable location for dust growth and eventually promote protoplanet formation at earlier times (see also \citealt{Koga2023}). Dust traps in gas meridional flows have been previously reported in simulations resulting from strong poloidal magnetic fields in protoplanetary discs \citep{Hu2022}, inclined isodensity and isotemperature (or isopressure) surfaces (i.e.~non-zero baroclinicity) in planetary envelopes \citep{Schulik2019, Schulik2020}, and planetary spiral wakes in circumplanetary discs \citep{Szulagyi2022}.

%%%%%%%%%%%%%%%%%%%%%%%%%%%%%%%%%%%%%%%%%%%%%%%%%%%%%%%%%%%%%%%%%%
\subsubsection{\textit{Long-term evolution: tracing an outflow}}
\label{sec:outflow1Msun}

\begin{figure}[!tp]
    \centering
    \includegraphics[width=0.9\linewidth]{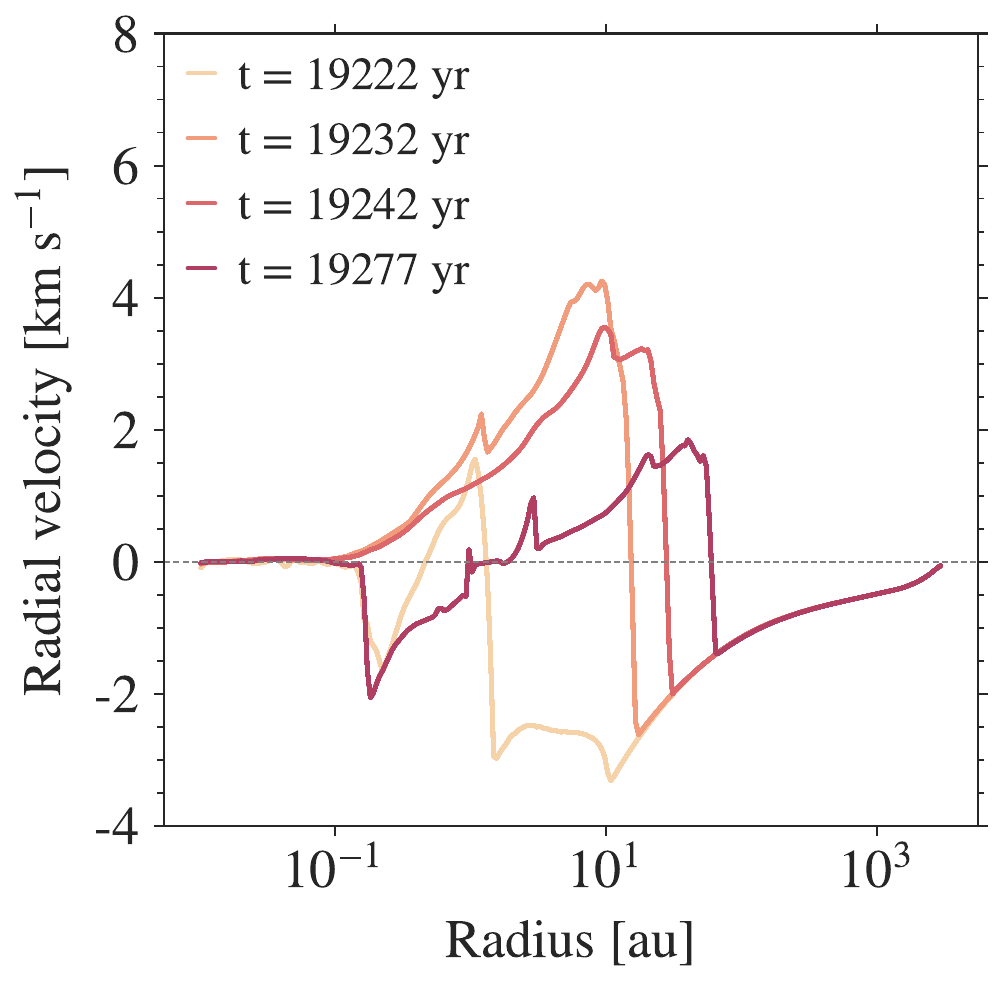} \\
    \caption{Polar-angle averaged radial velocity at different time snapshots showing the transient nature of the outflow during the 1~$M_{\odot}$ pre-stellar core collapse.}
    \label{fig:1Msun-outflowvr}
\end{figure}

\begin{figure}[!t]
    \centering
    \includegraphics[width=0.9\linewidth]{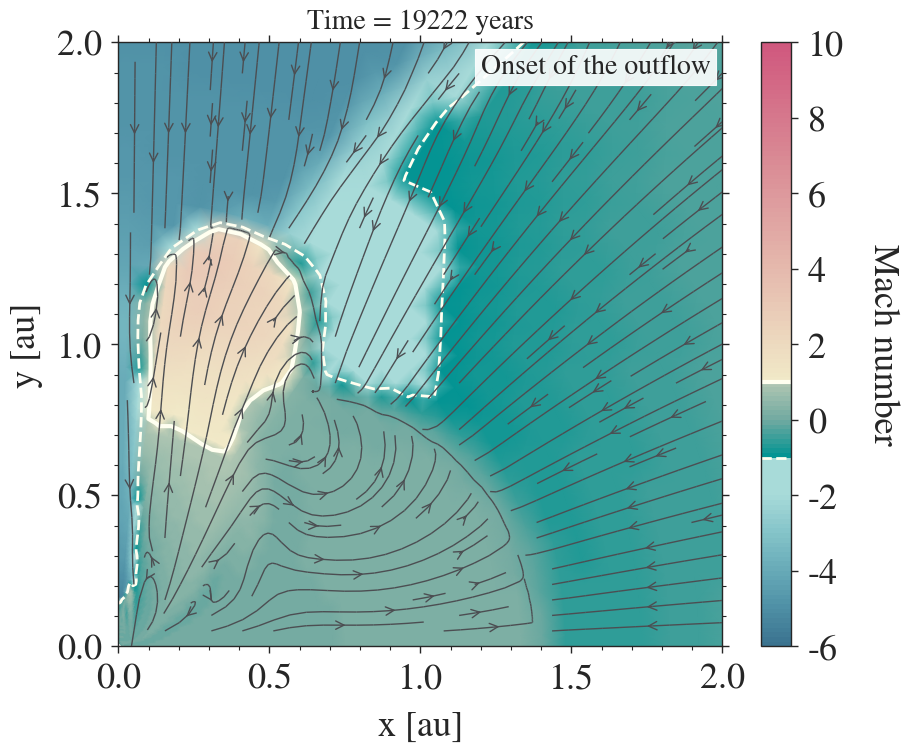} \\
    \vspace{0.5cm}
    \includegraphics[width=0.9\linewidth]{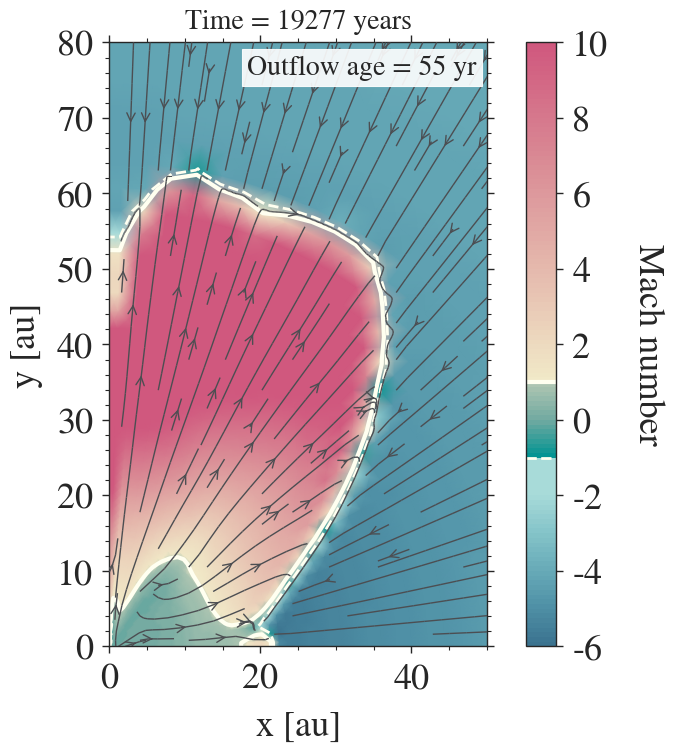} 
    \caption{Radial Mach number ($ v_{r} / c_\mathrm{s}$) at first and last time snapshots of outflow formation during the 1~$M_{\odot}$ pre-stellar core collapse. The gas velocity is indicated via the streamlines. The colour gradients in red and blue indicate the supersonic outflow and inflow regions, respectively. The colour gradient in green shows the subsonic regions mainly within the disc.}
    \label{fig:Mach}
\end{figure}

\begin{figure}[!tp]
    \centering
    \includegraphics[width=0.7\linewidth]{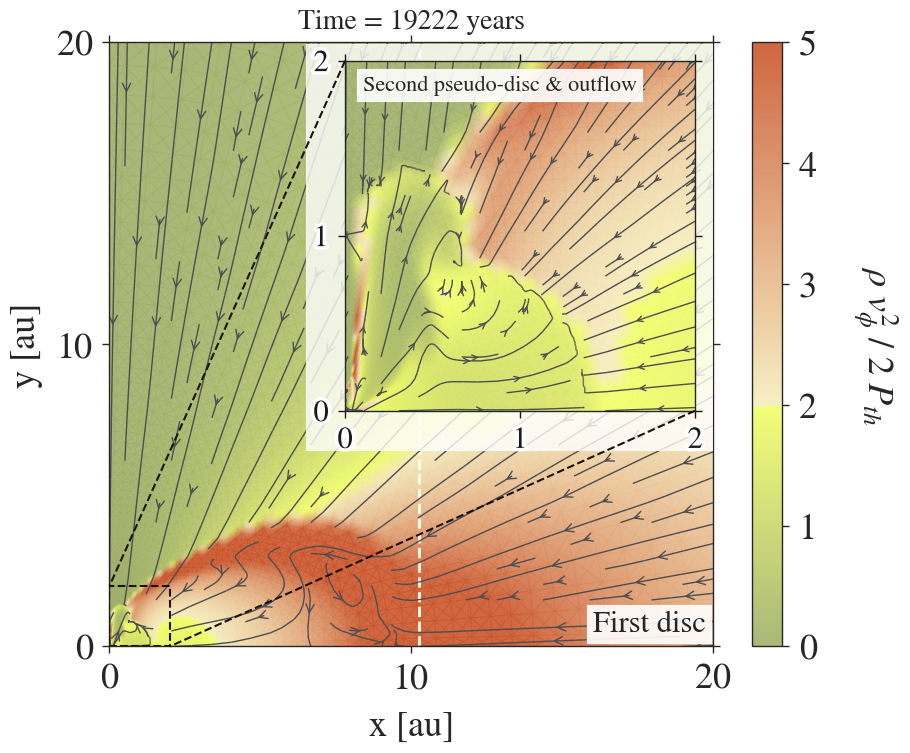}
    \includegraphics[width=0.7\linewidth]{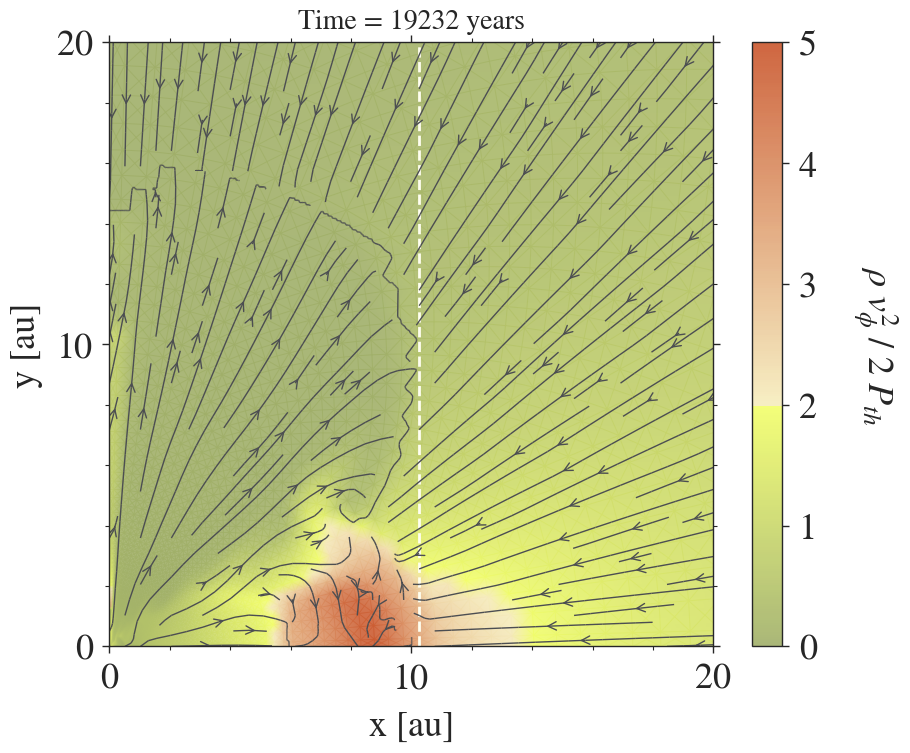}
    \includegraphics[width=0.7\linewidth]{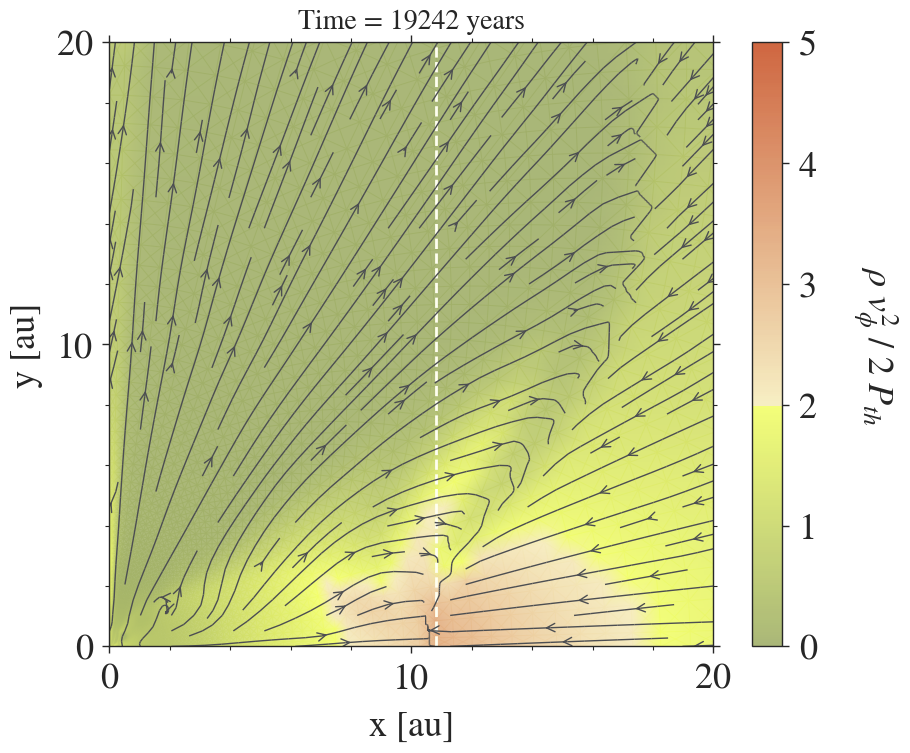}
    \includegraphics[width=0.7\linewidth]{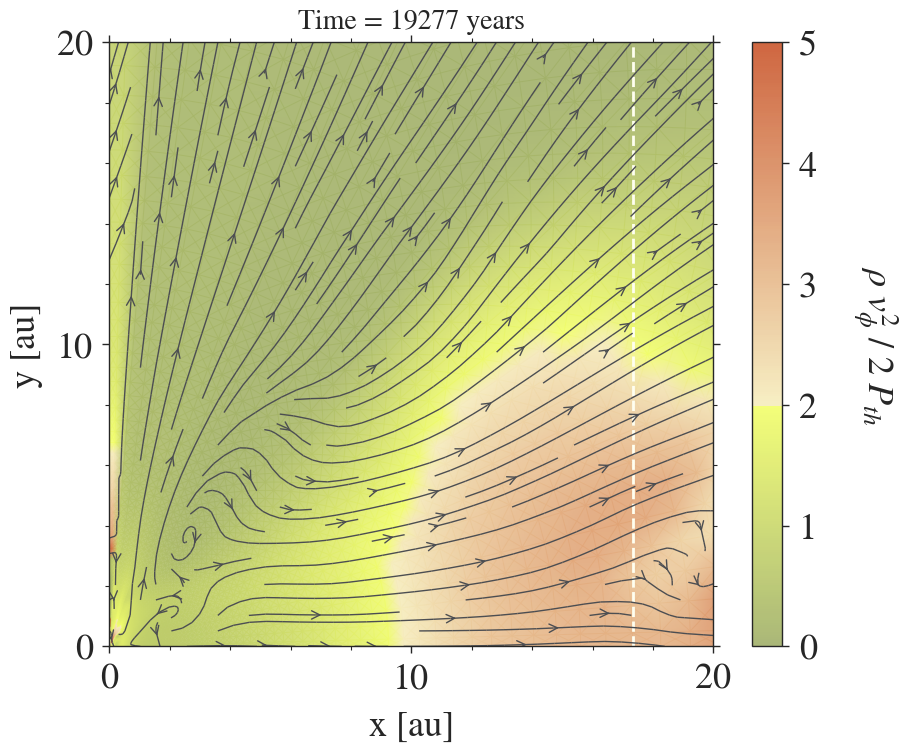}
    \caption{Rotational support (\mbox{$\rho v_{\phi}^2 / 2 > 2~P_\mathrm{th}$}) at different time snapshots highlighting the merger of the first core disc and pseudo-disc around second core formed during the 1~$M_{\odot}$ pre-stellar core collapse. The topmost plot shows the presence of both discs, which eventually merge after the outflow is launched. The white dashed vertical line marks the radius of the rotationally supported first core disc or merged disc using the definition detailed in Sect.~\ref{sec:hydrodisc}. The gas velocity streamlines show the infalling envelope, mixing within the discs, and the outflowing gas.} 
    \label{fig:discmerger}
\end{figure}

As the collapsing core evolves further, we find a thermal pressure driven outflow launched from close to the protostar at 19222~years after the onset of the 1~$M_{\odot}$ pre-stellar core collapse, which is four years after the second collapse and pseudo-disc formation. We follow its evolution for $\approx$~55 years limited by computational time expenses. An outflow is defined as the region where the radial velocity component becomes stronger than the sound speed (\mbox{$v_{r} > c_\mathrm{s}$}) as also used in \citet{Koga2022}. The polar-angle averaged radial velocity of the outflow reaches a maximum of $\approx9 ~\mathrm{km ~s^{-1}}$ and decreases over its lifetime. This transient nature of the outflow is shown in Fig.~\ref{fig:1Msun-outflowvr} over a few different time snapshots. Figure~\ref{fig:Mach} shows the radial Mach number during the first and last snapshots highlighting supersonic regions within the outflow. Regions with a stronger thermal pressure support defined when \mbox{$\rho v_{\phi}^2 / 2 < 2~P_\mathrm{th}$} within the outflow can be seen in Fig.~\ref{fig:discmerger} at different time snapshots from the onset until 55~years after the outflow is launched. Thermal pressure building up in the pseudo-disc triggers this outflow as seen in the topmost panel. This high-pressure, supersonic outflow destroys the vortical features in both the discs, while pushing gas and dust upwards from the midplane and radially outwards. As a result of the material being swept outwards we also find an eventual merger of the first core disc and pseudo-disc around the second core \citep[see also][]{Bate1998, Machida2011c}. During this period, thermal pressure becomes more dominant in most parts of the initially rotationally supported first core disc. The vertical white dashed line in Fig.~\ref{fig:discmerger} marks the gradually increasing radius of the first core disc or merged disc defined according to the criteria listed in Sect.~\ref{sec:hydrodisc}.

Thermal pressure driven outflows have also been reported in previous radiation hydrodynamic simulations \citep{Schoenke2011, Bate2011, Rodenkirch2022}. The closest comparison to our study are results from three-dimensional SPH simulations by \citet{Bate2010, Bate2011} and 2D grid-based hydrodynamic simulations by \citet{Schoenke2011}, including FLD and a realistic gas equation of state to accurately treat the effects of H$_2$ dissociation and ionisation. We focus on a qualitative comparison to these works and do not pursue a quantitative comparison due to differences in initial pre-stellar core properties, which have a strong effect on the protostellar and disc properties. \citet{Bate2010, Bate2011} investigated a 1~$M_{\odot}$ pre-stellar core collapse but with different initial size ($\approx$4679~au), rotation rates, and a uniform initial density profile. These studies report transient bipolar outflows launched after the second core formation with velocities reaching 5--$10~\mathrm{km ~s^{-1}}$ depending on the initial rotation rate. These outflows spreading to distances of about 30--60~au are stated to be an outcome of the release of accretion energy from the protostar. It also results in an eventual merger of the inner and outer discs and a shockwave propagating radially outwards. \citet{Bate2011} suggests that transient hydrodynamic outflows may reoccur if the accretion rate onto the stellar core increases and that this could be a source of episodic accretion and outburst events. Similarly, \citet{Schoenke2011} studied a 1~$M_{\odot}$ pre-stellar core collapse with an outer radius of 8700~au and different values of artificial viscosity to drive the angular momentum transport within the collapsing pre-stellar core. They also find an accretion energy driven outflow launched close to the protostar reaching distances of about 500~au. The material blown-out by the outflow is seen to be re-accreted onto the disc over an evolution period of 11590~years after the onset of second core formation. This long-term collapse calculation for a total of 56~kyr is achieved by restarting the simulation at 70~years after the protostar formation with a central sink neglecting the inner 0.7~au. Our findings of a short-lived outflow launched even in hydrodynamic simulations are in agreement with both these studies.

%%%%%%%%%%%%%%%%%%%%%%%%%%%%%%%%%%%%%%%%%%%%%%%%%%%%%%%%%%%%%%%%%%
%% Figure placement

\begin{figure*}[!htp]
    \centering
    \includegraphics[width=0.76\linewidth]{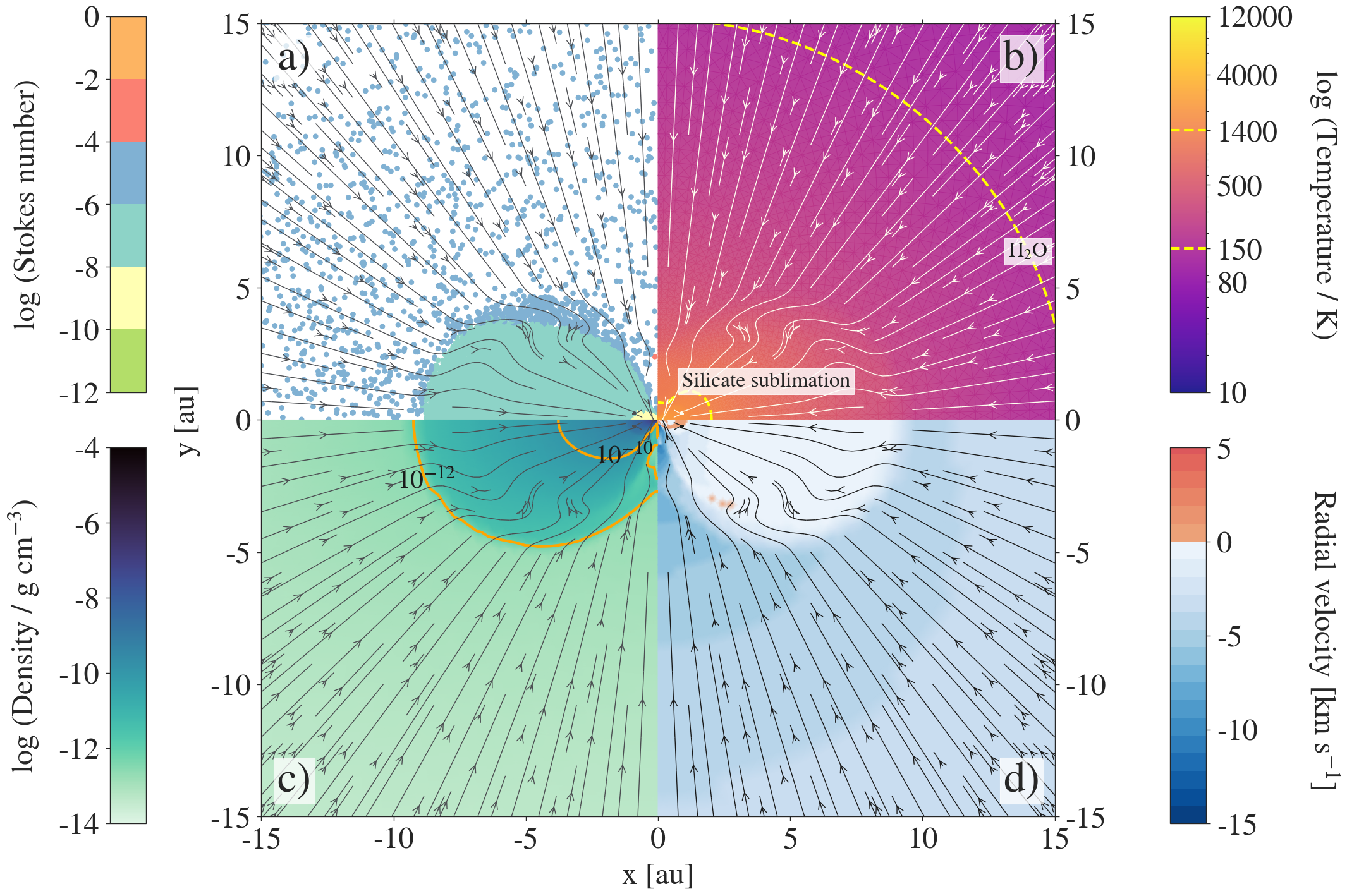}
    \caption{2D view of a first hydrostatic core evolved into a rotationally supported disc at 533~years after its formation as a result of the collapse of a 3~$M_{\odot}$ pre-stellar core with an initial rotation rate of \mbox{$\Omega_\mathrm{0} = 3.042 \times 10^{-13}$~rad~$\mathrm{s}^{-1}$} (same time snapshot as Fig.~\ref{fig:m3HD-seconddisc007}). The dust size is fixed to a constant value of 1~$\muup$m. The four panels show the \mbox{\bf a)}~Stokes number, \mbox{\bf b)}~gas temperature, \mbox{\bf c)}~gas density, and \mbox{\bf d)}~radial gas velocity within the inner 15~au of the 3000~au collapsing pre-stellar core. The gas velocity streamlines indicate the material falling onto the disc.} 
    \label{fig:m3HDdisc007}
\end{figure*}
\begin{figure*}[!htp]
    \centering
    \includegraphics[width=0.76\linewidth]{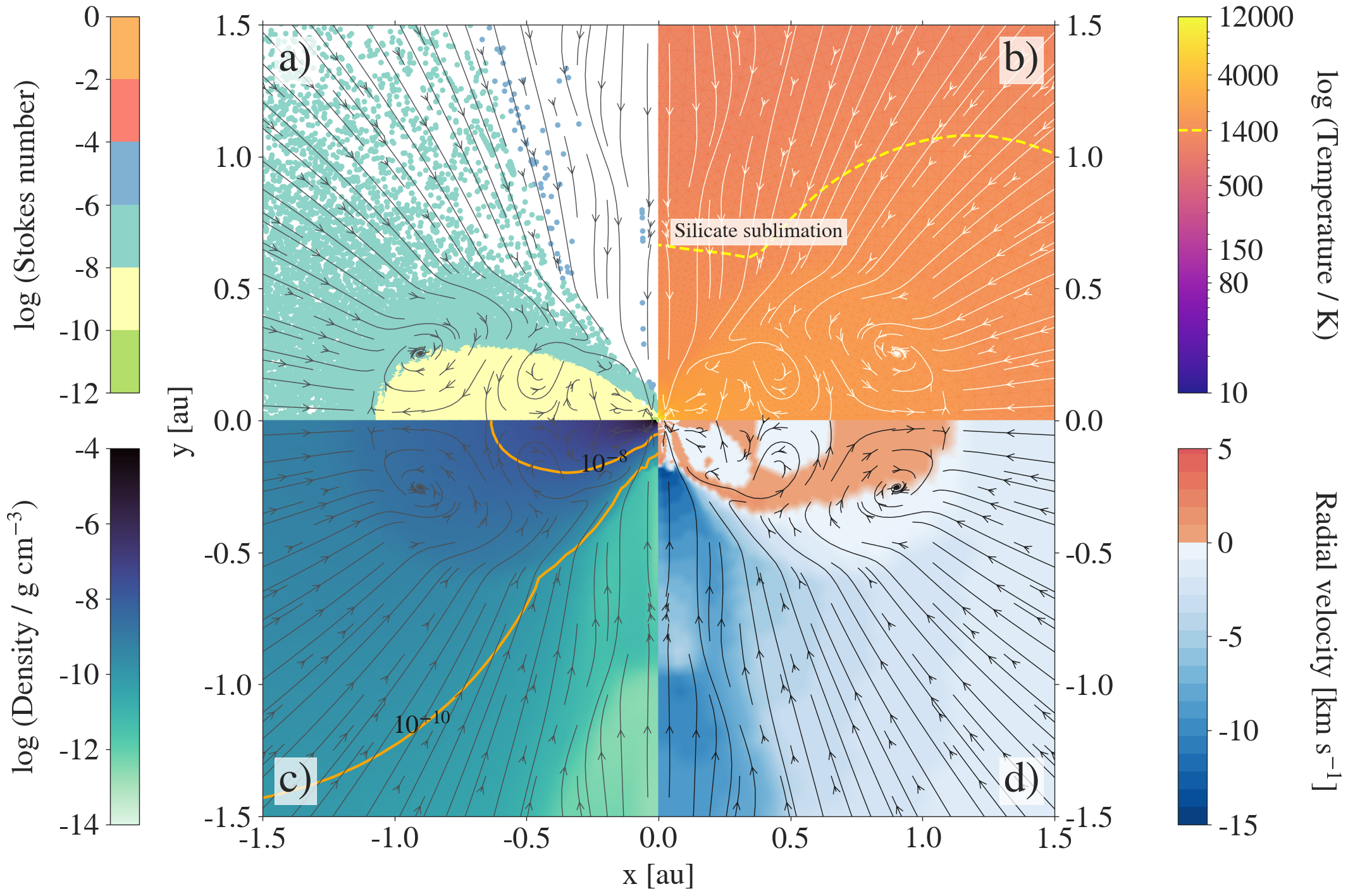}
    \caption{2D view of a pseudo-disc around the second core at two years after its formation resulting from the same initial conditions and at the same time snapshot as Fig.~\ref{fig:m3HDdisc007}. The vortical feature forming at sub-au scales in the vicinity of the unresolved protostar circulates dust in the high-temperature inner regions.}
    \label{fig:m3HD-seconddisc007}
\end{figure*}
\begin{figure}[t]
\centering
\includegraphics[width=0.9\linewidth]{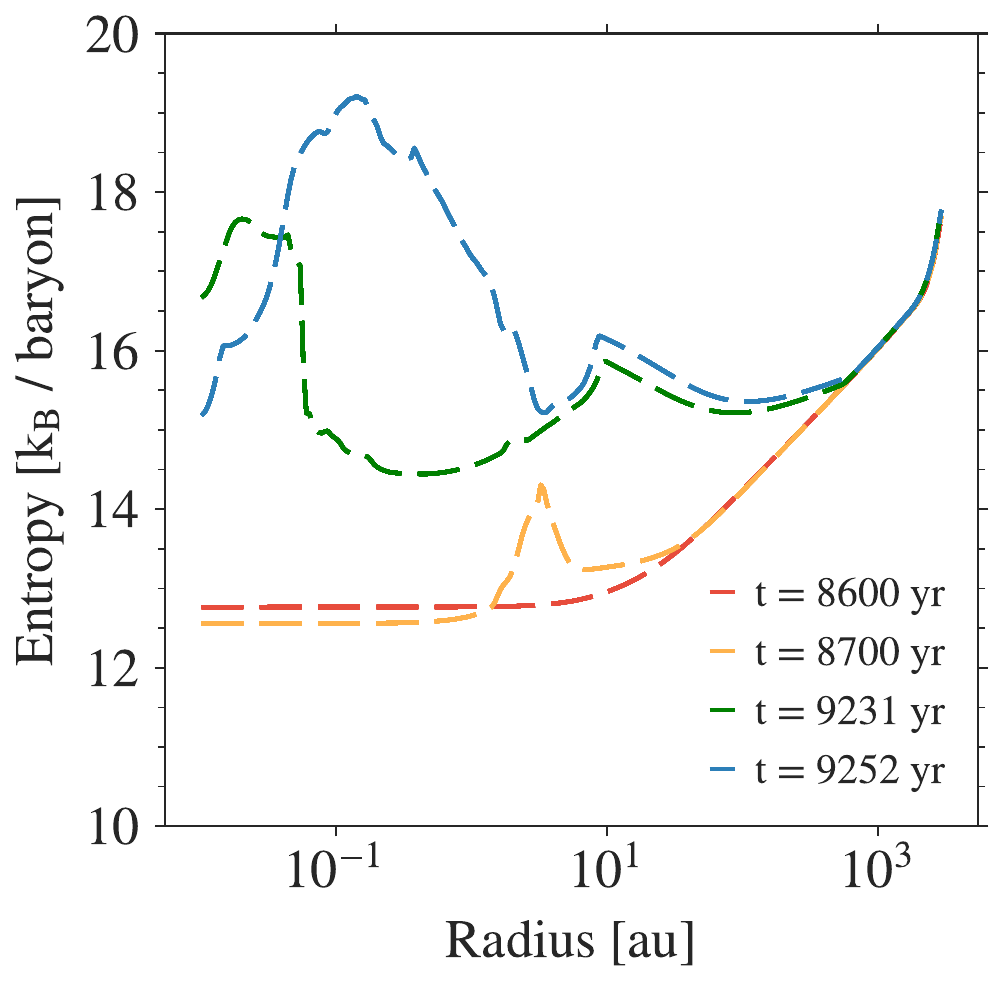}
\caption{Polar-angle averaged radial profile of the entropy at different time snapshots during the 3~$M_{\odot}$ pre-stellar core collapse. A negative radial gradient generates turbulent flows within the two discs.}
\label{fig:3Msunentropyvsr}
\end{figure}
\begin{figure}[ht!]
    \centering
    \includegraphics[width=0.9\linewidth]{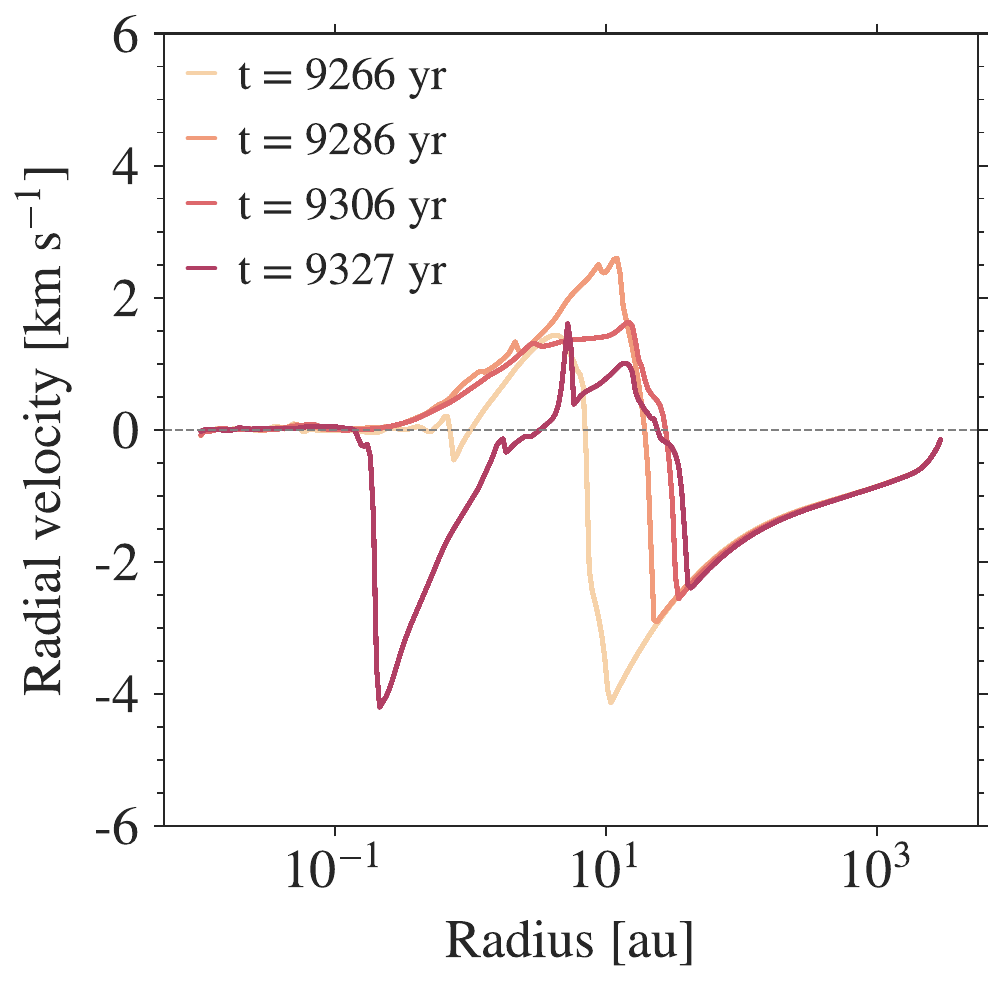} \\
    \caption{Polar-angle averaged radial velocity at different time snapshots showing the transient nature of the outflow during the 3~$M_{\odot}$ pre-stellar core collapse.}
    \label{fig:3Msun-outflowvr}
\end{figure}

%%%%%%%%%%%%%%%%%%%%%%%%%%%%%%%%%%%%%%%%%%%%%%%%%%%%%%%%%%%%%%%%%%
\subsection[Effect of initial mass: 3-Msol pre-stellar core collapse]{Effect of initial mass: 3~$M_{\odot}$ pre-stellar core collapse}
\label{sec:mass}

In this section we describe results from a comparison simulation starting with similar initial conditions to those described so far but with a more massive initial pre-stellar core of 3~$M_{\odot}$. The rotation rate fixed to 3.042~$\times 10^{-13}$~rad~$\mathrm{s}^{-1}$ gives the same $E_\mathrm{rot} / E_\mathrm{grav} =$~0.007 as for the 1~$M_{\odot}$ case. 1~$\muup$m dust is used throughout the evolution. The 3~$M_{\odot}$ pre-stellar core collapse proceeds alike the 1~$M_{\odot}$ but at a higher rate or a lower free-fall time. The first hydrostatic core transitions into a rotationally supported disc and a snapshot at 533~years after its onset is depicted in Fig.~\ref{fig:m3HDdisc007}. A further zoom-in within 1.5~au in Fig.~\ref{fig:m3HD-seconddisc007} magnifies the pseudo-disc around the second core at the same time snapshot and two years into its evolution. The four panels in both these plots show the same gas and dust properties as in Fig.~\ref{fig:HDdisc007} and Fig.~\ref{fig:HD-seconddisc007}. The structure and properties of both the discs are different in comparison to the 1~$M_{\odot}$ collapse. Dust with comparatively lower Stokes number is found at the outer edge of the first core disc whereas distribution in the pseudo-disc is similar to the 1~$M_{\odot}$ case. The temperature within the first core disc and pseudo-disc is much higher as well. For example, note the shift in the water iceline that sits further outside the first core shock and the silicate sublimation line, which has also moved to a larger radius. Both the discs are as dense as in the 1~$M_{\odot}$ collapse. The first core disc has a comparatively larger vertical spread and a slightly smaller radial extent. There is only a slight deviation from the radial inward motion of the gas and no presence of a meridional flow close to the first core shock. However, a turbulent flow is present in the pseudo-disc circulating material within the high-temperature interiors and is described in the next Sect.~\ref{sec:baro3Msun}.

\begin{figure}[t]
    \centering
    \includegraphics[width=0.82\linewidth]{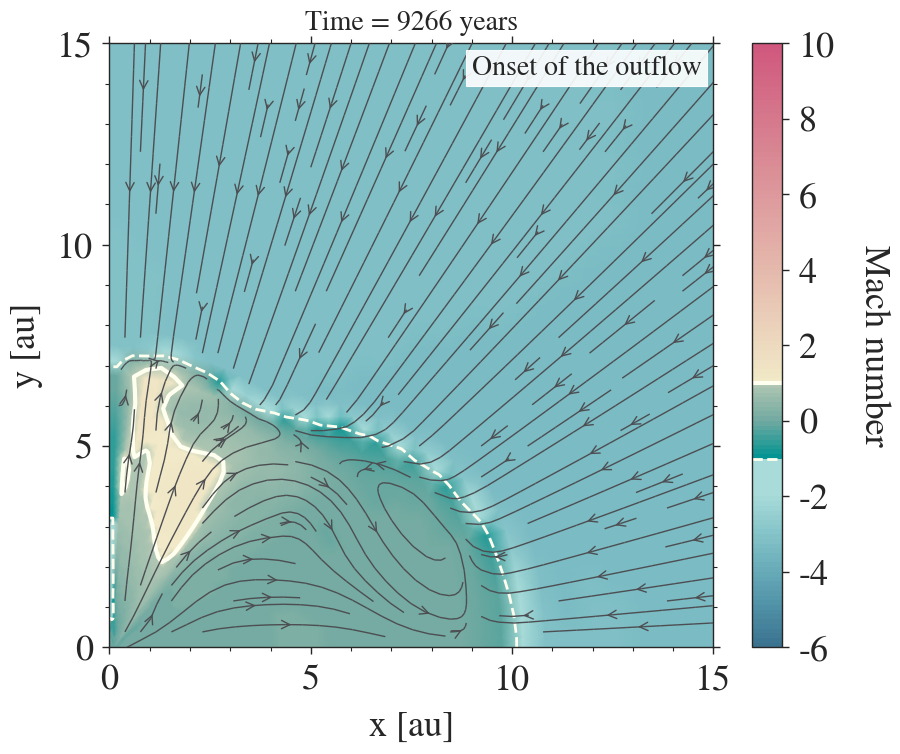} \\
    \vspace{0.2cm}
    \includegraphics[width=0.82\linewidth]{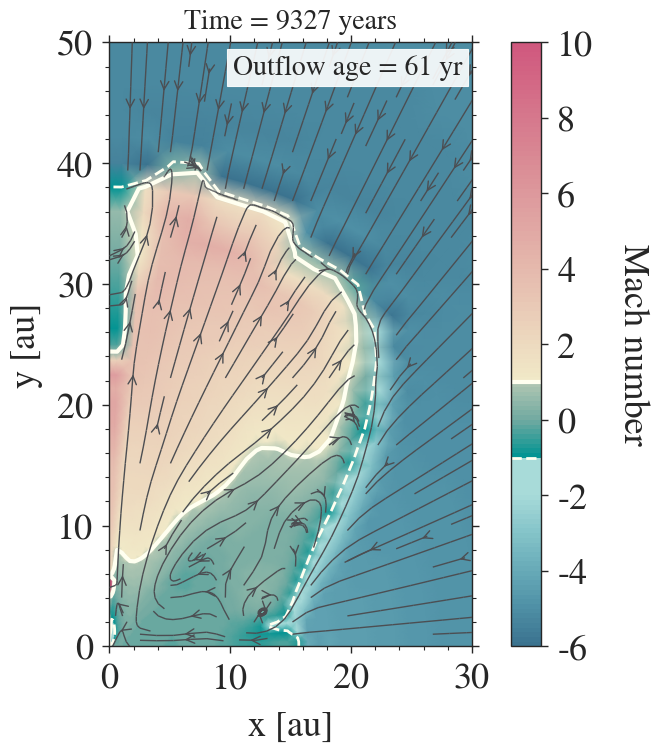} 
    \caption{Radial Mach number ($ v_{r} / c_\mathrm{s}$) at first and last time snapshots of outflow formation during the 3~$M_{\odot}$ pre-stellar core collapse. The gas velocity is indicated via the streamlines. The colour gradients in red and blue indicate the supersonic outflow and inflow regions, respectively. The colour gradient in green shows the subsonic regions mainly within the disc.}
    \label{fig:3MsunMach}
\end{figure}

%%%%%%%%%%%%%%%%%%%%%%%%%%%%%%%%%%%%%%%%%%%%%%%%%%%%%%%%%%%%%%%%%%
\subsubsection{\textit{Formation of dust traps in turbulent gas flows}}
\label{sec:baro3Msun}

In the case of a 3~$M_{\odot}$ pre-stellar core collapse we find the presence of turbulent gas flows within the pseudo-disc surrounding the second core and as the pseudo-disc mergers with the outer first core disc. Figure~\ref{fig:3Msunentropyvsr} shows the polar-angle averaged radial entropy over different time snapshots. The first core begins to form at 8700~years after the collapse is initiated. This is followed by the pseudo-disc building up around the second core at 9231~years, which evolves for a few years. The two nested discs merge by 9252~years. A negative entropy gradient generates turbulent gas flows within the two discs. In comparison with the 1~$M_{\odot}$ case, we do not find the development of a meridional flow at the outer edge of the first core disc. However, within the inner pseudo-disc, a negative entropy gradient leads to vortical features. During the merger of the two discs this feature is destroyed only to be followed by another vortical feature that develops in the upper layers. This then moves along the merged disc surface towards the midplane before being eventually destroyed by the outflow discussed next in Sect.~\ref{sec:outflow3Msun}. These transient features trap, circulate, and sweep dust outwards from the high-temperature interiors. We do not find a change in the baroclinicity at the locations of these vortical features.

%%%%%%%%%%%%%%%%%%%%%%%%%%%%%%%%%%%%%%%%%%%%%%%%%%%%%%%%%%%%%%%%%%
\subsubsection{\textit{Long-term evolution: tracing an outflow}}
\label{sec:outflow3Msun}

\begin{figure}[!htp]
    \centering
    \includegraphics[width=0.7\linewidth]{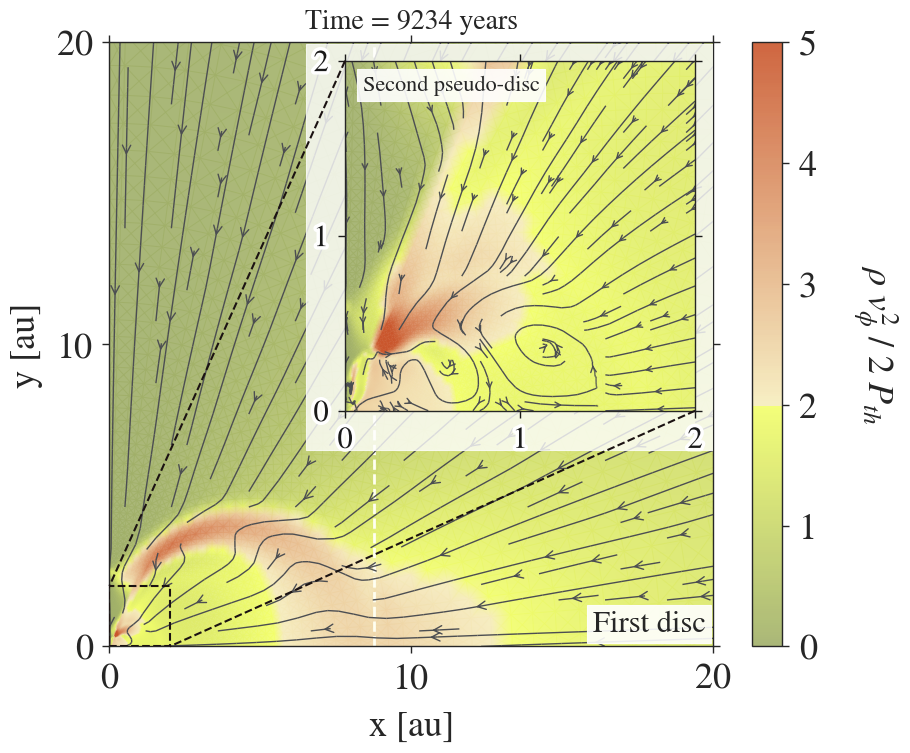}
    \includegraphics[width=0.7\linewidth]{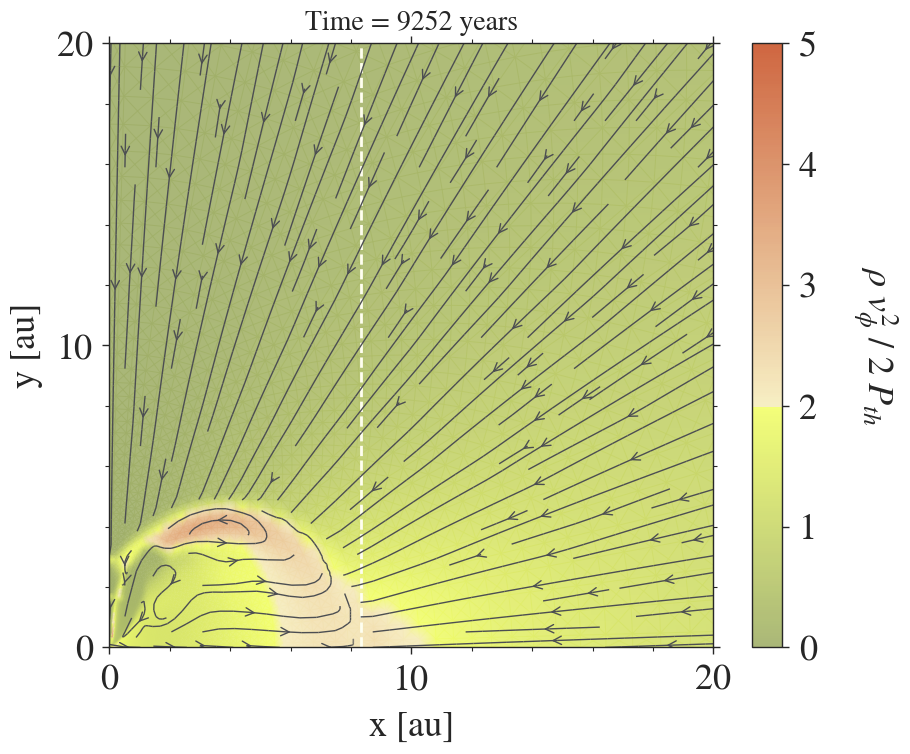}
    \includegraphics[width=0.7\linewidth]{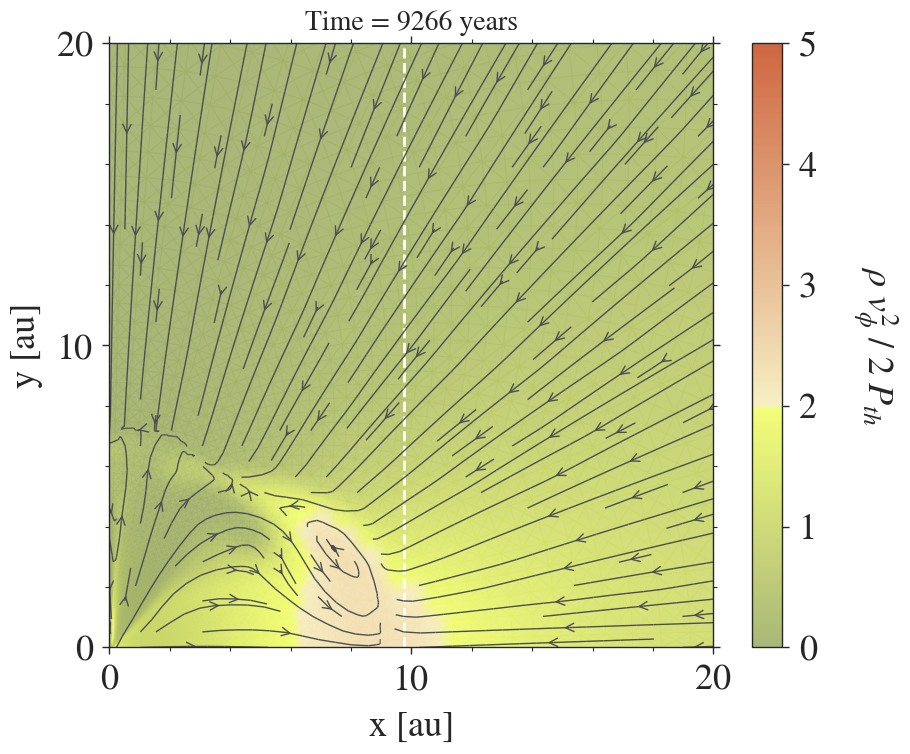}
    \includegraphics[width=0.7\linewidth]{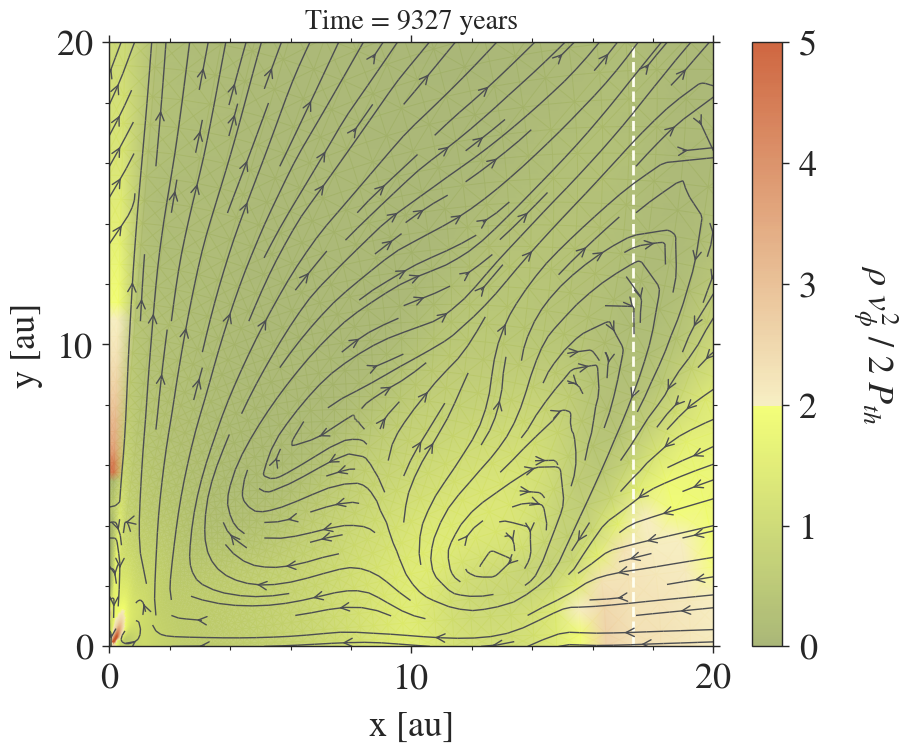}
    \caption{Rotational support (\mbox{$\rho v_{\phi}^2 / 2 > 2~P_\mathrm{th}$}) at different time snapshots highlighting the merger of the first core disc and pseudo-disc around second core formed during the 3~$M_{\odot}$ pre-stellar core collapse. The topmost plot shows the presence of both discs, which eventually merge before the outflow is launched (see third plot). The white vertical line marks the radius of the rotationally supported first core disc or merged disc using the definition detailed in Sect.~\ref{sec:hydrodisc}. The gas velocity streamlines show the infalling envelope, mixing within the discs, and the outflowing gas.}
    \label{fig:3Msundiscmerger}
\end{figure}

In comparison with the fiducial case, a thermal pressure driven outflow is launched after the two discs merge. The evolution of the outflow is followed for 61~years, reaches a maximum of $\approx6 ~\mathrm{km ~s^{-1}}$, and its velocity is seen to decrease over this time frame. The polar-angle averaged radial velocity profiles over a few different time snapshots are shown in Fig.~\ref{fig:3Msun-outflowvr}. The supersonic outflow region at its start and final simulation snapshot is shown in Fig.~\ref{fig:3MsunMach}. Thermal pressure builds up within the pseudo-disc at sub-au scales and the merged disc, which eventually triggers an outflow from the innermost regions. As in the fiducial case, the rotational and thermal support are compared in Fig.~\ref{fig:3Msundiscmerger} at different time snapshots. The vertical dashed white line marks the disc radius following the definition listed in Sect.~\ref{sec:hydrodisc}. The outflow launched at 9266~years eventually destroys the vortical flow. During the evolution of the outflow, high thermal pressure also reduces the rotational support within the merged disc. Circulation currents also occur within the outer wings of outflow as seen in the bottom-most panel. This replenishes the outer disc region with material that has been lifted up by the outflow. Once this transient outflow is quenched, the disc is expected to transition into a rotationally supported structure.

%%%%%%%%%%%%%%%%%%%%%%%%%%%%%%%%%%%%%%%%%%%%%%%%%%%%%%%%%%%%%%%%%%
\section{Results II: Dust dynamics} 
\label{sec:thermodynamics}

This section focuses on the behaviour of dust particles within transient meridional flows and outflow described in the previous section. We performed a simulation with the exact same initial conditions as for the fiducial 1~$M_{\odot}$ pre-stellar core collapse case discussed in Sect.~\ref{sec:hydrodisc} but with three different dust sizes of 1, 10, and 100~$\muup$m. We assign 50 particles per size within every grid cell to compare the behaviour of different sized dust. The three fixed dust sizes share the exact same initial location.

%%%%%%%%%%%%%%%%%%%%%%%%%%%%%%%%%%%%%%%%%%%%%%%%%%%%%%%%%%%%%%%%%%
%% Figure placement
\begin{figure}[!tp]
    \centering
    \includegraphics[width=0.78\linewidth]{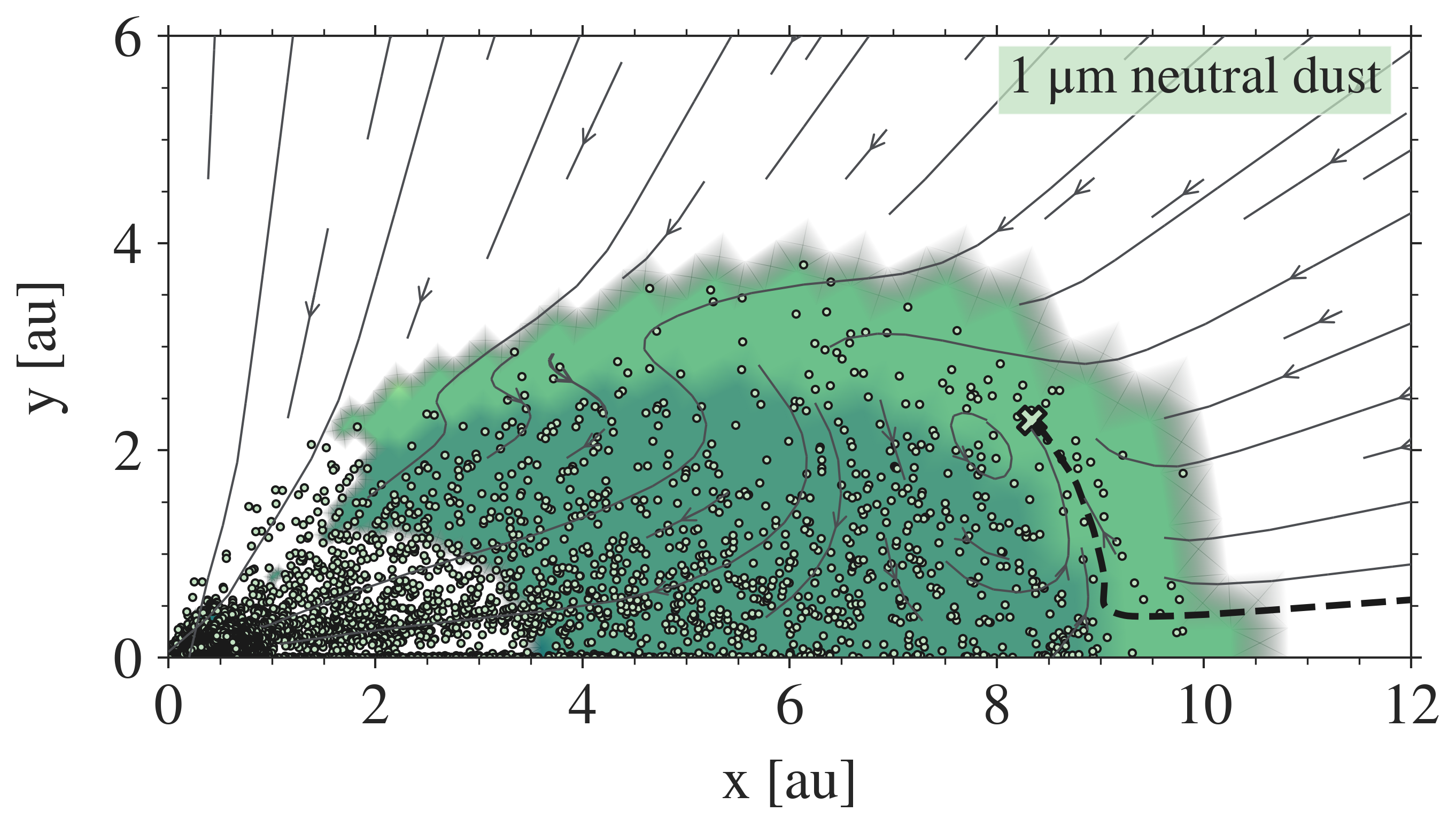} \\ \vspace{0.1cm}
    \includegraphics[width=0.78\linewidth]{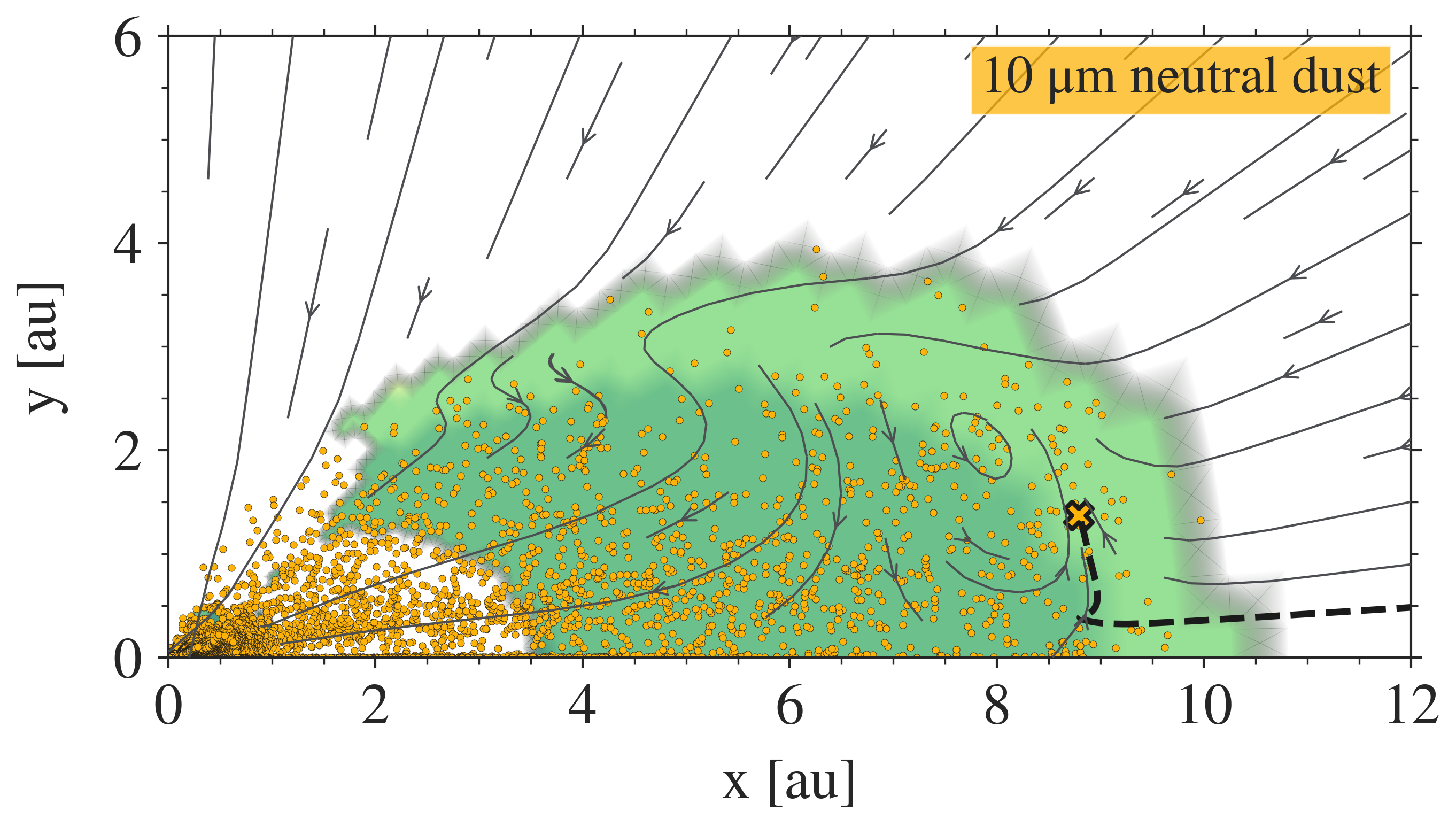} \\ \vspace{0.1cm}
    \includegraphics[width=0.78\linewidth]{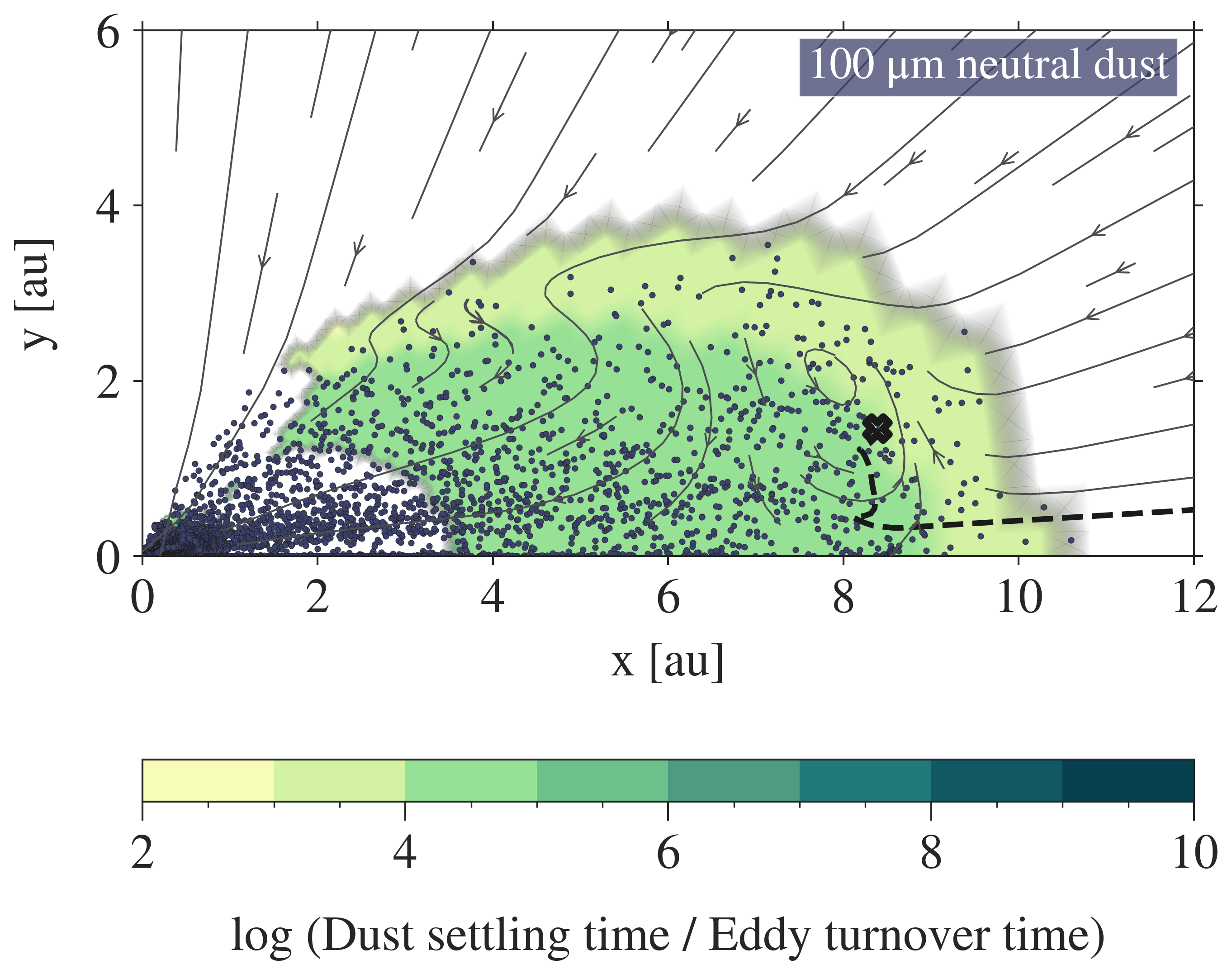}
    \caption{Ratio of dust settling to eddy turnover time only within the first core rotationally supported disc formed during the fiducial 1~$M_{\odot}$ pre-stellar core collapse. The three plots highlight the behaviour of different dust sizes. Also overplotted is a single particle trajectory as a black dashed line for three different sizes of dust particles starting at the same initial grid cell location. The gas velocity streamlines indicate the material falling onto the disc and the deviation of gas and dust flow from a standard radial inward motion.}
    \label{fig:settlingoverturnover}
\end{figure}

\begin{figure}[!hp]
    \centering
    \includegraphics[width=0.65\linewidth]{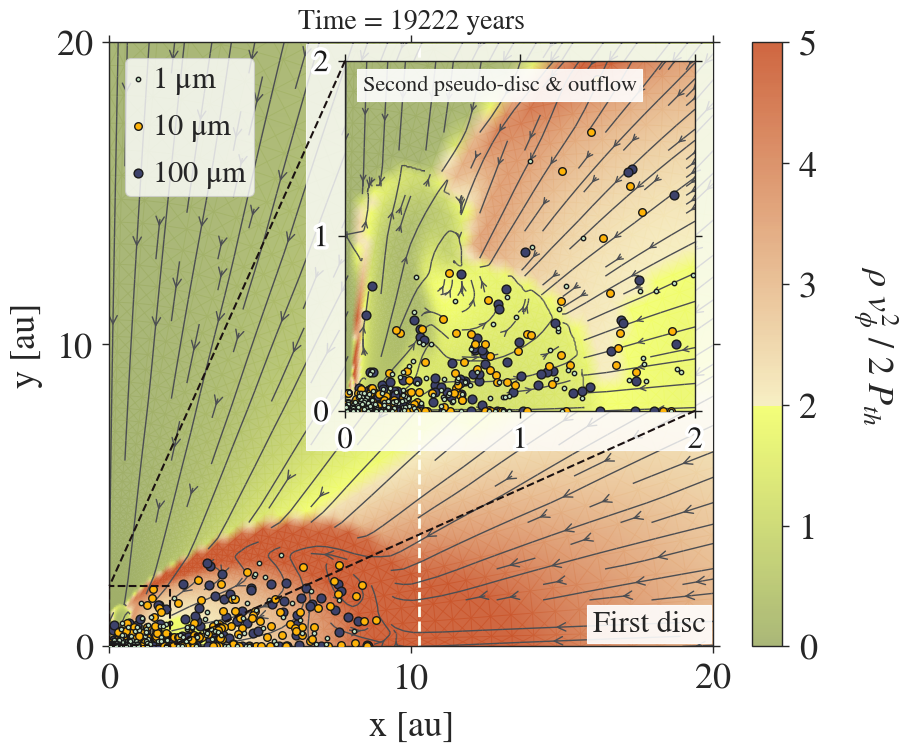}
    \includegraphics[width=0.65\linewidth]{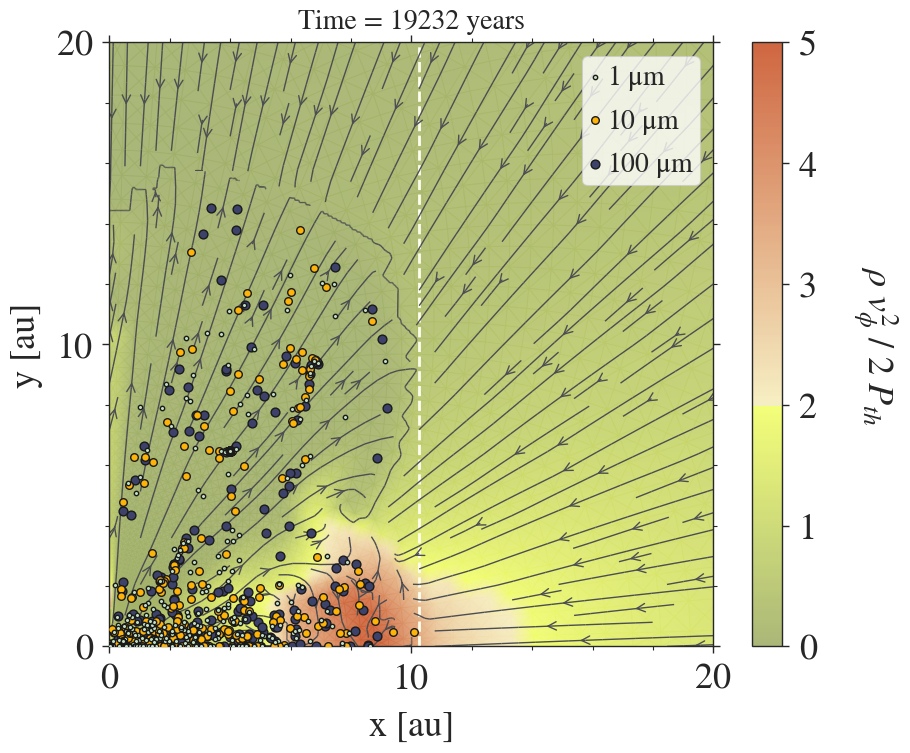}
    \includegraphics[width=0.65\linewidth]{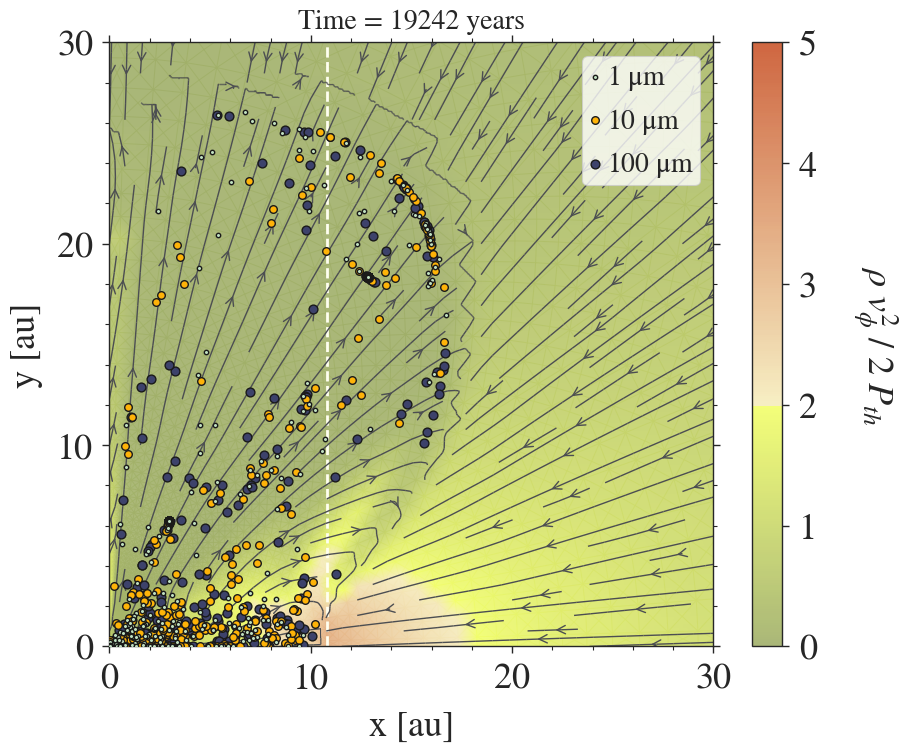}
    \includegraphics[width=0.6\linewidth]{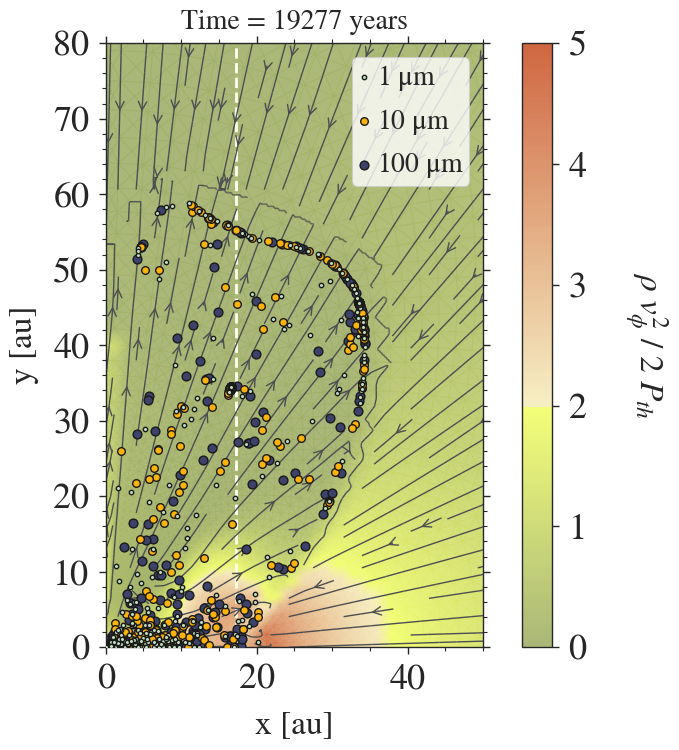}
    \caption{Rotational support (\mbox{$\rho v_{\phi}^2 / 2 > 2~P_\mathrm{th}$}) at different time snapshots during the 1~$M_{\odot}$ pre-stellar core collapse. The topmost plot shows the presence of both discs, which eventually merge after the outflow is launched. The white vertical line marks the radius of the rotationally supported first core disc or merged disc using the definition detailed in Sect.~\ref{sec:hydrodisc}. A thermal pressure driven outflow is also seen to transport dust from the inner regions. A selected sample of 1~$\muup$m (green), 10~$\muup$m (orange), and 100~$\muup$m (blue) dust from within the young protostellar disc that move radially outwards and/or above the midplane are overplotted. The gas velocity streamlines indicate the material falling onto the disc, mixing within the two discs, and outflowing. The size-independent well-coupled dust follows the gas motion in all the different regions.}
    \label{fig:dustinoutflow}
\end{figure}

\begin{figure*}[!hp]
    \centering
    \includegraphics[width=0.9\linewidth]{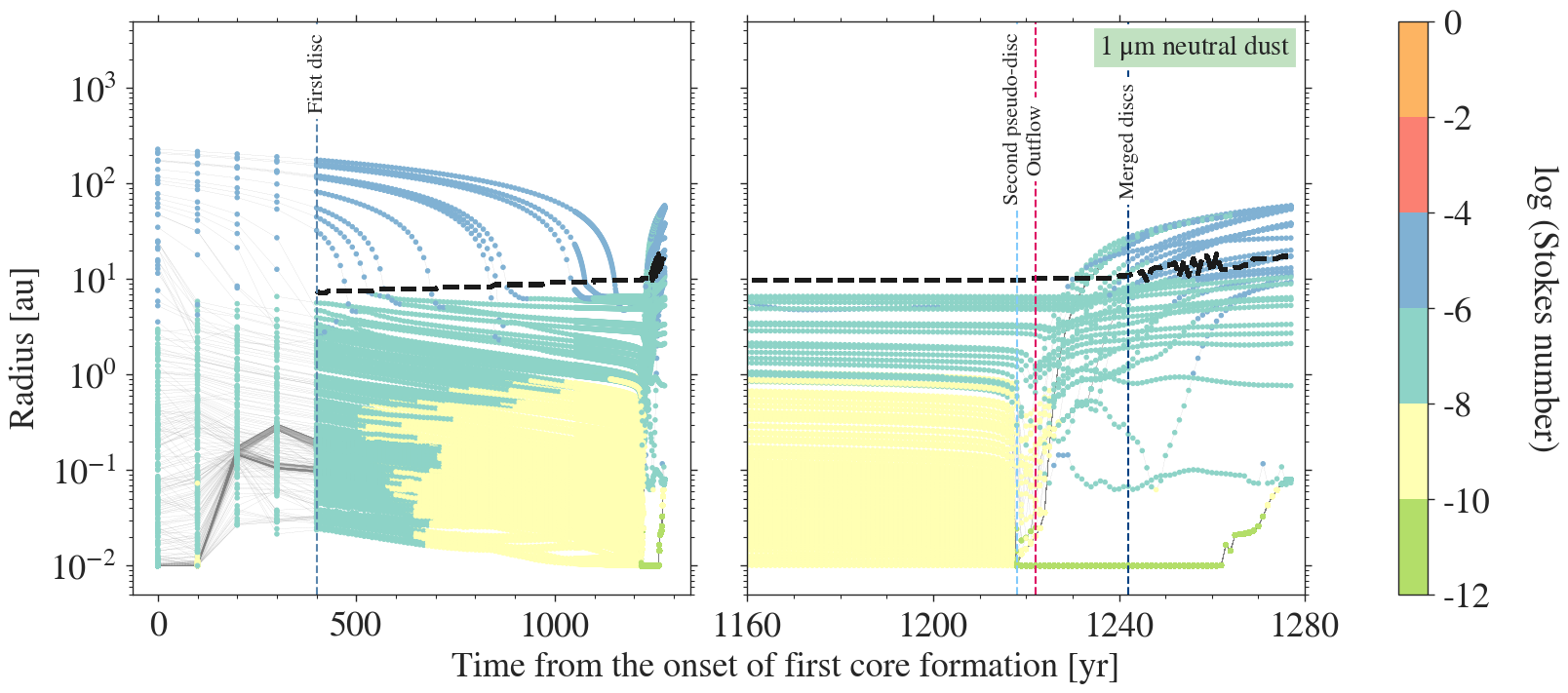} \\ \vspace{0.5cm}
    \includegraphics[width=0.9\linewidth]{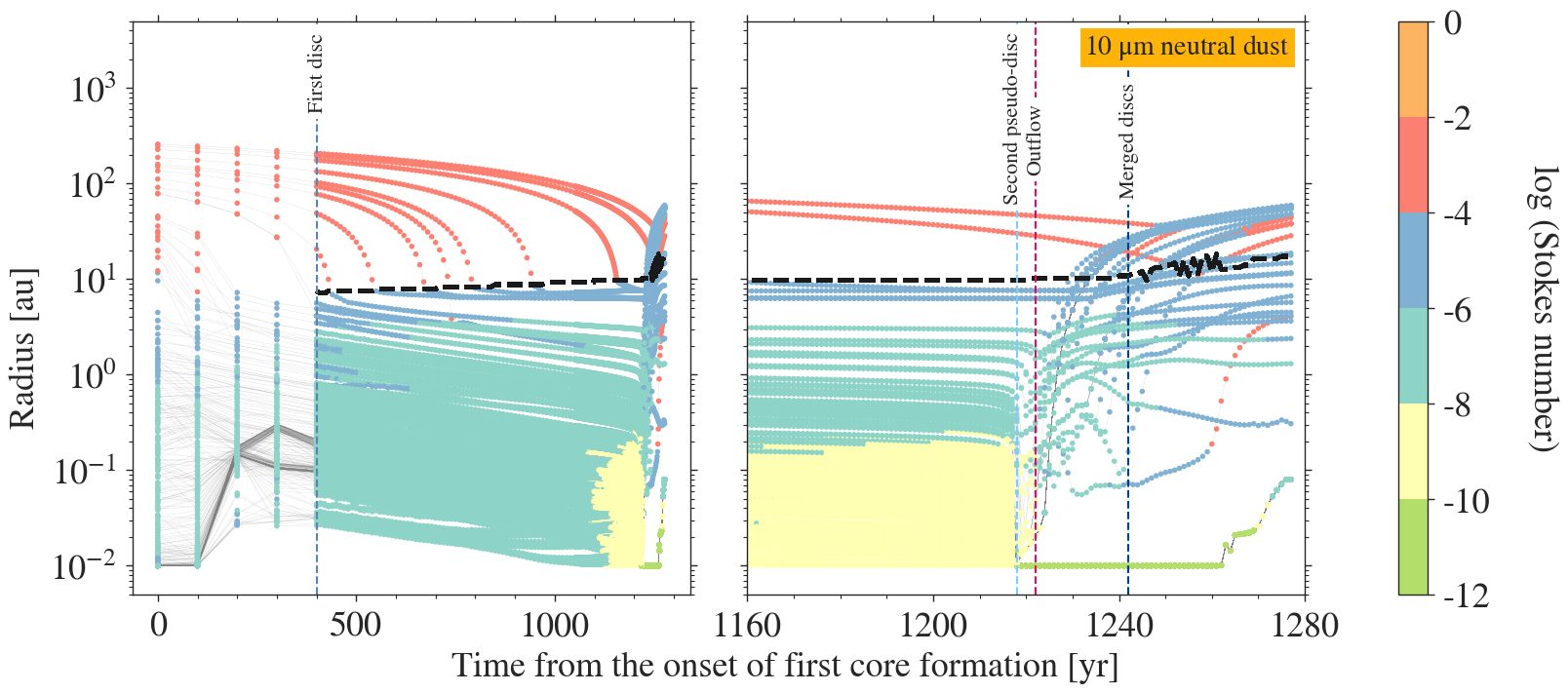} \\ \vspace{0.5cm}
    \includegraphics[width=0.9\linewidth]{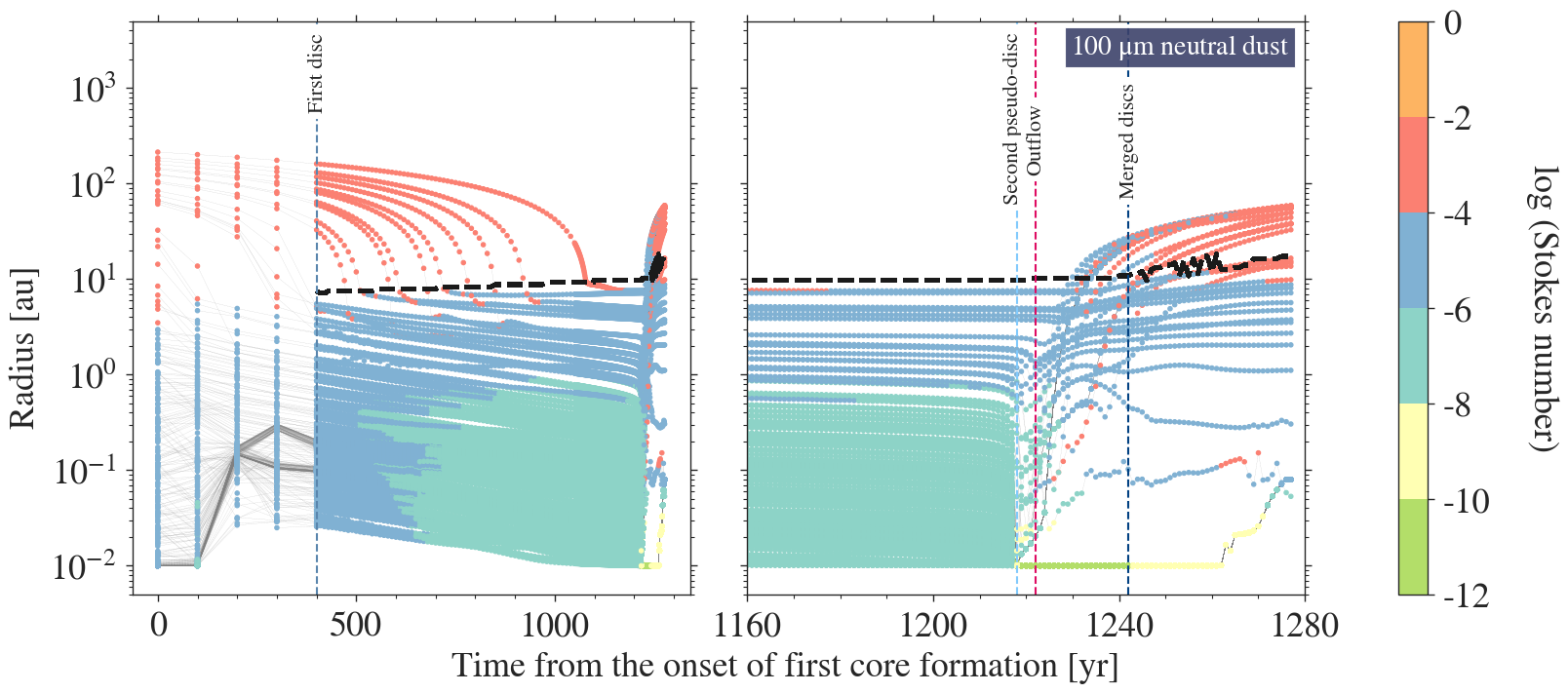}
    \caption{Tracks for a selected sample of 1~$\muup$m (top), 10~$\muup$m (middle), and 100~$\muup$m (bottom) dust that end up within the two discs during the 1~$M_{\odot}$ pre-stellar core collapse. The colourmap shows the Stokes number, which changes as the fixed-size dust moves through different density regimes within the envelope, discs, and outflow. The vertical lines indicate the formation times of the two discs, and the outflow, and merger of the discs from the onset of the first core formation ($t=0$). The horizontal dashed black line marks the first core disc or merged disc radius defined using the conditions listed in Sect.~\ref{sec:hydrodisc}. A Stokes number much smaller than one throughout the evolution in all three plots indicates the size-independent coupling to the gas. }
    \label{fig:particle_radius-Stokesvstime}
\end{figure*}

\begin{figure*}[ht]
    \centering
    \includegraphics[width=0.9\linewidth]{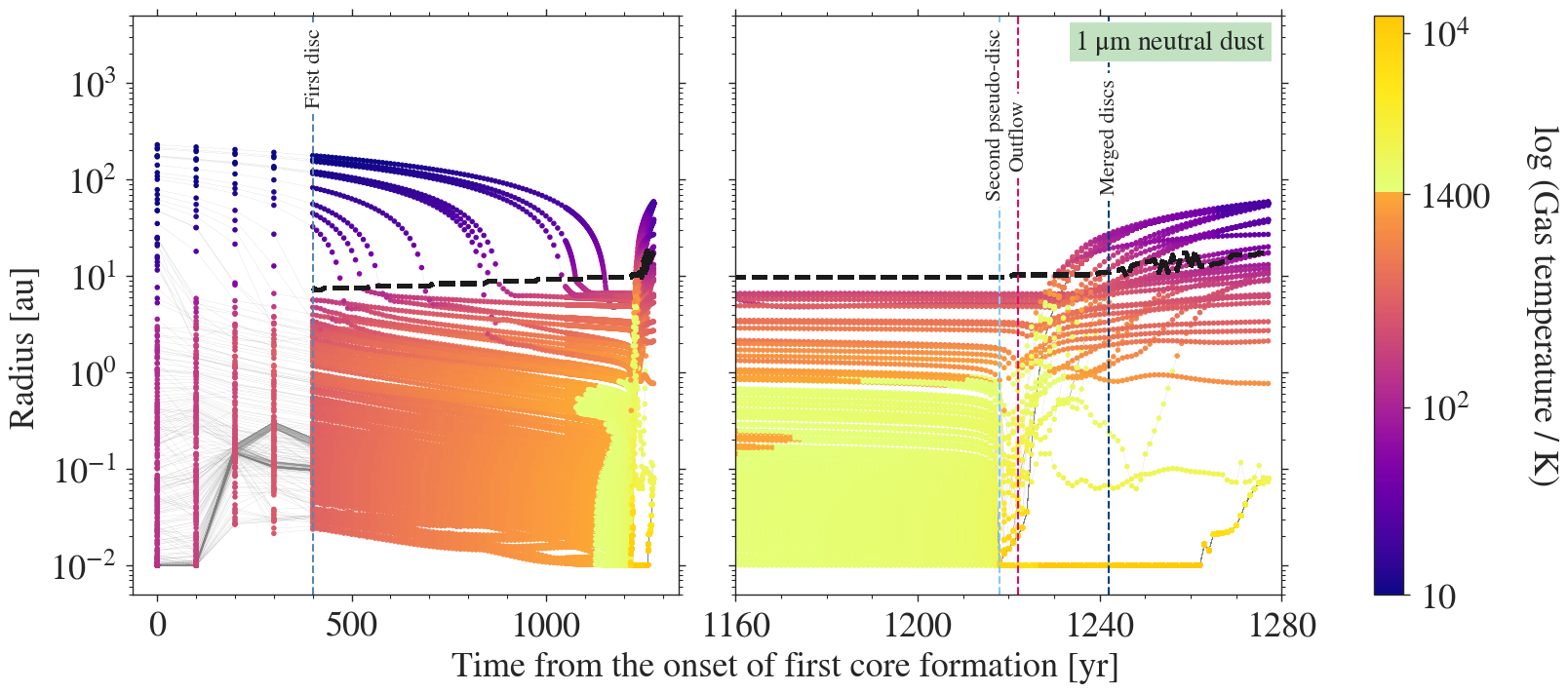} 
    \caption{Tracks for a selected sample of 1~$\muup$m dust that end up within the two discs during the 1~$M_{\odot}$ pre-stellar core collapse. The colourmap is split below and above silicate sublimation at 1400~K and indicates the gas temperature at dust location. The vertical lines indicate the formation times of the two discs, and the outflow, and merger of the discs from the onset of the first core formation ($t=0$). The horizontal dashed black line marks the first core disc or merged disc radius defined using the conditions listed in Sect.~\ref{sec:hydrodisc}. The temporal evolution of the dust provides a clear indication of thermal reprocessing within the young protostellar disc. Tracks for 10 and 100~$\muup$m can be found in Appendix~\ref{sec:tracks-allsizes} for comparison.}
    \label{fig:particle_radius-tempvstime}
\end{figure*}

%%%%%%%%%%%%%%%%%%%%%%%%%%%%%%%%%%%%%%%%%%%%%%%%%%%%%%%%%%%%%%%%%%
\subsection{Mixing and transport within high-pressure and high-temperature regions}
\label{sec:dustinflow}

Dust settling ($t_{\mathrm{dust,\,sett}}$) and eddy turnover times ($t_{\mathrm{eddy}}$) within the first core disc are compared in Fig.~\ref{fig:settlingoverturnover} and computed as 
%\begin{linenomath}
\begin{align}
\label{eq:tdustsetll}
\mbox{$ t_{\mathrm{dust,\,sett}} = \dfrac{1 + \mathrm{St}^2}{\mathrm{St} ~\Omega_{\mathrm{K}}} $},
\end{align}
%\end{linenomath}
%\begin{linenomath}
\begin{align}
\mbox{$ t_{\mathrm{eddy}} = 
\dfrac{1}{\Omega_{\mathrm{K}}} $}, 
\end{align}
%\end{linenomath}
where Stokes number $\mathrm{St}$ is previously defined in Sect.~\ref{sec:hydrodisc} and $\Omega_{\mathrm{K}} = \sqrt{G M(r)/(r\sin\theta)^3}$ \citep{Youdin2007}. The infalling gas and dust from the surrounding envelope have been masked out in Fig.~\ref{fig:settlingoverturnover}. The three plots show the behaviour of three dust sizes. As expected 100~$\muup$m dust settles faster compared to the 10~$\muup$m and 1~$\muup$m dust. However, owing to the high gas density within the disc, all three sizes have a Stokes number lower than $10^{-2}$ and can be trapped within the meridional flow instead of moving radially inwards. As an example, trajectories for the spatial motion of different sized dust initially located in the same grid cell are marked in each plot using a black dashed line. 

Regions with a stronger rotational support defined when \mbox{$\rho v_{\phi}^2 / 2 > 2~P_\mathrm{th}$} within the disc can be seen in Fig.~\ref{fig:dustinoutflow} at different time snapshots from the start of the outflow until 55~years after it is launched. Given the exact same initial conditions as used for the fiducial case, gas shows the same behaviour as in Fig.~\ref{fig:discmerger}. A selected sample of 1~$\muup$m (green), 10~$\muup$m (orange), and 100~$\muup$m (blue) dust located within the very young disc are overplotted. Dust grains are well-coupled to the dense gas within both the discs irrespective of their size and hence follow the gas motion instead of gradually moving towards or accreting onto the central protostar. The innermost disc regions experience an influx of dust either directly from regions closer to the pole or via a sliding effect over the disc surface. Once the outflow is launched dust particles from the innermost region are lifted upwards and/or swept radially outwards. The outflow velocity decreases over time (see Fig.~\ref{fig:1Msun-outflowvr}) and will eventually be quenched, returning the entrained dust back in the disc and potentially at larger radii. Similar to our findings, previous works have also suggested \citep{Bate2010} and shown protostellar disc outflows \citep{Wong2016, Tsukamoto2021b, Koga2023} and magneto-centrifugal disc winds \citep{Giacalone2019} to be carriers of $\geq$~10~$\muup$m dust.

The Lagrangian particle treatment allows us to trace the thermal history of different dust grains. Figure~\ref{fig:particle_radius-Stokesvstime} and Fig.~\ref{fig:particle_radius-tempvstime} display tracks for a selected sample of dust that end up within the two discs during the collapse and their dynamical changes due to the turbulent flows and outflow. The time $t=0$ in these figures marks the beginning of the first hydrostatic core formation. Different vertical dashed lines mark the times of onset of the two discs, outflow, and merger of the discs. 

Figure~\ref{fig:particle_radius-Stokesvstime} highlights the change in the Stokes number as the dust grains move through the envelope and mix within the disc and outflow regions with different densities. The Stokes number is found to be well below unity throughout the evolution, even for the largest 100~$\muup$m grains, which indicates a size independent strong coupling to the gas.

During post-processing we locate particles closest to a grid cell centre and tag them with gas properties such as temperature from the corresponding cell. In Fig.~\ref{fig:particle_radius-tempvstime} and Appendix~\ref{sec:tracks-allsizes}, gas temperature at the dust location is shown in colour with two different colourmaps differentiating between the temperature below and above silicate sublimation at 1400~K. A thermal pressure driven outflow and outward gas motion during the merger of the nested discs transports high-temperature sublimated dust from the innermost sub-au parts of the disc to outer regions, while cooling it down to below 1400~K. The hotter innermost regions of the disc provide ideal conditions to condensate CAIs and AOAs. Subsequent mixing and outward transport of dust via vortical gas flows and outflow could act as potential mechanisms that enable the presence of CAIs and AOAs embedded within carbonaceous chondrites found in the outer Solar System \citep{Morfill1978, Tscharnuter2009, Haugbolle2019, Tsukamoto2021b, Jongejan2023}. 

%%%%%%%%%%%%%%%%%%%%%%%%%%%%%%%%%%%%%%%%%%%%%%%%%%%%%%%%%%%%%%%%%%
\subsection{Dust mass to gas mass variations during collapse}
\label{sec:dust2gas}

\begin{figure}[!hp]
	\centering
	\includegraphics[width=0.8\linewidth]{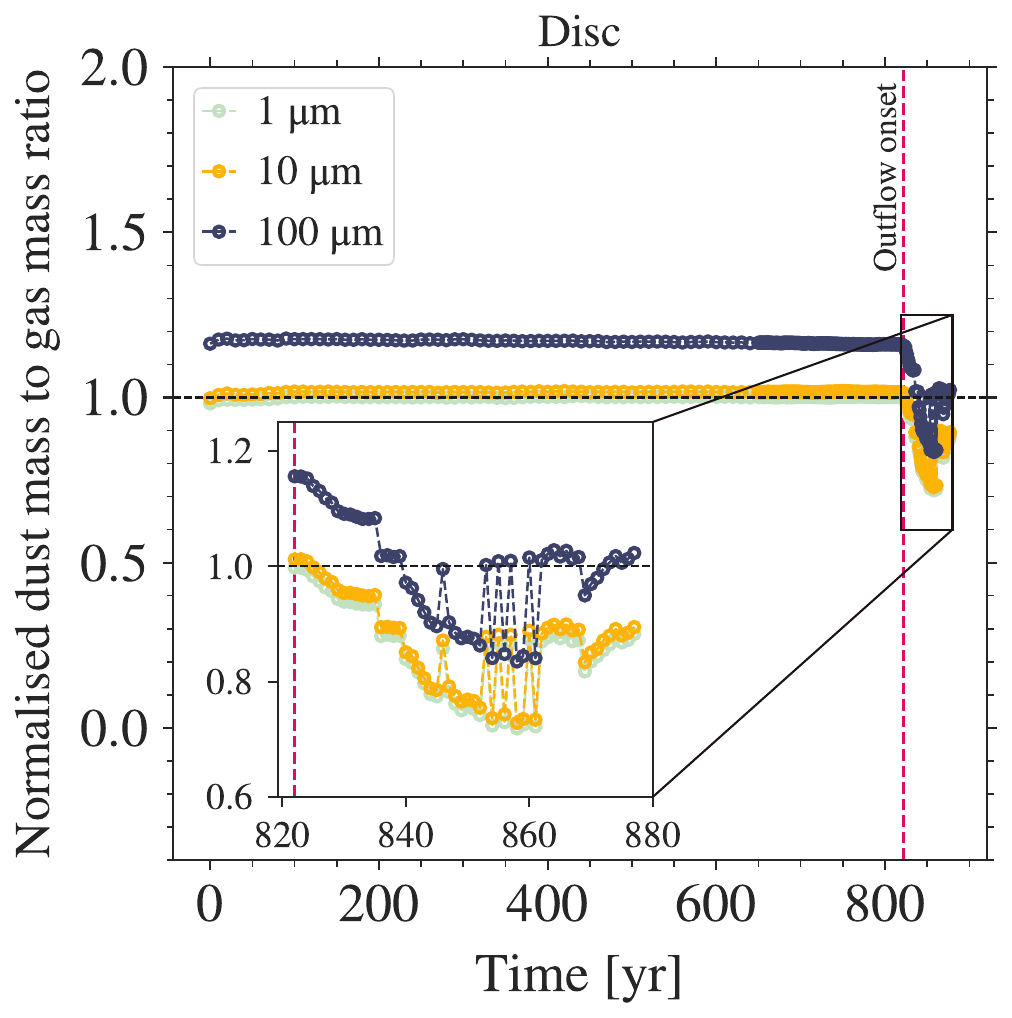} \\ 
	\vspace{0.5cm}
	\includegraphics[width=0.8\linewidth]{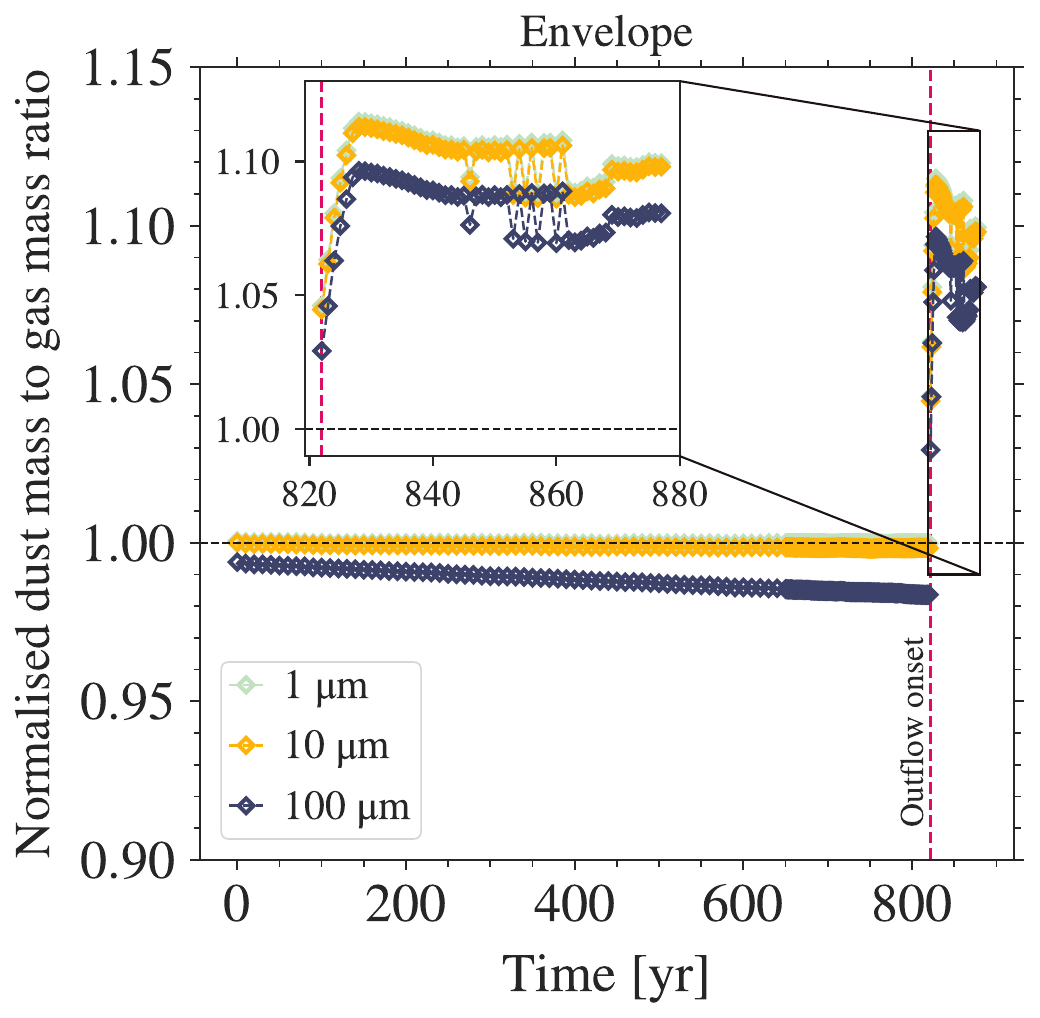} \\
	\vspace{0.5cm}
	\includegraphics[width=0.8\linewidth]{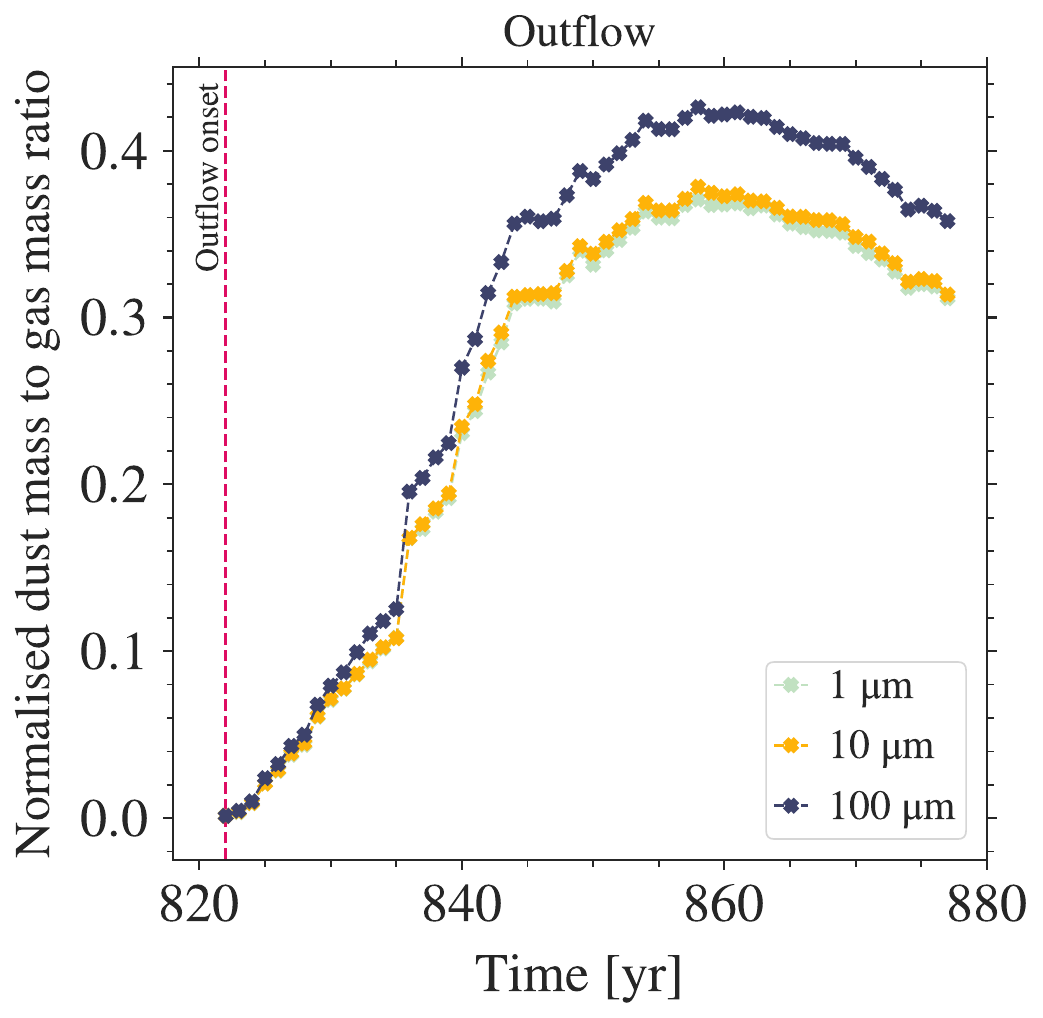} 
	\caption{Time evolution of the dust mass to gas mass ratio normalised to its initial value for the three dust sizes within the two nested and merged discs (top), infalling envelope (middle), and outflow (bottom) formed during the collapse of a 1~$M_{\odot}$ pre-stellar core. Time $t=0$ indicates the beginning of transition of the first hydrostatic core into a rotationally supported disc. The vertical dashed magenta line marks the launch of the outflow.}  %onset at 18400
	\label{fig:dustmasstogasmass}
\end{figure}

\begin{figure}[!htp]
	\centering
	\includegraphics[width=0.75\linewidth]{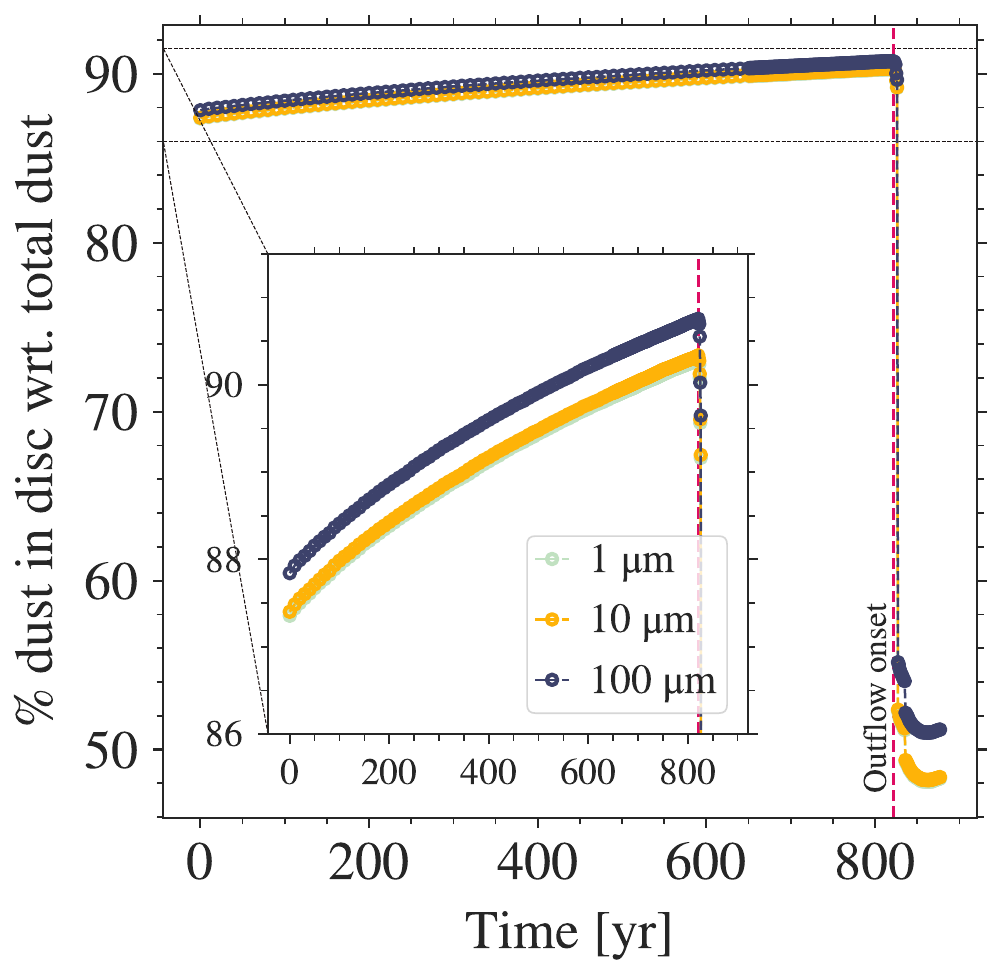} \\ 
	\vspace{0.5cm}
	\includegraphics[width=0.75\linewidth]{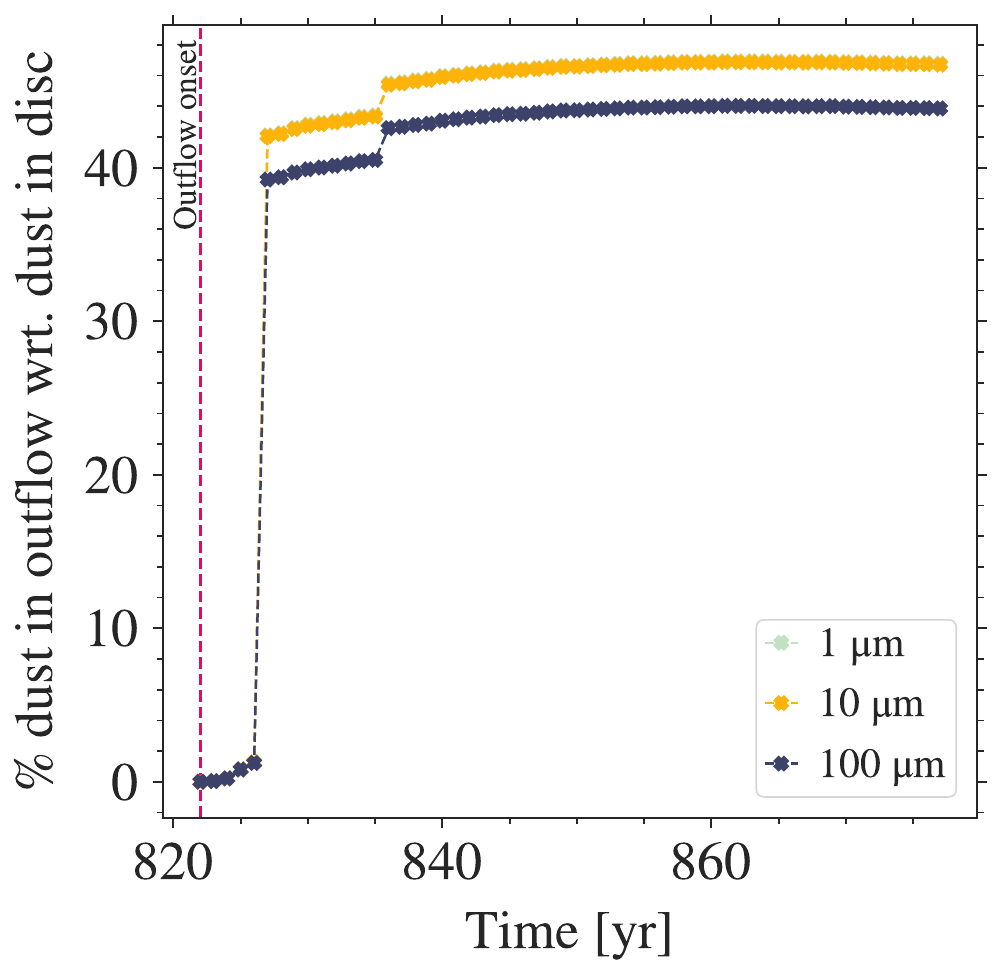} \\
	\caption{Fraction of dust particles within the disc and outflow over time with respect to its reservoir. Top: Time evolution of the fraction of dust particles within the disc with respect to the total number of particles used in the simulation. Bottom: Time evolution of the fraction of dust particles entrained in the outflow with respect to those present in the disc just before the outflow is launched. Time $t=0$ indicates the beginning of transition of the first hydrostatic core into a rotationally supported disc formed during the collapse of a 1~$M_{\odot}$ pre-stellar core. The vertical dashed magenta line marks the launch of the outflow.}  %onset at 18400
	\label{fig:numberofdust}
\end{figure}

In Fig.~\ref{fig:dustmasstogasmass} we plot the ratio of the dust mass and enclosed gas mass summed within the nested and merged discs, infalling envelope, and the outflow, normalised to its initial value. We use global values within different regions since we observe that the local dust density is highly sensitive to the particle sampling and grid resolution for a Lagrangian treatment \citep{Commercon2023}. Time $t=0$ in these plots indicates the beginning of transition of the first core into a rotationally supported disc. As the collapse forms a protostar-disc-outflow system, we find an increase in the normalised dust mass to gas mass ratio within the disc until the outflow is launched. As more of the 100~$\muup$m (blue) dust settles in the disc, an increase in the ratio is slightly higher compared to the 10~$\muup$m (orange) and 1~$\muup$m (green) dust (see also Fig.~\ref{fig:numberofdust}). During the same period, dust is depleted in the envelope and the ratio slowly decreases over time until the outflow is triggered. As the outflow evolves while lifting dust from the disc there is a slight enhancement of the ratio in the outflow and a corresponding depletion in the disc. As mentioned earlier in Sect.~\ref{sec:outflow1Msun}, the outflow velocity decreases over time marking its transient nature, which enables a replenishment of dust back into the disc. This is seen as a decrease in the normalised dust mass to gas mass ratio within the outflow and a subsequent increase in the disc. The outflow shock front also prevents the infalling dust to be transported in the disc and creates a short-term increase in the normalised dust mass to gas mass ratio in the surrounding envelope. 

%%%%%%%%%%%%%%%%%%%%%%%%%%%%%%%%%%%%%%%%%%%%%%%%%%%%%%%%%%%%%%%%%%
\subsection{Comparisons with previous work}
\label{sec:dustcomparison}

Recent work by \citet{Koga2022} and \citet{Koga2023} made use of a fluid method for the gas evolution and a Lagrangian particle treatment for the dust (without dust feedback) on a three-dimensional nested grid. This study treats the effects of non-ideal magneto-hydrodynamics via the Ohmic dissipation term and uses a barotropic equation of state. They follow the evolution of the protostar--disc--outflow system for 85000 years and track the motion of neutral dust particles (0.01--1000~$\muup$m) throughout the collapse spanning the 6130~au-wide envelope down to 1~au. During a 1.25~$M_{\odot}$ pre-stellar core collapse they also find $\leq$~10~$\muup$m dust to be very well coupled to the gas, in agreement with \citet{Bate2017} and \citet{Lebreuilly2020}. Larger $\geq$~100~$\muup$m dust concentrates comparatively faster in the central highest density region and within the rotationally supported $\sim$20~au disc due to a high Stokes number and hence weak coupling in the low-density envelope. Additionally, dust up to 100~$\muup$m is entrained in the outflow driven from the disc surface reaching $\sim$1000~au. They report a size-independent dust motion within the young dense protostellar disc controlled by the gas flow, similar to our findings. In their three-dimensional simulation, a radial inward drift outside of the spiral structure found in the outer disc is prevented as this region receives angular momentum from the inner part via the gravitational instability. On the other hand, in our two-dimensional simulation, we find dust concentration in the outer disc via meridional gas flows.

In \citet{Koga2022} and \citet{Koga2023}, dust-to-gas mass ratio normalised to the initial value is computed by introducing gas tracer particles and mass weighting. In order to overcome the limitations of the Lagrangian particle treatment, the dust-to-gas mass ratio is summed up within specific regions (such as disc, outflow, envelope) and, similarly to our treatment, is not evaluated at each location. In agreement with the results presented here, a variation within 10$\%$ from the initial value is seen in the dust-to-gas mass ratio for $\leq$~10~$\muup$m grains. On the other hand, 100~$\muup$m grains show a larger increase in the dust-to-gas mass ratio within the disc and outflow and a stronger depletion in the envelope. In both studies using a Lagrangian particle treatment, dust-to-gas mass ratio variation in different regions is a few times lower than that reported in \citet{Lebreuilly2020} who treat the dust using a fluid method. 

%%%%%%%%%%%%%%%%%%%%%%%%%%%%%%%%%%%%%%%%%%%%%%%%%%%%%%%%%%%%%%%%%%
\section{Future additions to the collapse models}
\label{sec:limitations}

The gravitational collapse scenario of a pre-stellar core is outlined through the formation stages of the first core transitioning into a disc, protostar (or second core), a second pseudo-disc, and an outflow driven by thermal pressure. Here we focus on the effects of self-gravity, radiation transport, an accurate gas equation of state to model $\mathrm{H_2}$ dissociation, and solid-body rotation during these different evolutionary phases. The influence of the fluid gas drag on different sized dust particles is also detailed. Future improvements to our current set-up will be able to provide further understanding of complex physical processes affecting gas and dust dynamics within young protostellar discs. 

The importance of non-ideal magneto-hydrodynamics effects \citep{Lesur2023, Tsukamoto2023PPVII} on gas and dust interaction in two- and three-dimensional simulations, especially on formation of vortical gas features and outflows, will be studied in our follow-up work. Modelling long-term disc evolution \citep{Vorobyov2018} through the episodic accretion phase is essential to study the origin of FU Orionis and EX Orionis outbursts that have been widely observed in low-mass star--disc systems \citep[e.g.][]{Cieza2018} and their impact on dust dynamics. Additionally, the consequences of aerodynamic drag forces including dust back-reaction \citep{Hennebelle2023} and the Lorentz force for charged dust on formation of substructures and instabilities through disc evolution stages and for a dust size distribution need to be investigated. 

A further step to interpret the influence of dust evolution \citep{Birnstiel2012, Testi2014, Birnstiel2023} will be to combine ongoing independent efforts to include charged dust species \citep[e.g.][]{Tsukamoto2021a}, dust growth \citep[e.g.][]{Tsukamoto2021b, Beitia-Antero2021, Tu2022, Tsukamoto2023, Vorobyov2024}, coagulation \citep[e.g.][]{Guillet2020, Tominaga2021, Marchand2022, Bate2022}, fragmentation \citep[e.g.][]{Kawasaki2022, Kawasaki2023}, evaporation \citep[e.g.][]{Grassi2017}, and also the porosity \citep[e.g.][]{Ormel2009, Ormel2011} of the dust in the current collapse models. These effects can significantly alter the coupling between magnetic fields and gas flow via changes in opacity, resistivity, and ionisation and in turn impact disc formation, disc evolution, and dust dynamics. 

%%%%%%%%%%%%%%%%%%%%%%%%%%%%%%%%%%%%%%%%%%%%%%%%%%%%%%%%%%%%%%%%%%
\section{Summary}
\label{sec:summary}

A self-consistent numerical analysis of the role of gas and dust dynamics in the context of gravitational pre-stellar core collapse has received attention only over the last few years and is slowly gaining momentum. In this study we shed some light on the effect of gas drag on dust transport during protostar and protostellar disc formation stages. We explored different phases of a collapsing pre-stellar core with a microphysical lens accounting for interactions between fluid gas and Lagrangian dust. We performed two-dimensional radiation hydrodynamics simulations for different initial pre-stellar core masses, sizes, and rotation rates. We followed the evolution for a few tens of years after the protostellar formation, while resolving the innermost sub-au regions (i.e.~no sink). Spatial and temporal evolution of neutral, spherical dust with three fixed sizes of 1, 10, and 100~$\muup$m was traced through the first core, disc, protostar, and outflow formation stages. Our main findings are summarised below:   
\begin{itemize}
\item Pressure and temperature within young protostellar discs are much higher than the values assumed for an isolated cold nebula, thereby suggesting revisions to cosmochemical findings.
\item Qualitatively, for all initial pre-stellar core configurations discussed here, disc formation starts even `before' the formation of the protostar. Following the transition of the first core into a rotationally supported disc, another pseudo-disc builds up around the forming protostar. The nested inner and outer discs eventually merge into a single disc. During further evolution, this protostar--disc system is also disrupted by a transient outflow, launched in the vicinity of the protostar and driven by thermal pressure.
\item The order of the disc merger and outflow launching depends on the initial pre-stellar core mass. The evolutionary pathway for a 1~$M_{\odot}$ collapse unfolds as follows: First core; first core disc; protostar and second inner pseudo-disc; outflow; inner and outer disc merger. Using the same initial pre-stellar core properties but with a higher mass of 3~$M_{\odot}$ the two discs merge before the onset of the outflow. The sequence then becomes first core; first core disc; protostar and second inner pseudo-disc; inner and outer disc merger; outflow.
\item The size, mass, and lifetime of the young protostellar disc have a strong dependence on the initial pre-stellar core size, mass, and ratio of rotational-to-gravitational energy. 
\item The early stages of disc formation and evolution are extremely dynamic during which the dust mostly follows the gas flow. The low Stokes number (far below unity) due to a high gas density within the disc keeps even large dust (up to 100~$\muup$m) very well coupled to the gas. As the pre-stellar core evolution proceeds through the first and second collapse stages, we find the presence of two transient gas features, namely meridional gas flows and outflow, that enable mixing and transport of dust and gas within the nested discs. \\
\begin{itemize}
\vspace{-0.2cm} 
\item[$\bullet$] Meridional gas flows: In the case of a 1~$M_{\odot}$ collapse 
a meridional flow is generated at the outer edge of the first core disc and another vortical feature within the inner pseudo-disc. On the other hand, in the 3~$M_{\odot}$ case, this transient dust trap only appears in the inner pseudo-disc. These turbulent flows are a consequence of a negative radial entropy gradient. The meridional gas flow at the outer edge of the first core disc is further enhanced due to a non-zero baroclinicity that develops over its evolution. Such meridional gas flows circulating material for a few hundred years in the outer disc and vortical features in the inner sub-au regions lasting for a few years can act as ideal locations to concentrate and mix well-coupled dust. This can provide building material for early dust growth at varied locations within the protostellar disc leading to a heterogeneous planet population. \\
\vspace{-0.2cm} 
\item[$\bullet$] Outflow: A thermal pressure driven outflow is launched a few years after the protostellar birth. This outflow is seen to transport up to 100~$\muup$m dust upwards to distances of few tens of au and radially outwards until the outer edge of the first core disc. The entrained dust can be resupplied back into the very young disc due to the short lifetime of the outflow. This provides a possible mechanism to transport dust from the hotter ($\geq$1400~K) innermost sub-au regions to the  colder ($\geq$150~K) outer parts of the disc. An analogue of this remixing could be a solution to condensate, transport, and store CAIs and AOAs mostly found in the outer Solar System. 
\end{itemize}
\end{itemize}

Our results showcase fresh insights from pre-stellar core collapse for gas evolution and its influence on dust dynamics during the birth of a protostar and protostellar disc that could have significant implications for thermal reprocessing, cosmochemical evolution, and early protoplanet formation.

%%%%%%%%%%%%%%%%%%%%%%%%%%%%%%%%%%%%%%%%%%%%%%%%%%%%%%%%%%%%%%%%%%
\begin{acknowledgements} 
The authors thank the referee for their positive and constructive feedback. AB would like to thank Manuel Riener, Hans Baehr, Francesco Lovascio, David Melon Fuksman, Hubert Klahr, Yusuke Tsukamoto, Kengo Tomida, Yves Marrocchi, Enrico Ragusa, and Benedetta Veronesi for helpful comments and discussions during this project. We gratefully acknowledge support from the PSMN (P\^{o}le Scientifique de Mod\'{e}lisation Num\'{e}rique) of the ENS de Lyon for the computing resources. This project was partly supported by the IDEX-Lyon Turbullet project (contract ANR-16-IDEX-0005). AB and GL acknowledge funding from the ERC CoG project PODCAST No. 864965. AB also acknowledges funding by the Deutsche Forschungsgemeinschaft (DFG, German Research Foundation) under Germany's Excellence Strategy - EXC-2094 - 390783311. BC acknowledges funding from the French Agency Nationale de la Recherche (ANR) through the projects DISKBUILD (ANR-20-CE49-0006) and PROMETHEE (ANR-22-CE31-0020). GDM acknowledges the support of the DFG priority program SPP 1992 ``Exploring the Diversity of Extrasolar Planets'' (MA~9185/1) and from the Swiss National Science Foundation under grant 200021\_204847 ``PlanetsInTime''. Parts of this work have been carried out within the framework of the NCCR PlanetS supported by the Swiss National Science Foundation. \\
\\
\textit{Data availability:} \mbox{\url{https://asmitabhandare.github.io/##tara}} \\
\\
\textit{Software:} Matplotlib \citep{Hunter2007}, SciPy \citep{Virtanen2020}, NumPy \citep{Harris2020}, seaborn \citep{Waskom2021}, and Jupyter Notebook \citep{Kluyver2016}. 

\end{acknowledgements} 

%%%%%%%%%%%%%%%%%%%%%%%%%%%%%%%%%%%%%%%%%%%%%%%%%%%%%%%%%%%%%%%%%%
% \section*{Data availability}

% \mbox{\url{https://asmitabhandare.github.io/##tara}}

%%%%%%%%%%%%%%%%%%%%%%%%%%%%%%%%%%%%%%%%%%%%%%%%%%%%%%%%%%%%%%%%%%
% \section*{Software}

% Matplotlib \citep{Hunter2007}, SciPy \citep{Virtanen2020}, NumPy \citep{Harris2020}, seaborn \citep{Waskom2021} and Jupyter Notebook \citep{Kluyver2016}. 

%%%%%%%%%%%%%%%%%%%%%%%%%%%%%%%%%%%%%%%%%%%%%%%%%%%%%%%%%%%%%%%%%%
\bibliographystyle{yahapj}

\bibliography{Bibliography}

%%%%%%%%%%%%%%%%%%%%%%%%%%%%%%%%%%%%%%%%%%%%%%%%%%%%%%%%%%%%%%%%%%
\begin{appendix}

\section{Dependence on initial pre-stellar core properties}

\subsection[Disc properties for the fiducial 1-Msol case]{Disc properties for the fiducial 1~$M_{\odot}$ case}
\label{sec:discproperties}

Protostellar disc properties are compared for cases of a 1~$M_{\odot}$ collapse with different initial ratio of rotational-to-gravitational energy (or rotation rate) and pre-stellar core size until they reach the same central density of 0.01~$\mathrm{g~{cm^{-3}}}$. The evolution is shown from the beginning ($t=0$) of transition of the first hydrostatic core into a rotationally supported disc. For the same initial conditions except the ratio of rotational-to-gravitational energy (compare red and yellow lines), the disc radius, lifetime, and the final enclosed disc mass are larger for a faster rotating pre-stellar core in agreement with results from \citet{Machida2010}, \citet{Bate2011}, and \citet{Machida2011c}. 

A larger initial pre-stellar core with the same ratio of rotational-to-gravitational energy (but different rotation rate) also leads to longer lived and larger disc (compare blue and yellow lines). However, a slower collapse for an initially larger pre-stellar core leads to a slower build up of disc mass. These comparisons emphasise the strong dependence of protostellar disc properties on the initial physical conditions. 

\begin{figure}[tp]
    \centering
    \includegraphics[width=0.9\linewidth]{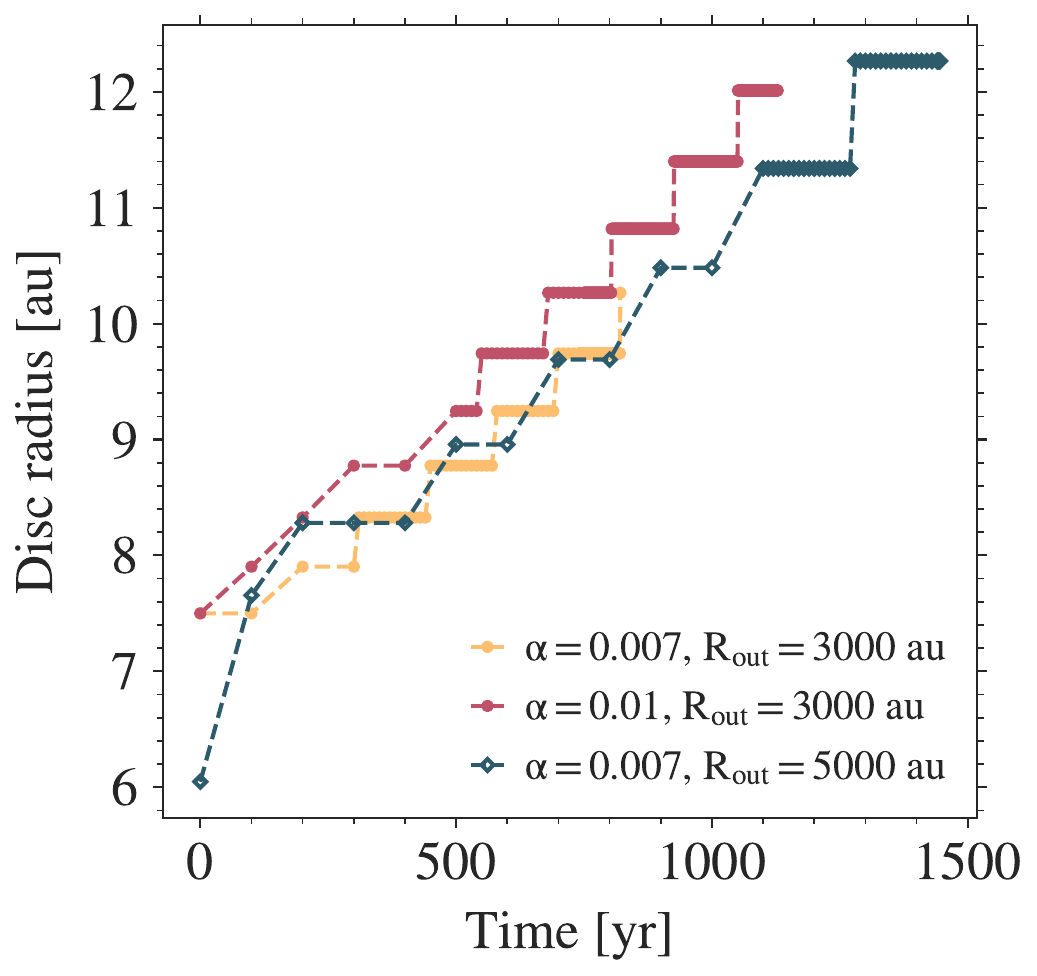} \\ \vspace{0.5cm}
    \includegraphics[width=0.9\linewidth]{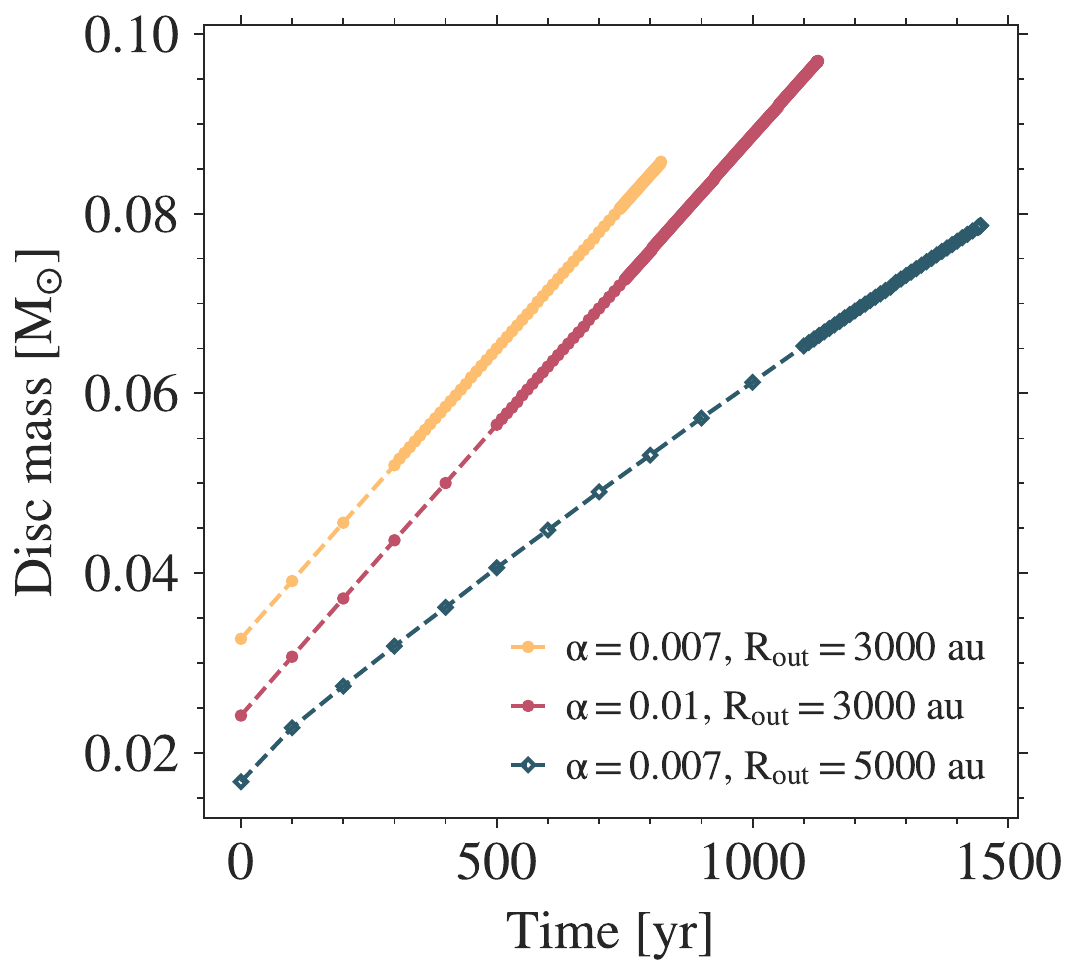}
    \caption{Time evolution of the nested protostellar disc radius (top) and enclosed mass (bottom) formed during the collapse of a 1~$M_{\odot}$ pre-stellar core. Comparisons between two different initial rotation rates of \mbox{$\alpha = $ 0.007} and 0.01 and an initial pre-stellar core radius of 3000~au vs 5000~au are shown. Time $t=0$ indicates the beginning of transition of the first hydrostatic core into a rotationally supported disc. The three cases are compared until central density reaches 0.01 $\mathrm{g~{cm^{-3}}}$ and before the launch of an outflow.}
    %onset at 18400 for fiducial case
    \label{fig:discradiusvstime}
\end{figure}

\subsection{Dependence on initial rotation rate}
\label{sec:rotation}

\begin{figure*}[t]
    \centering
    \includegraphics[width=0.76\linewidth]{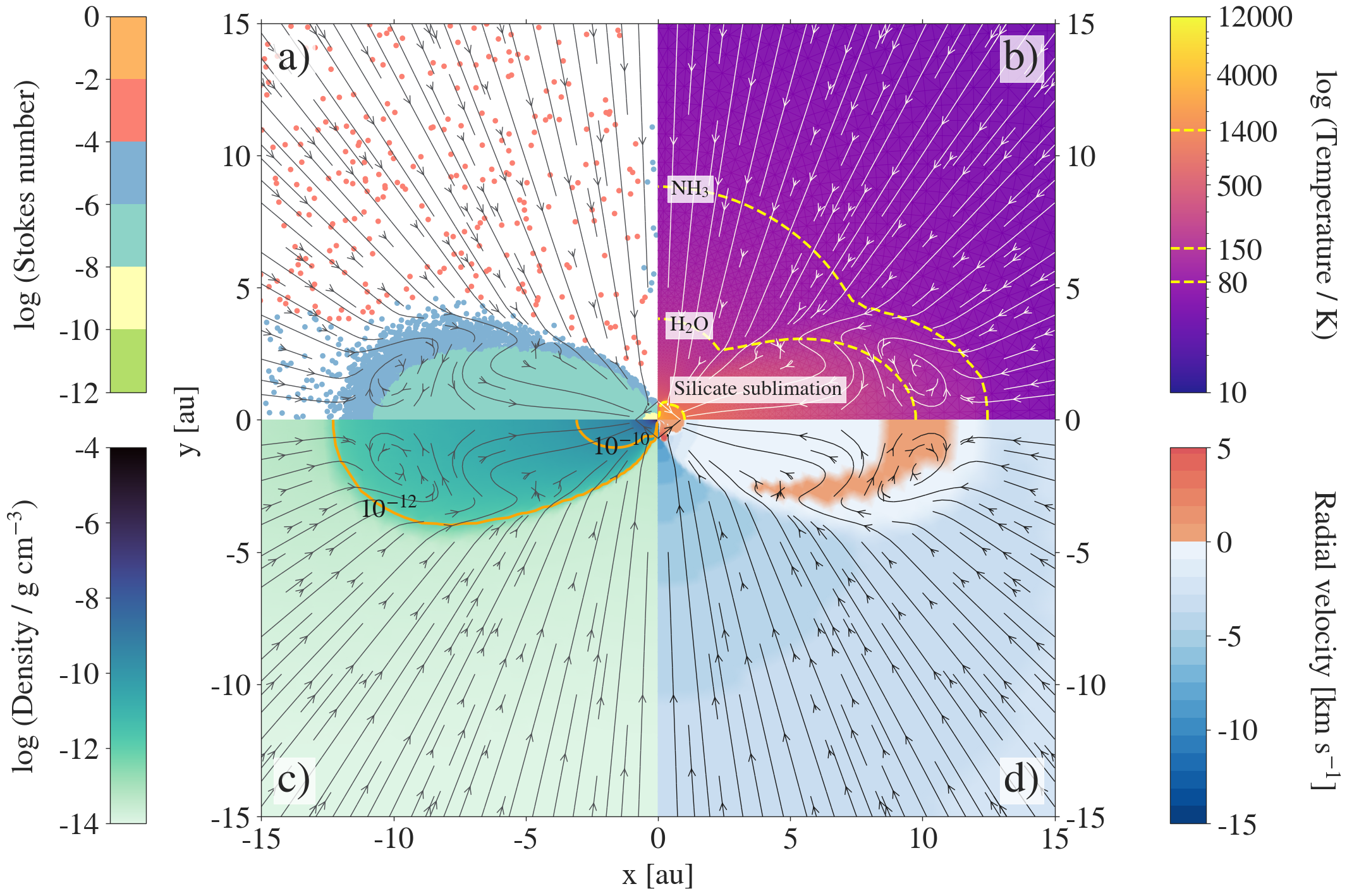}
    \caption{2D view of a first hydrostatic core evolved into a rotationally supported disc at 1427~years after its formation as a result of the collapse of a 1 $M_{\odot}$ pre-stellar core with an initial rotation rate of \mbox{$\Omega_\mathrm{0} = 2.099 \times 10^{-13}$ rad $\mathrm{s}^{-1}$} (same time snapshot as Fig.~\ref{fig:HD-seconddisc001}). The dust size is fixed to a constant value of 1 $\muup$m. The four panels show the \mbox{\bf a)}~Stokes number, \mbox{\bf b)}~gas temperature, \mbox{\bf c)}~gas density, and \mbox{\bf d)}~radial gas velocity within the inner 15~au of the 3000~au collapsing pre-stellar core. The gas velocity streamlines indicate the material falling onto the disc and the mixing within the meridional flow located at around 10~au.} %t0 = 18000
    \label{fig:HDdisc001}
\end{figure*}

\begin{figure*}[b]
    \centering
    \includegraphics[width=0.76\linewidth]{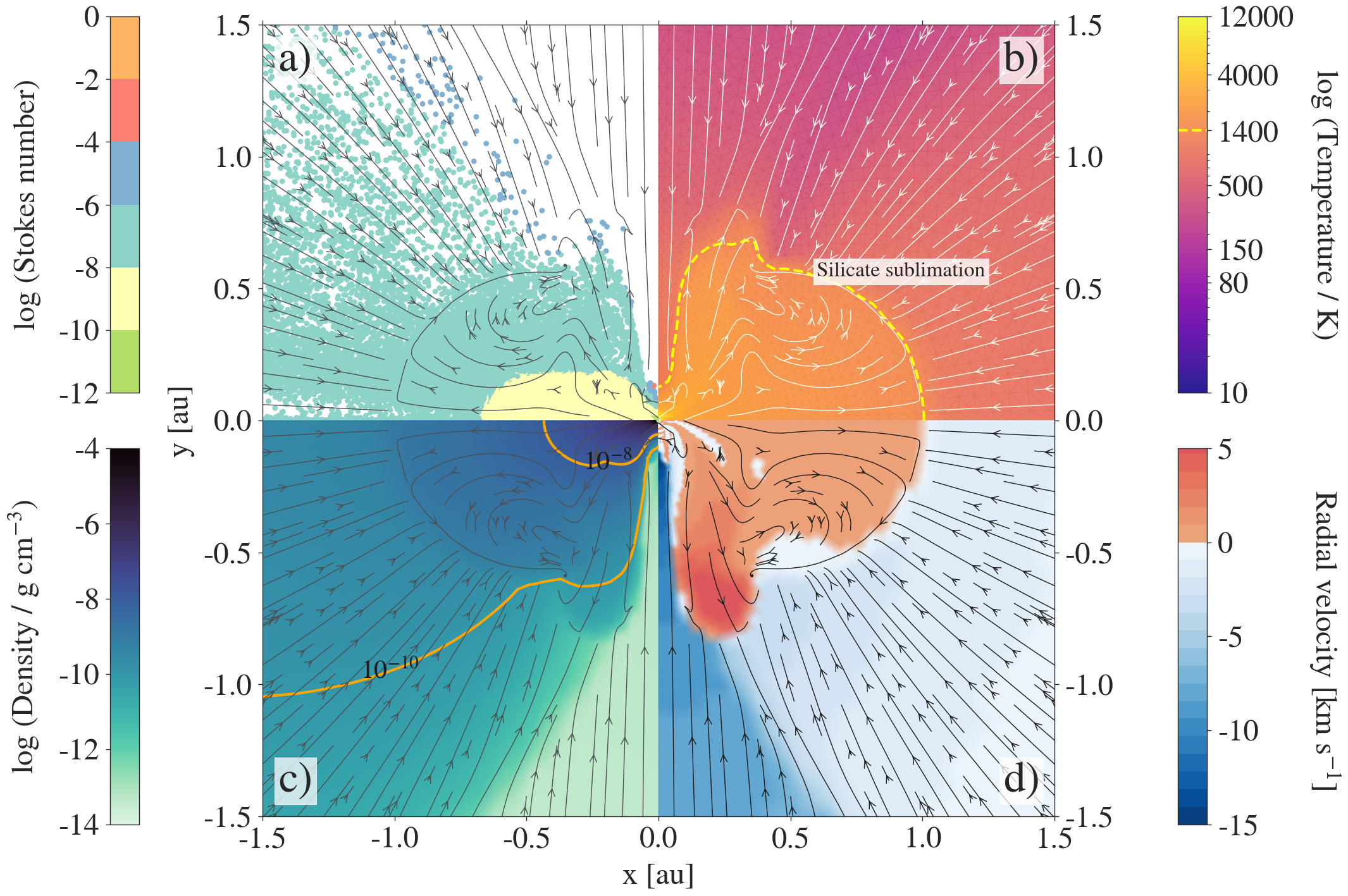}
    \caption{2D view of a pseudo-disc forming around the second core at three years after its formation as a result of the same initial conditions and at the same time snapshot as Fig.~\ref{fig:HDdisc001}. The gas velocity streamlines indicate material falling onto the pseudo-disc, circulation within the pseudo-disc via a turbulent flow, and outflowing gas from the vicinity of the unresolved protostar.} %t0 = 18000
    \label{fig:HD-seconddisc001}
\end{figure*}

In order to investigate the dependence on the initial rotation rate we performed a simulation with the same initial conditions as for the fiducial case discussed in Sect.~\ref{sec:hydrodisc} but with a faster initial rotation \mbox{$\Omega_\mathrm{0} = 2.099 \times 10^{-13}$ rad $\mathrm{s}^{-1}$}. The ratio of rotational-to-gravitational energy was fixed to 0.01. Dust size was fixed to 1~$\muup$m throughout the evolution. Figure~\ref{fig:HDdisc001} shows a 2D view of the first hydrostatic core evolved into a rotationally supported disc at 1427~years after the onset of first core formed during a 1~$M_{\odot}$ collapse. Both Fig.~\ref{fig:HDdisc007} and Fig.~\ref{fig:HDdisc001} are shown at the snapshot when the central density reaches $\approx$~0.01 $\mathrm{g~{cm^{-3}}}$. An initially faster rotating pre-stellar core leads to a comparatively flattened and slightly larger disc. The disc radius and mass for the two runs are compared in the previous Appendix~\ref{sec:discproperties}. However, different gas and dust properties within the first core disc in both cases are similar, with the meridional flow seen at the outer edge. Figure~\ref{fig:HD-seconddisc001} is a further zoom-in within 1.5~au at the same time snapshot as Fig.~\ref{fig:HDdisc001} magnifying the turbulent flow within the pseudo-disc and the onset of a thermal pressure driven outflow. We do not pursue a further evolution of the outflow but expect a similar behaviour as in the fiducial case.

\subsection{Dependence on initial pre-stellar core radius}
\label{sec:Outerradius}

All the simulations discussed in the main text were performed with an outer pre-stellar core radius of 3000~au. Here we discuss the effects of initiating a 1~$M_{\odot}$ collapse with a 5000~au outer radius but setting all other cloud core properties to be the same as the fiducial case (see Sect.~\ref{sec:hydrodisc}). The radial grid extends from $10^{-2}$~au to 5000~au and consists of 167 logarithmically spaced cells. We used 20 uniformly spaced cells in the polar direction stretching from the pole ($\theta = 0^\circ$) to the midplane ($\theta = 90^\circ$). In this case, we used 50 dust particles per cell (167000 particles in total) with a fixed size of 1 $\muup$m. 

\begin{figure*}[t]
    \centering
    \includegraphics[width=0.76\linewidth]{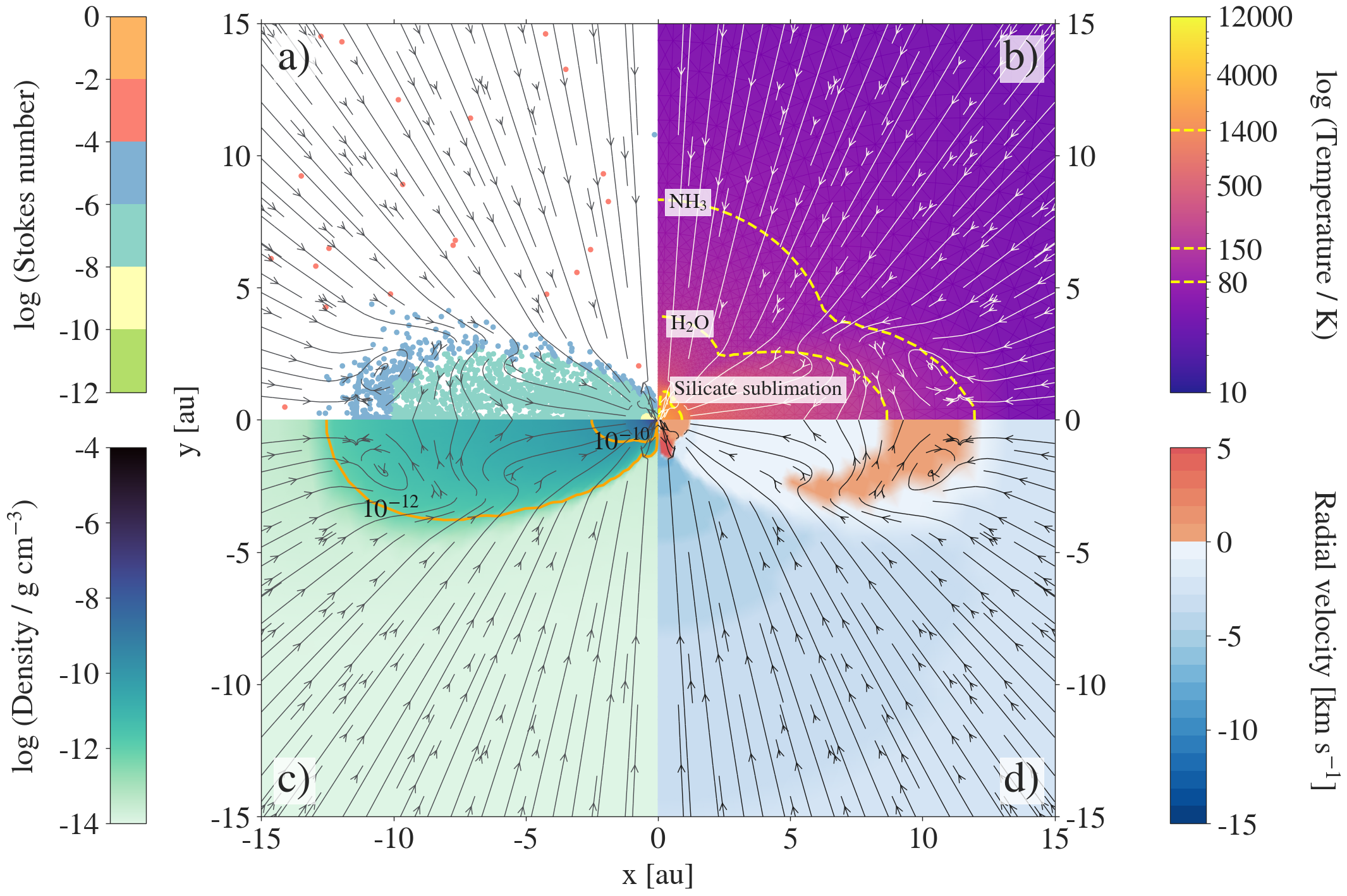}
    \caption{2D view of a first hydrostatic core evolved into a rotationally supported disc at 1646~years after its formation as a result of the collapse of a 1 $M_{\odot}$ pre-stellar core with an initial rotation rate of \mbox{$\Omega_\mathrm{0} = 8.16 \times 10^{-14}$ rad $\mathrm{s}^{-1}$} or \mbox{$E_\mathrm{rot} / E_\mathrm{grav} =$ 0.007} (same time snapshot as Fig.~\ref{fig:HD-seconddisc007R5000}). The dust size is fixed to a constant value of 1 $\muup$m. The four panels show the \mbox{\bf a)}~Stokes number, \mbox{\bf b)}~gas temperature, \mbox{\bf c)}~gas density, and \mbox{\bf d)}~radial gas velocity within the inner 15~au of the 5000~au collapsing pre-stellar core. The gas velocity streamlines indicate the material falling onto the disc and the mixing within the meridional flow located at around 10~au.} 
    %t0 = 48000, 50 particles per cell  
    \label{fig:HDdisc007R5000}
\end{figure*}

\begin{figure*}[b]
    \centering
    \includegraphics[width=0.76\linewidth]{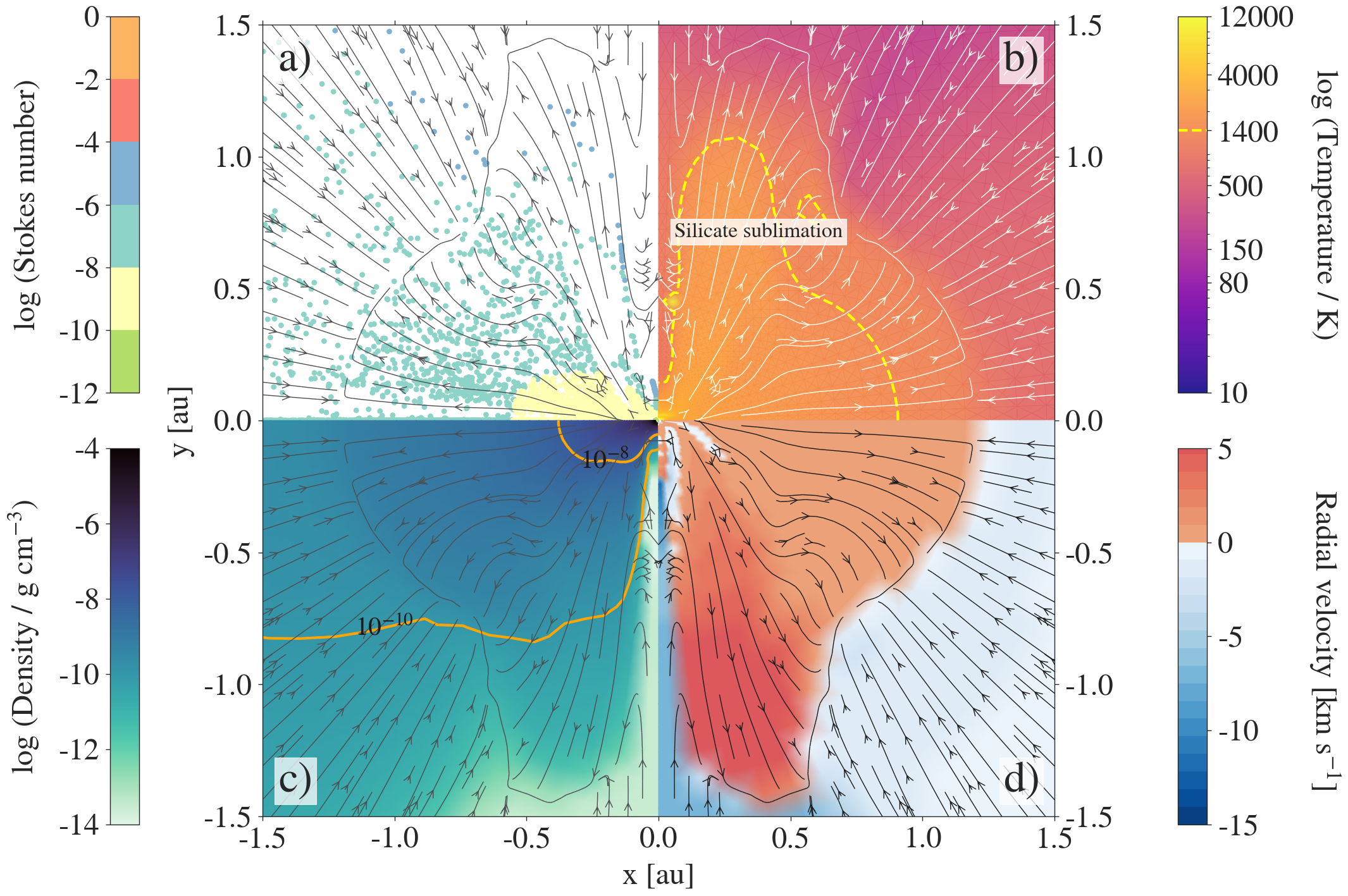}
    \caption{2D view of a pseudo-disc forming around the second core at three years after its onset as a result of the same initial conditions and at the same time snapshot as Fig.~\ref{fig:HDdisc007R5000}. The gas velocity streamlines indicate material falling onto the pseudo-disc and outward transport within the pseudo-disc, due to the outflowing gas from the vicinity of the unresolved protostar. At this stage the vortical flow within the pseudo-disc has already been destroyed by the outflow. } 
    %t0 = 48000, 50 particles per cell  
    \label{fig:HD-seconddisc007R5000}
\end{figure*}

Figure~\ref{fig:HDdisc007R5000} shows different quantities at 1646~years after the onset of the first core. The time snapshot is selected such that the central density is 0.01~$\mathrm{g~{cm^{-3}}}$ to compare with the fiducial case shown in Fig.~\ref{fig:HDdisc007}. Similar meridional gas flow acting as a dust trap at the outer disc edge is visualised in the radial velocity panel. Figure~\ref{fig:discradiusvstime} highlights the dependence of the disc radius, mass, and lifetime on the initial pre-stellar core size. A larger initial pre-stellar core leads to a slower collapse due to a longer free-fall time. The resulting protostellar disc is relatively larger and flattened. Figure~\ref{fig:HD-seconddisc007R5000} is a further zoom-in within 1.5~au at the same time snapshot as Fig.~\ref{fig:HDdisc007R5000} magnifying the second pseudo-disc and the thermal pressure driven outflow. The vortical feature within the inner disc is already destroyed by the outflow at this time snapshot. We do not pursue a further evolution of the outflow but expect a similar behaviour as in the fiducial case. 

\clearpage

%%%%%%%%%%%%%%%%%%%%%%%%%%%%%%%%%%%%%%%%%%%%%%%%%%%%%%%%%%%%%%%%%%
\section{Dust temporal evolution}
\label{sec:tracks-allsizes}

\begin{figure*}[!b]
    \centering
    \includegraphics[width=0.8\linewidth]{particle_radius-tempvstime_multisize-1micron_200ppc_subplot_discs+outflow.png} \\
    \vspace{0.5cm}
    \includegraphics[width=0.8\linewidth]{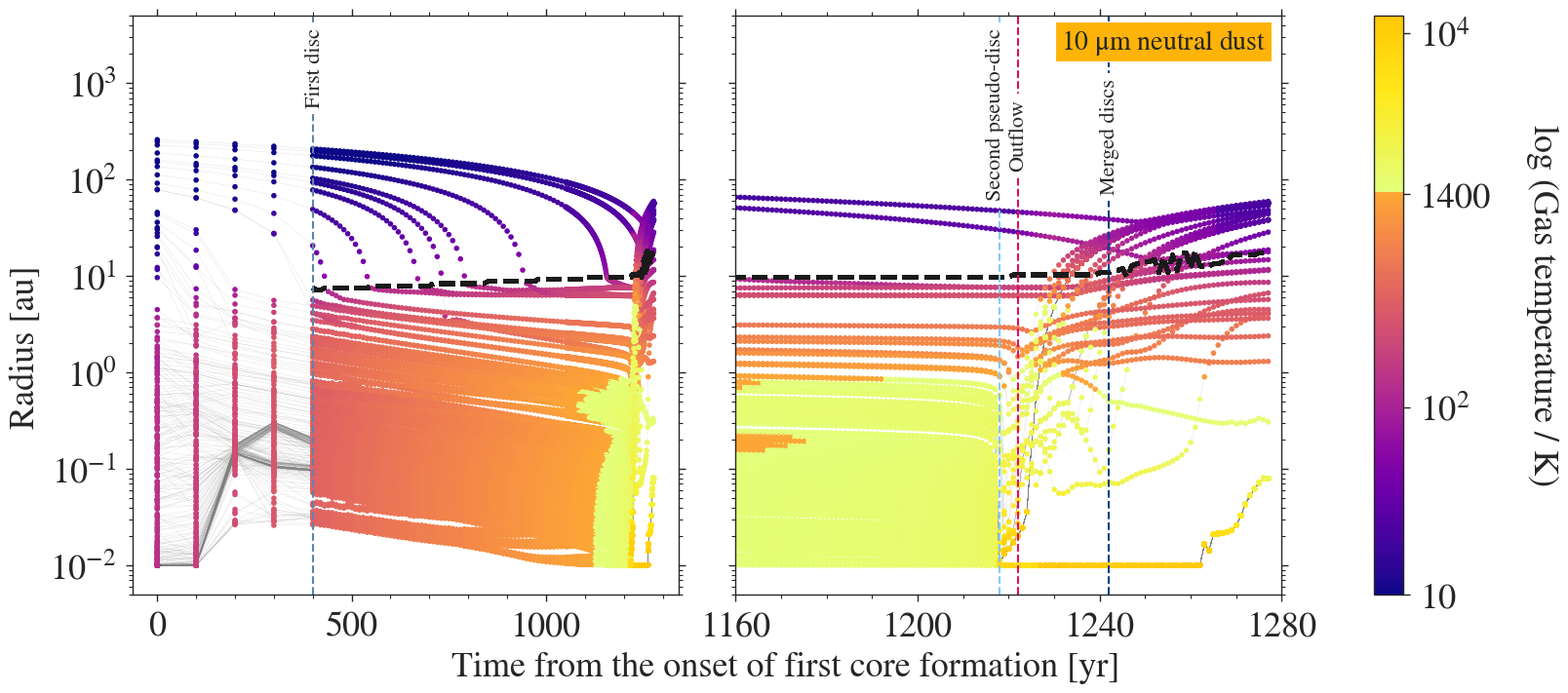} \\
    \vspace{0.5cm}
    \includegraphics[width=0.8\linewidth]{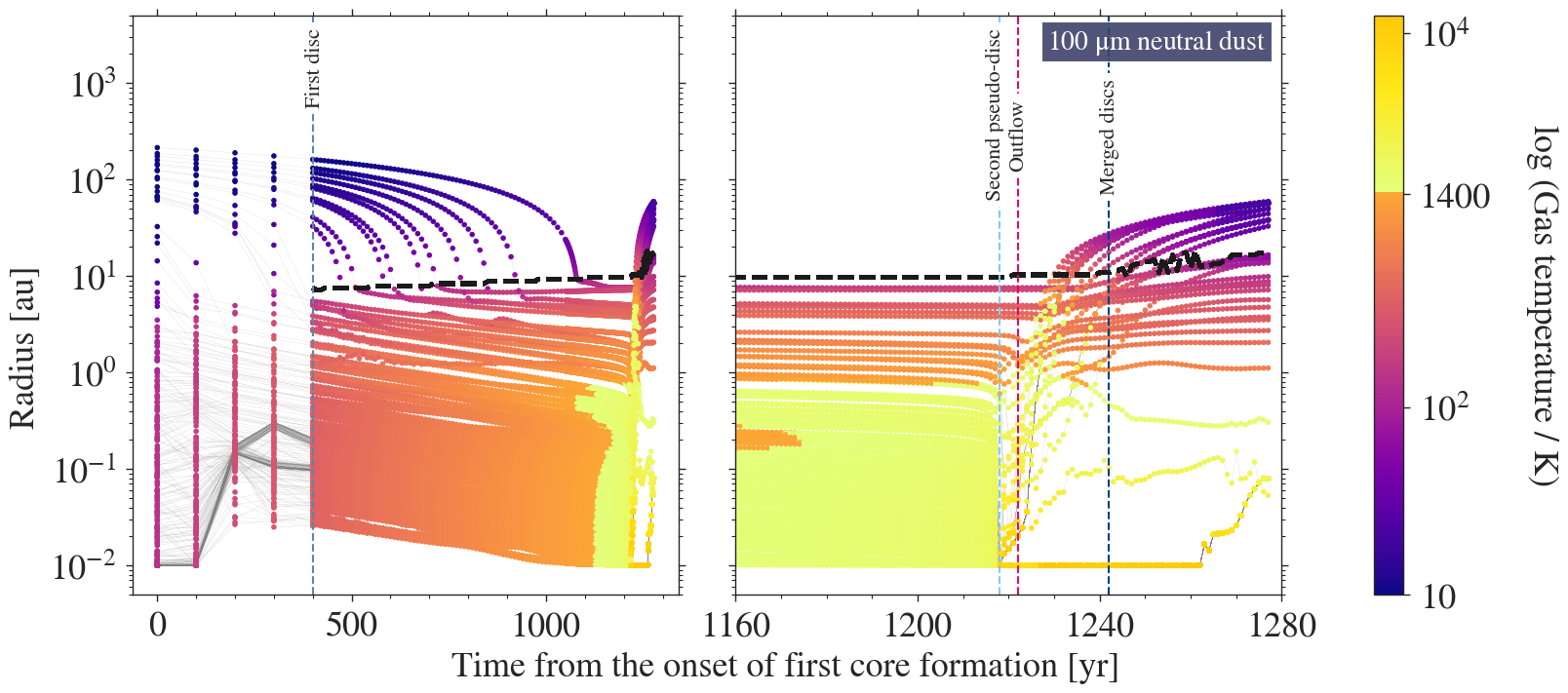}
    \caption{Tracks for a selected sample of 1~$\muup$m (top), 10~$\muup$m (middle), and 100~$\muup$m (bottom) dust that end up within the two discs during the 1~$M_{\odot}$ pre-stellar core collapse. The colourmap is split below and above silicate sublimation at 1400~K and indicates the gas temperature at dust location. The vertical lines indicate the formation times of the two discs, and the outflow, and merger of the discs from the onset of the first core formation ($t=0$). The horizontal dashed black line marks the first core disc or merged disc radius defined using the conditions listed in Sect.~\ref{sec:hydrodisc}. The temporal evolution of the dust provides a clear indication of thermal reprocessing within the very young disc.}
    \label{fig:particle_radius-tempvstime-allsizes}
\end{figure*}

In Figure~\ref{fig:particle_radius-tempvstime-allsizes}, we display gas temperature at the location of a selected sample of 10 and 100~$\muup$m dust for comparison with the 1~$\muup$m case previously also shown in Fig.~\ref{fig:particle_radius-tempvstime}. The strong gas--dust coupling within the dense disc leads to a similar temporal evolution for the three different dust sizes. We find regions (see yellow part of the tracks) where some dust particles could undergo sublimation/evaporation. This will be investigated in our follow-up work. Note that although dust evaporation is not included in our current treatment for Lagrangian particles, it is considered in the opacity tables used in the radiative transfer module.

\end{appendix}

\end{document}